\documentclass[11pt]{article}

\usepackage[letterpaper,margin=1in]{geometry}

\usepackage[T1]{fontenc}
\usepackage{lmodern}
\usepackage{amsthm,amsmath,amsfonts,amssymb}
\usepackage{mathtools}

\usepackage[authoryear,round]{natbib}

\usepackage[protrusion=true,expansion=false]{microtype}
\usepackage{url}

\usepackage{graphicx}
\usepackage[font=small]{caption}
\usepackage{subcaption}

\usepackage{booktabs}
\usepackage{multirow}
\usepackage{placeins}
\usepackage{tabularx}
\newcolumntype{Y}{>{\centering\arraybackslash}X}

\usepackage[group-separator={,},group-minimum-digits={3}]{siunitx}

\usepackage[dvipsnames]{xcolor}

\usepackage{dsfont}

\usepackage[colorlinks,citecolor=blue,urlcolor=blue,linkcolor=blue]{hyperref}

\usepackage[capitalize,nameinlink,noabbrev]{cleveref}

\theoremstyle{plain}

\newtheorem{theorem}{Theorem}[section]

\theoremstyle{definition}

\theoremstyle{remark}

\newcommand{\relmiddle}[1]{\mathrel{}\middle#1\mathrel{}}
\DeclareMathOperator*{\argmin}{\arg\!\min}

\theoremstyle{definition}
\newtheorem{condition}{Condition}

\theoremstyle{plain}
\newtheorem{proposition}{Proposition}

\crefname{condition}{Condition}{Conditions}

\begin{document}

\title{Regularized Exponentially Tilted Empirical Likelihood for Bayesian Inference}

\author{%
 Eunseop Kim\thanks{\texttt{markean@pm.me}} \quad
 Steven N.~MacEachern\thanks{\texttt{snm@stat.osu.edu}} \quad
 Mario Peruggia\thanks{\texttt{peruggia@stat.osu.edu}} \\[4pt]
 Department of Statistics, The Ohio State University
}
\date{}

\maketitle

\begin{abstract}
Bayesian inference with empirical likelihood faces a challenge as the posterior domain is a proper subset of the original parameter space due to the convex hull constraint.
We propose a regularized exponentially tilted empirical likelihood to address this issue.
Our method removes the convex hull constraint using a novel regularization technique, incorporating a continuous exponential family distribution to satisfy a Kullback--Leibler divergence criterion.
The regularization arises as a limiting procedure where pseudo-data are added to the formulation of exponentially tilted empirical likelihood in a structured fashion.
We show that this regularized exponentially tilted empirical likelihood retains certain desirable asymptotic properties with improved finite-sample performance.
Simulation and data analysis demonstrate that the proposed method provides a suitable pseudo-likelihood for Bayesian inference.
\end{abstract}

\medskip
\noindent\textbf{Keywords:} Bernstein--von Mises theorem; convex hull; entropy balancing; Kullback--Leibler divergence; pseudo-data.

\section{Introduction}\label{sec:intro}
Statistical models defined through estimating equations and moment conditions allow semiparametric inferences on quantities of interest without distributional assumptions.
Empirical likelihood (EL) \citep{owen1988empirical, qin1994empirical}, a popular approach in the frequentist setting, enables nonparametric but still likelihood-style inference.
It shares many desirable properties with parametric likelihood, exhibiting Wilks' phenomenon under mild conditions and allowing for the Bartlett correction \citep{diciccio1991empirical}.
EL is a member of the class of generalized EL \citep{smith1997alternative, newey2004higher}, which includes the exponential tilting of \citet{efron1981nonparametric}.
\citet{newey2004higher} showed a duality between generalized EL and the class of minimum discrepancy methods \citep{cressie1984multinomial,corcoran1998bartlett}.
In this context, EL is formulated by finding a distribution supported on the sample that minimizes the Kullback--Leibler divergence to the empirical distribution, subject to moment constraints.
Exponentially tilted empirical likelihood (ETEL) \citep{efron1981nonparametric, jing1996exponential,schennach2005bayesian} is obtained by combining exponential tilting and EL, which minimizes the reverse Kullback--Leibler divergence.

Bayesian methods are fundamentally based on probability, with inference proceeding from the prior distribution to the posterior distribution via conditioning on the observed data.
Bayesian analysis of EL poses a challenge, as posterior inference via Bayes' theorem requires a complete specification of the sampling distribution or the likelihood function.
\citet{lazar2003bayesian} proposed using EL as a replacement for the likelihood function in Bayesian inference.
\citet{schennach2005bayesian} strengthened the case by showing that ETEL arises as the limit of nonparametric Bayesian procedures with a particular type of prior favoring entropy-maximizing distributions.
\citet{chib2018bayesian} established a Bernstein--von Mises theorem for the Bayesian ETEL posterior distribution.
Bayesian versions of EL and ETEL place a prior distribution on a finite number of features of a nonparametric (and hence infinite-dimensional) distribution and regard the remainder of the distribution itself as a nuisance parameter.
The lack of a full probability model prevents one from integrating over the nuisance parameter.
EL and ETEL replace the integration with a maximization, and this replacement produces artifacts that clash with known properties that all Bayesian methods must have.

The most striking departure from Bayesian behavior is the zeroing out of regions of parameter space as one moves from a prior distribution to a posterior distribution, with the expectation that, as more data are collected, the zeroed-out regions will reappear and be assigned positive probability.
The issue is caused by the convex hull constraint or the empty set problem \citep{grendar2009empty}, and thus both EL and ETEL are only defined on a proper subset of the parameter space.
The zeroed-out regions concern the main parameters of interest---those that are represented by the estimating equations that give rise to EL and ETEL.
The regions and behavior follow from the convex hull constraint.
This is conceptually unsatisfactory as, with a larger sample size, the convex hull may expand and the likelihood become positive.
Additionally, as the restricted posterior domain is often nonconvex \citep{chaudhuri2017hamiltonian}, a more sophisticated posterior sampling scheme may be needed to fit the model.

Various adjustments have been suggested \citep{bartolucci2007penalized,chen2008adjusted,tsao2013empirical} to address the convex hull constraint.
Most relevant to our work, \citet{chen2008adjusted} proposed the adjusted empirical likelihood (AEL), which adds a pseudo-observation in a way that satisfies the convex hull constraint for a given parameter value.
This approach has been further developed by \citet{emerson2009calibration} and \citet{liu2010adjusted}, and has been adapted for ETEL by \citet{zhu2009adjusted} as the adjusted exponentially tilted empirical likelihood (AETEL).
However, these adjustments inevitably rely on pseudo-data that vary with the parameter, introducing a certain degree of irregularity and deviating from Bayesian methods (as well as most other statistical methods).

In this paper, we propose a method to address the convex hull constraint for Bayesian ETEL.
While previous proposals have primarily focused on EL and frequentist inference, our approach builds upon the AEL framework while introducing two notable distinctions.
First, we extend the method to accommodate multiple pseudo-observations with fractional weights, combining the logic of weighted EL \citep{glenn2007weighted} with an entropy balancing scheme \citep{hainmueller2012entropy}.
Second, by passing to the limit of these pseudo-observations, we move beyond the need to satisfy the convex hull constraint with finite data.
Instead, our formulation induces a form of regularization that ensures the dual optimization problem is well-posed with a unique, finite solution for all parameter values.
Consequently, the resulting likelihood is free from the constraint even when the observed data do not satisfy it.
Our method's main contributions encompass the following: (i) addressing the convex hull constraint for ETEL while retaining desirable asymptotic properties; (ii) enhancing stability and robustness of small-sample performance compared to existing methods; (iii) providing flexibility in Bayesian modeling and allowing one to incorporate a novel form of prior information.

This paper is organized as follows.
In \cref{sec:preliminary}, we introduce the notation used in the paper and provide a brief overview of ETEL.
In \cref{sec:wetel}, we propose a weighted version of ETEL that incorporates fractional pseudo-data with the maximum entropy reweighting scheme.
In \cref{sec:retel}, we propose inducing regularization on the formulation of ETEL, exploring two equivalent approaches: (i) a limiting procedure with fractional pseudo-data and (ii) direct incorporation of a continuous exponential family distribution in the minimization of the Kullback--Leibler divergence.
We derive asymptotic properties of the proposed method.
In \cref{sec:simulations}, we evaluate the performance of the methods through simulation studies.
In \cref{sec:application}, we present applications to datasets on median income for four-person families and Roman-era Egyptian lifespan.
Finally, we conclude with a discussion of the directions for future research in \cref{sec:discussion}.

\section{Preliminaries}\label{sec:preliminary}
Let \({\mathcal{D}_n} = {\{X_i\}_{i = 1}^n}\) denote independent and identically distributed \({d_x}\)-dimensional observations from a complete probability space \({(\mathcal{X}, \mathcal{F}, P)}\) satisfying the moment condition:
\begin{equation*}
E_P\{g(X_i, \theta)\} = 0,
\end{equation*}
where \({g}: {\mathbb{R}^{d_x} \times \Theta} \to {\mathbb{R}^p}\) is an estimating function with the true parameter value \({\theta_0} \in {\Theta} \subset \mathbb{R}^p\).
Consider a discrete probability distribution \({P_0}\) that is absolutely continuous with respect to the empirical distribution \({P_n}\).
The Kullback--Leibler divergence from \({P_n}\) to \({P_0}\) is
\begin{equation*}
{D_{KL}(P_0\ \Vert\ P_n) 
= \sum_{i = 1}^n p_i \log (np_i)},
\end{equation*}
where \({p_i}\) are probabilities attached to the observations by \({P_0}\).
By minimizing the Kullback--Leibler divergence subject to the constraints in the moment condition, we obtain a unique set of \({p_i}\) and the associated distribution.
For a given \({\theta}\), the maximization problem
\begin{equation*}
\max_{p_1, \dots, p_n}\left\{
\sum_{i = 1}^n \left(-p_i \log \left(np_i\right)\right)
\relmiddle | \sum_{i = 1}^n p_i g\left(X_i, \theta\right) = 0,\quad p_i \geq 0,\quad \sum_{i = 1}^n p_i = 1
\right\}
\end{equation*}
yields a unique solution \({(p_1(\theta), \dots, p_n(\theta))}\), and ETEL is defined as
\begin{equation*}
L_{ET}(\theta) =
\prod_{i = 1}^n p_i(\theta).
\end{equation*}
By applying the method of Lagrange multipliers, we obtain
\begin{equation*}
p_i\left(\theta\right) =
\frac{\exp\left({\lambda_{ET}}^\top g\left(X_i, \theta\right)\right)}
{\sum_{j = 1}^n \exp\left({\lambda_{ET}}^\top g\left(X_j, \theta\right)\right)},
\end{equation*}
where \({\lambda_{ET}} \equiv {\lambda_{ET}(\theta)}\) solves the equation \({n^{-1}\sum_{i = 1}^n \exp({\lambda}^\top g(X_i, \theta) )g(X_i, \theta) =
0}\).
The dual problem provides the solution:
\begin{equation*}
\lambda_{ET} =
\argmin_{\lambda \in \mathbb{R}^p} \sum_{i = 1}^n \exp(\lambda^\top g(X_i, \theta)).
\end{equation*}
By construction, any \({Z}\)-estimator \({\widehat{\theta}}\) that solves \({n^{-1}\sum_{i = 1}^ng(X_i, \theta)} = {0}\) maximizes ETEL \citep{yiu2020inference}.

In the Bayesian framework, ETEL can be used with a prior \({\pi(\theta)}\) to define the exponentially tilted posterior distribution \(\pi(\theta \mid \mathcal{D}_n)
\propto
\pi(\theta)
L_{ET}(\theta)\).
\citet{schennach2005bayesian} showed that when all observations are distinct, \({L_{ET}(\theta)}\) can be obtained as the limit of a nonparametric Bayesian procedure.
Her procedure involves assigning a mixture of uniform densities as a nonparametric prior on \({P}\) that satisfies the moment condition and then marginalizing over the nuisance parameters.
The convex hull constraint serves as the implicit constraint in the primal optimization problem, indicating that the interior of the convex hull of \({\{g(X_i, \theta)\}_{i = 1}^n}\), denoted by \({\textnormal{Conv}_n(\theta)}\), must contain \({0}\).
Consequently, the (posterior) domain of ETEL is restricted to \({\Theta_n} =
{\{\theta \in \Theta: 0 \in \textnormal{Conv}_n(\theta)\}}\) so that even a \({100\%}\) credible set may fail to contain \({\theta_0}\).
In general, \({\Theta_n}\) is nonconvex and is challenging to identify.
Simulation methods to fit the models, such as Markov chain Monte Carlo or Hamiltonian Monte Carlo, require long runs and may or may not be effective \citep{chaudhuri2017hamiltonian,yu2024variational}, leading to potential undercoverage issues and unreliable inference.

To address the convex hull constraint for EL, the AEL approach introduces a pseudo-observation that depends on \({\theta}\).
Here and throughout, we use \({g_i(\theta)} = {g(X_i, \theta)}\), \({i} = {1, \dots, n}\), for notational convenience.
The pseudo-observation has
\begin{equation}\label{eq:ael pseudo observation}
g_{n + 1}(\theta)
=
-\frac{a_n}{n}\sum_{i = 1}^n g_i(\theta),
\end{equation}
where \({a_n} > {0}\).
The properties of the sequence \({a_n}\) are used to establish asymptotic results.
The addition of \({g_{n + 1}(\theta)}\) ensures that the convex hull constraint is satisfied for each \({\theta} \in {\Theta}\).

\citet{emerson2009calibration} and \citet{liu2010adjusted} proposed adding two pseudo-observations to improve the coverage accuracy of confidence regions obtained from AEL.
\citet{yu2024variational} established a Bernstein--von Mises theorem for Bayesian AEL.
While the AEL approach is directly applicable to ETEL for fixing the convex hull constraint for a particular \({\theta}\), it may introduce irregularities throughout \({\Theta}\) in the resulting posterior distribution when applied to Bayesian analysis, since it involves a preliminary entropy maximization step in constructing the likelihood function.
Incorporating one or two pseudo-observations, specific to each \({\theta}\), and treating them on par with actual observations may contribute to these irregularities.

\section{Weighted Exponentially Tilted Empirical Likelihood with Fractional Pseudo-Data}\label{sec:wetel}
As an initial step towards addressing the convex hull constraint for ETEL and establishing a connection with the regularization method discussed in \cref{sec:retel}, we propose a weighted exponentially tilted empirical likelihood (WETEL) approach with fractional pseudo-data.
Our approach extends the AEL method by incorporating multiple pseudo-observations, in combination with the entropy balancing scheme of \citet{hainmueller2012entropy}.
Entropy balancing is a data preprocessing technique used to achieve covariate balance in observational studies with a binary treatment and in survey sampling.
The preprocessing step involves applying a maximum-entropy reweighting scheme to ensure that the reweighted data satisfy a set of moment conditions.
In the context of our framework, the pseudo-data can be seen as providing additional information for the analysis.

We introduce a fixed number, \({m} \in \mathbb{N}\), of pseudo-data denoted as \({g_{n + j}(\theta)} \in {\mathbb{R}^p}\) for \({j} = {1, \dots, m}\).
The use of the estimating function \({g}\) for the pseudo-data is for notational consistency.
Apart from their dependence on \({\theta}\), they need not necessarily be related to the observed data or estimating function.
At this stage, we do not discuss any specific strategy for creating the pseudo-data.
Instead, for our current purposes, we simply assume that the augmented data, comprising both the observed data and pseudo-data, satisfy the convex hull constraint.

Let \({w_i}\) be the base weight for the \({i}\)th observation in the augmented data, such that \({\sum_{i = 1}^N w_i} = {1}\), with \({N} = {n + m}\).
We consider the following maximum-entropy reweighting scheme:
\begin{equation*}
\max_{p_1, \dots, p_N}
\left\{
\sum_{i = 1}^N \left(-p_i \log \left(\frac{p_i}{w_i}\right)\right) \relmiddle |
\sum_{i = 1}^N p_i g_i\left(\theta\right) = 0,\quad p_i \geq 0,\quad \sum_{i = 1}^N p_i = 1
\right\}.
\end{equation*}
This scheme is equivalent to minimizing \({D_{KL}(P_0\ \Vert \ P_w)}\) subject to the constraints above, where \({P_w}\) is the weighted empirical distribution.
Both \({P_0}\) and \({P_w}\) are now supported on the augmented data.
The objective function is modified to account for the weights and pseudo-data, and the moment condition is matched by the augmented data.
The method of Lagrange multipliers yields
\begin{equation*}
p_i\left(\theta\right) =
\frac{w_i\exp\left({\lambda_{WET}}^\top g_i\left(\theta\right)\right)}
{\sum_{j = 1}^N w_j\exp\left({\lambda_{WET}}^\top g_j\left(\theta\right)\right)},
\end{equation*}
where \({\lambda_{WET}} =
{\argmin_{\lambda \in \mathbb{R}^p}
\sum_{i = 1}^N w_i\exp({\lambda}^\top g_i(\theta))}\).

Next, building upon the weighted EL approach proposed by \citet{glenn2007weighted}, we formulate the likelihood function as \({L_{WET}(\theta)} = {\prod_{i = 1}^N {p_i(\theta)}^{Nw_i}}\).
Based on the inequality \({\prod_{i = 1}^N {p_i(\theta)}^{Nw_i}} \leq {\prod_{i = 1}^N w_i^{Nw_i}}\) for any solution \({p_i(\theta)}\), the likelihood ratio function of WETEL can be defined as
\begin{equation*}
R_{WET}\left(\theta\right) = \prod_{i = 1}^N \left(\frac{p_i(\theta)}{w_i}\right)^{Nw_i}.
\end{equation*}
Consequently, the maximum WETEL estimator \({\widehat{\theta}_{w}}\) is obtained by solving \({\sum_{i = 1}^N w_i g_i(\theta)} = {0}\).
When using uniform weights with \(w_i = {1 / N}\), the resulting WETEL reduces to ETEL with the pseudo-data included.
However, in finite-sample settings, the size of \({m}\) relative to \({n}\), the pseudo-data specification, and the choice of weights can lead to substantial differences between WETEL and ETEL.
To prevent this, we treat all pseudo-data as a single observation and assign fractional weights to them.
Specifically, we set the weights as follows:
\begin{equation}\label{eq:fractional weights}
w_i =
\begin{dcases*}
\frac{1}{n + 1} & \(\left(i = 1, \dots, n\right)\),\\[1ex]
\frac{1}{m\left(n + 1\right)} & \(\left(i = n + 1, \dots, n + m\right)\).
\end{dcases*}
\end{equation}
This weight specification balances the contribution from the pseudo-data with the modified multiplier:
\begin{equation*}
\lambda_{WET}
=
\argmin_{\lambda \in \mathbb{R}^p}
\left\{
\sum_{i = 1}^n \exp({\lambda}^\top g_i(\theta))
+m^{-1}
\sum_{i = n + 1}^{n + m}
\exp
({\lambda}^\top g_i(\theta) )
\right\}.
\end{equation*}
Since WETEL is a generalization of ETEL with finite pseudo-data, it preserves the major asymptotic properties of ETEL as \({n} \to {\infty}\).
Let \({G} = {E_P\{\partial_{\theta}g_i(\theta_0)\}}\), \({V} = {E_P\{g_i(\theta_0){g_i(\theta_0)}^\top\}}\), and \({\Omega} = {({G}^\top{V}^{-1}G)^{-1}}\), where
\({\partial_{\theta}g_i(\theta_0)}\) denotes the Jacobian matrix of \({g_i(\theta)}\) evaluated at \({\theta_0}\).
Moreover, the Euclidean norm for vectors is denoted by \({\vert\cdot\vert}\), and the Frobenius norm for matrices is denoted by \({\Vert\cdot\Vert}\).
We also use \({N(\mu, \Sigma)}\) to represent a \({p}\)-dimensional multivariate normal distribution with mean \({\mu}\) and covariance matrix \({\Sigma}\), and \({\chi^2_p}\) to represent a chi-square distribution with \({p}\) degrees of freedom.
\begin{condition}\label{condition:parameter space}
The parameter space \({\Theta}\) is compact, with \({\theta_0}\) an interior point of \({\Theta}\) and the unique solution to \({E_P\{g_i(\theta)\}} = {0}\).
\end{condition}
\begin{condition}\label{condition:continuity}
With probability~\({1}\), \({g_i(\theta)}\) is continuous at each \({\theta} \in {\Theta}\) and continuously differentiable in a neighborhood \({\mathcal{N}}\) of \({\theta_0}\), and \({E_P\{\sup_{\theta \in \mathcal{N}} \Vert \partial_{\theta}g_i(\theta) \Vert\}} < {\infty}\).
\end{condition}
\begin{condition}\label{condition:full rank}
\({\textnormal{rank}({G})} = {\textnormal{rank}({V})} = {p}\).
\end{condition}
\begin{condition}\label{condition:uniform boundedness}
For some \({\alpha} > {3}\), \({E_P\{\sup_{\theta \in \Theta} \vert g_i(\theta) \vert^\alpha\}} < {\infty}\).
\end{condition}
These are standard regularity conditions used to study the asymptotic behavior of generalized EL.
We establish that the discrepancies between ETEL and WETEL, in terms of estimators and Lagrange multipliers, become asymptotically negligible.
\begin{proposition}\label{prop:negligible difference}
Under \cref{condition:parameter space,condition:continuity,condition:full rank,condition:uniform boundedness}, \({\widehat{\theta}_{w} - \widehat{\theta}} = {o_P(n^{-1/2})}\) and \({\lambda_{WET}(\theta_0)} - {\lambda_{ET}(\theta_0)} = {O_P(n^{-1})}\).
\end{proposition}
Consequently, WETEL shares with generalized EL first-order asymptotic properties.
\begin{theorem}
Under \cref{condition:parameter space,condition:continuity,condition:full rank,condition:uniform boundedness}, \({n^{1/2}(\widehat{\theta}_{w} - \theta_0)}\) converges in distribution to \({N(0, \Omega)}\) as \({n} \to {\infty}\), and \({-2\log R_{WET}(\theta_0)}\) converges in distribution to \({\chi^2_p}\) as \({n} \to {\infty}\).
\end{theorem}

\section{Regularized Exponentially Tilted Empirical Likelihood}\label{sec:retel}
When \({m}\) is fixed, the convex hull constraint issue may arise in WETEL, where \({0} \notin {\textnormal{Conv}_N(\theta)}\) for certain values of \({\theta}\), unless the pseudo-data are carefully specified.
Even if we adopt a strategy like the one in \eqref{eq:ael pseudo observation}, the limitation remains because the specification of pseudo-data, regardless of careful selection or the magnitude of \({m}\), still needs to depend on the observed data and parameter values.
In this sense, the pseudo-data approach can be viewed as an ad hoc solution that pragmatically addresses the issue but does not fully resolve the underlying challenge associated with a finite \({m}\).

In this section, we consider a procedure where \({m}\) tends to infinity, enabling the pseudo-data to represent a continuous distribution in the limit.
Since ETEL induces an exponential family of distributions supported on the data, a natural choice for the pseudo-data is a continuous exponential family distribution.
To accomplish this, we introduce an auxiliary random variable \({\widetilde{g}} \sim {N(\mu, \Sigma)}\) with known \({\mu}\) and \({\Sigma}\), where \({\Sigma}\) is assumed to be of full rank.
The pseudo-data \({\{\widetilde{g}_1, \widetilde{g}_2, \dots \}}\) may be selected as appropriate quantiles of \({N(\mu, \Sigma)}\), aiming to approximate the distribution as \({m}\) increases.
For the purposes of our discussion, we assume that the pseudo-data are independent samples from \({N(\mu, \Sigma)}\), while treating the sample size \({n}\) and the parameter \({\theta}\) as fixed.

Using the fractional weights in \eqref{eq:fractional weights}, we introduce a sequence of stochastic minimization problems for WETEL: \({\min_{\lambda \in \mathbb{R}^p}
c_m(\lambda)}
=
{\min_{\lambda \in \mathbb{R}^p}
 \{d_n(\theta, \lambda) +
p_m(\lambda)
\}}\), where \({d_n(\theta, \lambda)} = {\sum_{i = 1}^n \exp(\lambda^\top g_i(\theta))}\) and \({p_m(\lambda)} = {m^{-1}\sum_{j = 1}^m \exp(\lambda^\top \widetilde{g}_j)}\).
It follows from the independent sampling that \({p_m(\lambda)} \to {p(\lambda)}\) with probability~\({1}\) as \({m} \to {\infty}\), where \({p(\lambda)} = {\exp(\lambda^\top\mu + \lambda^\top \Sigma\lambda/2)}\) is the moment-generating function of \({\widetilde{g}}\).
This suggests directly considering the following minimization problem:
\begin{equation}\label{eq:RETEL_minimization}
\min_{\lambda \in \mathbb{R}^p}
c\left(\lambda\right)
=
\min_{\lambda \in \mathbb{R}^p}
\left\{d_n\left(\theta, \lambda\right) +
p\left(\lambda\right)
\right\},
\end{equation}
with the minimization performed after taking the limit.
Then, the sequence of minimization problems can be viewed as a discretization of the population version of the minimization problem.
Such a setting can be commonly found in applications of stochastic programming \citep{wets1974stochastic,dupavcova1992epi}, equipped with epi-convergence \citep{dupacova1988asymptotic,king1991epi,rockafellar2009variational}.
We refer to the method as regularized exponentially tilted empirical likelihood (RETEL) and define the corresponding multiplier \({\lambda_{RET} =
\argmin_{\lambda \in \mathbb{R}^p}
\{d_n(\theta, \lambda) +
p(\lambda)
\}}\).

\begin{figure}[!t]
\centering
\begin{subfigure}{0.495\linewidth}
\includegraphics[width=\linewidth]{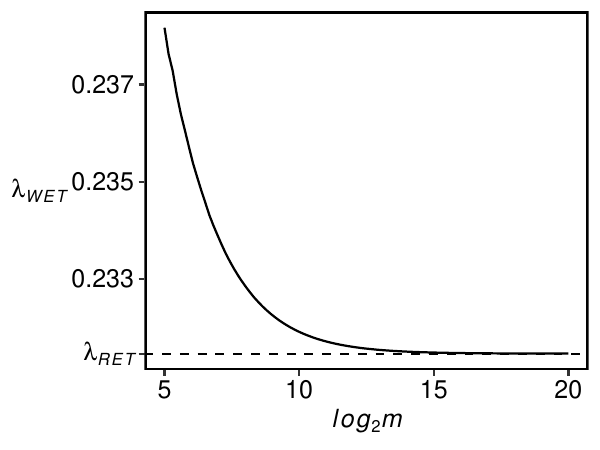}
\caption{}
\end{subfigure}
\hfill
\begin{subfigure}{0.495\linewidth}
\includegraphics[width=\linewidth]{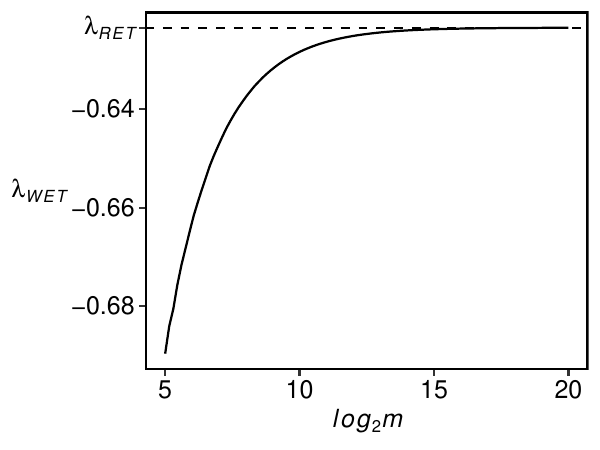}
\caption{}
\end{subfigure}
\caption{
Plots of \({\lambda_{WET}(\theta)}\) versus \({\log_2{m}}\) for the mean parameter \({\theta}\).
With two observations \({-2}\) and \({2}\) fixed, the convex hull constraint is satisfied at \({\theta} = {1}\) in (a) and violated at \({\theta} = {3}\) in (b).
For each \({m}\), the pseudo-data are generated as the \({k/(m + 1)}\) quantiles of the standard normal distribution for \({k} = {1, \dots, m}\).
Quantiles are used here instead of random samples to provide a clearer illustration without affecting the main points.
When the convex hull constraint is satisfied, \({\lambda_{WET}}\) converges faster to the respective \({\lambda_{RET}}\) (horizontal dashed lines).
}
\label{fig:discretization}
\end{figure}

From the convexity and lower semicontinuity of \({\exp(\lambda^\top \widetilde{g})}\), it is shown that \({p_m(\lambda)}\) epi-converges to \({p(\lambda)}\) as \({m} \to {\infty}\) with probability~\({1}\) \citep[see, for example,][Theorem~2.3]{artstein1995consistency}.
This establishes the consistency of the minimizers with the following intermediate result:
\begin{proposition}\label{prop:epiconvergence}
For each \({\theta} \in {\Theta}\), the minimization problem in \eqref{eq:RETEL_minimization} has a unique global minimizer \({\lambda_{RET}}\).
With probability~\({1}\), for any \({\epsilon_m} \downarrow {0}\),
\begin{equation*}
\lim_{m \to \infty}\left\{\epsilon_m\textnormal{-}\argmin_{\lambda \in \mathbb{R}^p} c_m\left(\lambda\right)\right\}
=
\left\{\lambda_{RET}\right\},
\end{equation*}
where \({\{\epsilon_m\textnormal{-}\argmin_{\lambda \in \mathbb{R}^p} c_m(\lambda)\}} = {\{\lambda \mid c_m(\lambda) \leq \inf_{\lambda^\prime \in \mathbb{R}^p} c_m(\lambda^\prime) + \epsilon_m\}}\).
\end{proposition}
With probability~\({1}\), \({\lambda_{RET}}\) is a limit point of the sequence of approximate solutions to the minimization problems.
For any finite \({m}\), \({\argmin_{\lambda \in \mathbb{R}^p}c_m(\lambda)}\) may not exist with positive probability.
However, the existence and uniqueness of \({\lambda_{RET}}\) are guaranteed by the strict convexity and coercivity of \({p(\lambda)}\), which acts as a penalty that regularizes \({\lambda}\) and prevents \({\vert\lambda\vert}\) from diverging.
Consequently, the RETEL minimization objective function is strictly convex and coercive, ensuring a unique global minimum regardless of the data, rank conditions of \(g_i(\theta)\), or whether the convex hull constraint \({0} \in {\textnormal{Conv}_n(\theta)}\) is satisfied.
\cref{fig:discretization} shows an example where \({\lambda_{WET}}\) converges to \({\lambda_{RET}}\) as a sequence of pseudo-data approximates a normal distribution.

More generally, we can consider
\begin{equation}\label{eq:penalty}
p_n\left(\theta, \lambda\right)
=
\tau_n
\exp\left(
\lambda^\top\mu_{n, \theta} + \frac{1}{2}\lambda^\top\Sigma_{n, \theta}\lambda\right),
\end{equation}
and the corresponding minimization problem in \eqref{eq:RETEL_minimization} becomes
\begin{equation}\label{eq:RETEL_minimization2}
\min_{\lambda \in \mathbb{R}^p}
c_n\left(\theta, \lambda\right)
=
\min_{\lambda \in \mathbb{R}^p}
\left\{
d_n\left(\theta, \lambda\right) + p_n\left(\theta, \lambda\right)
\right\},
\end{equation}
with the solution still denoted by \({\lambda_{RET}}\).
Here, \({\tau_n}\) is a positive parameter that controls the strength of \({p_n(\theta, \lambda)}\) as a penalty.
The parameters \({\mu_{n, \theta}}\) and \({\Sigma_{n, \theta}}\), which may vary with \({n}\) and \({\theta}\), can be informed by prior information, allowing for greater flexibility in the regularization.
Throughout, \({\theta} \mapsto {\mu_{n, \theta}}\) and \({\theta} \mapsto {\Sigma_{n, \theta}}\) are assumed measurable with probability~\({1}\), as holds in particular whenever they are continuous in \({\theta}\).
The choices of \({\mu_{n, \theta}}\), \({\Sigma_{n, \theta}}\), and \({\tau_n}\) depend on the requirements of a specific application, and each configuration uniquely determines the shape and curvature of \({p_n(\theta, \lambda)}\).
These choices and their practical implications are discussed in detail later in this section.

From an operational perspective, any function \({p(\cdot)}:{\mathbb{R}^p} \to {\mathbb{R}_{>0}}\) that increases superlinearly with \({\vert\lambda\vert}\) can be considered to ensure a finite \({\lambda_{RET}}\).
This penalty method can also be extended to other generalized EL methods that share the Cressie--Read family of discrepancies.
However, we focus on ETEL due to its connection to the exponential family it generates \citep{yiu2020inference} and to the auxiliary continuous exponential family distribution that is naturally introduced.

In the following, we present an alternative approach to formulating the minimization problem in \eqref{eq:RETEL_minimization2}.
This approach does not involve the concept of a sequence of procedures with pseudo-data but instead directly considers a mixture of a normal and a multinomial distribution supported on the data.
For a given \({\lambda}\), we apply exponential tilting to the \({N(\mu_{n, \theta}, \Sigma_{n, \theta})}\) distribution of \({\widetilde{g}}\), resulting in the \({\lambda}\)-tilted distribution \({N(\mu_{n, \theta} + \Sigma_{n, \theta}\lambda, \Sigma_{n, \theta})}\).
We denote the corresponding random variable as \({\widetilde{g}_{\lambda}}\).
To formalize this, we define the reference mixture \({\widetilde{P}_n}\) and its tilted counterpart \({\widetilde{P}_{\lambda}}\) as follows:
\begin{equation*}
\widetilde{P}_n
=
\frac{n}{n + \tau_n} P_n + \frac{\tau_n}{n + \tau_n}N\left(\mu_{n, \theta}, \Sigma_{n, \theta}\right),\quad
\widetilde{P}_{\lambda}
=
\left(1 - p_c\right) P_0 + p_c N\left(\mu_{n, \theta} + \Sigma_{n, \theta}\lambda, \Sigma_{n, \theta}\right),
\end{equation*}
where each distribution is a convex mixture of a discrete and a continuous component.
The constant \({p_c}\) in \({\widetilde{P}_{\lambda}}\) represents the probability assigned to the tilted distribution.
The following result parallels the idea that \({D_{KL}({P}_{0}\ \Vert\ {P}_n)}\) is minimized by ETEL.
\begin{proposition}\label{prop:kl}
For any \({\theta} \in {\Theta}\), the minimization problem in \eqref{eq:RETEL_minimization2} is the dual problem of minimizing \({D_{KL}(
\widetilde{P}_{\lambda} \Vert\ \widetilde{P}_n
)}\) with respect to \({p_i}\), \({i} = {1, \dots, n}\), and \({p_c}\), subject to
\(\sum_{i = 1}^n p_i g_i(\theta) + p_c E_{\widetilde{P}_{\lambda}}\{{\widetilde{g}_{\lambda}}\} = {0}\), \({p_c} \geq {0}\), \({p_i} \geq {0}\), and \({\sum_{i = 1}^n p_i + p_c} = {1}\).
\end{proposition}
As a consequence, the optimal values of \({p_i(\theta)}\) and \({p_c(\theta)}\) can be expressed as follows:
\begin{equation*}
p_i\left(\theta\right) = \frac{\exp\left({\lambda_{RET}}^\top g_i\left(\theta\right)\right)}{c_n\left(\theta, \lambda_{RET}\right)} \quad
\left(i = 1, \dots, n\right),\quad
p_c\left(\theta\right) = \frac{p_n\left(\theta, \lambda_{RET}\right)}{c_n\left(\theta, \lambda_{RET}\right)},
\end{equation*}
where \({\lambda_{RET}}\) is the solution to the equation:
\begin{equation}\label{eq:RETEL_equation}
\sum_{i = 1}^n \exp\left(\lambda^\top g_i\left(\theta\right)\right) g_i\left(\theta\right)
+
p_n\left(\theta, \lambda\right)
\left(\mu_{n, \theta} + \Sigma_{n, \theta}\lambda\right) =
0.
\end{equation}
Once we have determined \({\lambda_{RET}}\), we define the likelihood and likelihood ratio functions as follows:
\begin{equation}\label{eq:RETEL_f}
L_{RET}\left(\theta\right) = p_c\left(\theta\right) \prod_{i = 1}^n p_i\left(\theta\right),\quad
R_{RET}\left(\theta\right) = \left(\frac{n + \tau_n}{\tau_n}p_c\left(\theta\right)\right) \prod_{i = 1}^n \left(n + \tau_n\right)p_i\left(\theta\right).
\end{equation}
We note that RETEL differs from penalty approaches for EL \citep{tang2010penalized,leng2012penalized,chang2018new,chang2025bayesian}, where a penalty term is added to the empirical log-likelihood to induce sparsity in the solution \({\widehat{\theta}}\) or perform moment selection.
Instead, RETEL aims to regularize the behavior of the multiplier \({\lambda}\) before constructing the likelihood to ensure the preservation of posterior support.
In standard EL-based methods, the optimization for \({\lambda}\) is restricted to a data-dependent feasible set, which arises from the convex hull constraint.
When this requirement is not satisfied---a frequent occurrence during Bayesian posterior sampling---the feasible set for \({\lambda}\) is effectively truncated, and the likelihood is zeroed out.
By deriving regularization from the Kullback--Leibler divergence dual problem in \cref{prop:kl}, RETEL ensures unconstrained optimization for \({\lambda}\) across \({\mathbb{R}^p}\).
This avoids the logical inconsistency where data-dependent geometric constraints restrict the posterior support while the prior remains unrestricted.

With \({\lambda}\) having a concrete interpretation as a tilting parameter in minimizing the Kullback--Leibler divergence, RETEL is also distinct from the penalized EL approach of \citet{bartolucci2007penalized}.
While RETEL shares some connection with hybrid EL approaches that combine EL with a parametric likelihood \citep{qin1994semi,hjort2018hybrid}, it takes a more indirect approach by employing \({p_c(\theta)}\) instead of directly multiplying by a parametric likelihood.
This captures the information-theoretic tilting effect from the assumed auxiliary distribution to ensure a robust and logically consistent transition from prior to posterior.

To make RETEL more closely reflect the observed data, we can drop \({p_c(\theta)}\) from \eqref{eq:RETEL_f} and define another version of RETEL with the following functions:
\begin{equation}\label{eq:RETEL_r}
\widetilde{L}_{RET}\left(\theta\right) = \prod_{i = 1}^n p_i\left(\theta\right),\quad
\widetilde{R}_{RET}\left(\theta\right) = \prod_{i = 1}^n \left(n + \tau_n\right)p_i\left(\theta\right).
\end{equation}
Dropping \({p_c(\theta)}\) does not mean reverting to ETEL since \({p_c(\theta)}\) affects the other \({p_i(\theta)}\) such that \({\sum_{i = 1}^n p_i(\theta) + p_c(\theta)} = {1}\).
The impact of \({p_c(\theta)}\) and the underlying auxiliary distribution remains embedded in the procedure and cannot be entirely removed, although \({\tau_n}\) can control the degree of this effect.
A larger value of \({\tau_n}\) assigns more probability to \({p_c(\theta)}\) relative to the other \({p_i(\theta)}\), resulting in a greater reliance on the \({\lambda}\)-tilted distribution for inference.
We refer to the approaches in \eqref{eq:RETEL_f} and \eqref{eq:RETEL_r} as the full and reduced versions of RETEL, denoted by \({\textnormal{RETEL}_f}\) and \({\textnormal{RETEL}_r}\) respectively.

\begin{figure}[!t]
\centering
\includegraphics[width=\linewidth,keepaspectratio]{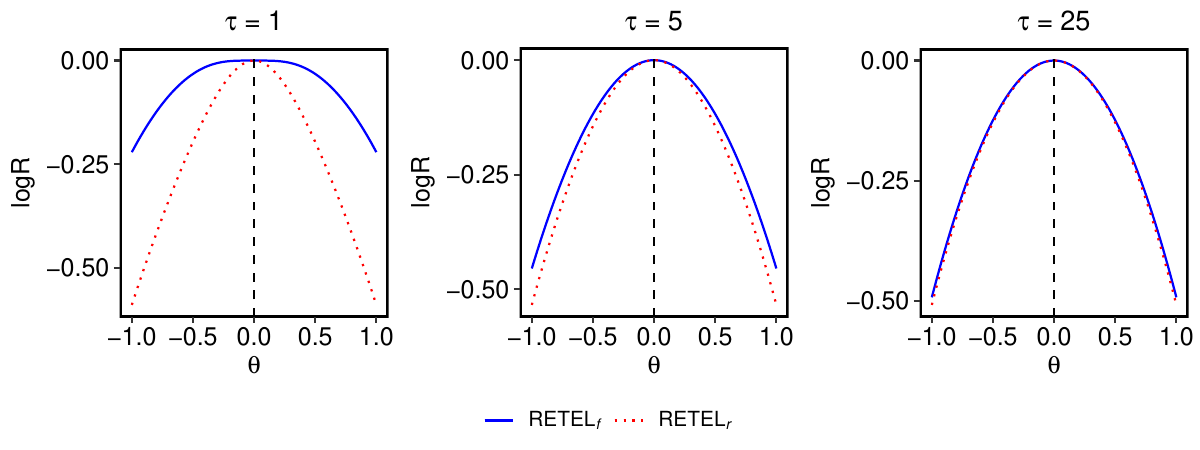}
\caption{
Plots of \({\log R_{RET}(\theta)}\) (solid blue lines) and \({\log \widetilde{R}_{RET}(\theta)}\) (dashed red lines) for the mean parameter with varying \({\tau_n} \in \{1, 5, 25\}\).
Both \({\textnormal{RETEL}_f}\) and \({\textnormal{RETEL}_r}\) achieve their maximum at the single data point \({0}\) (vertical dashed line).
Here,~\({\mu_{n, \theta}}\) and \({\Sigma}_{n, \theta}\) are set to \({-\theta}\) and \({1}\), respectively.
}
\label{fig:tau}
\end{figure}

Contrary to other methods that add finite pseudo-data, both \({\textnormal{RETEL}_f}\) and \({\textnormal{RETEL}_r}\) can naturally preserve the same \({Z}\)-estimator \({\widehat{\theta}}\) of ETEL.
This can be achieved by setting \({\mu_{n, \theta}} = {0}\) or \({\mu_{n, \theta}} = {n^{-1}\sum_{i = 1}^n g_i(\theta)}\) in \eqref{eq:RETEL_equation}, resulting in \({\lambda_{RET}(\widehat{\theta})} = {0}\) and \({R_{RET}(\widehat{\theta})} = \widetilde{R}_{RET}(\widehat{\theta}) = {1}\).
\cref{fig:tau} illustrates, with a single observation, the difference between \({\log R_{RET}(\theta)}\) and \({\log \widetilde{R}_{RET}(\theta)}\) as \({\tau_n}\) increases.

Beyond preserving the ETEL estimator, the specification of the auxiliary parameters \({\mu_{n, \theta}}\), \({\Sigma_{n, \theta}}\), and \({\tau_n}\) is related to the interpretation and the practical considerations of RETEL.
Often, the choice of \({\mu_{n,\theta}}\) implicitly defines the relationship between the regularization and the parameter \({\theta}\).
For example, when considering the mean parameter \({\theta}\), setting \({\mu_{n, \theta} = 0}\) corresponds to an anchor approach, effectively assuming a latent distribution \({N(\theta, \Sigma_{n, \theta})}\) at each \({\theta}\) and obviating the need for further tilting from the auxiliary component.
In contrast, a bridge specification, such as setting \({\mu_{n, \theta}} = {\overline{X} - \theta}\), introduces a distribution \({N(\overline{X}, \Sigma_{n, \theta})}\), inducing a \({\theta}\)-dependent tilt to reconcile the moment condition.
Analogous to the bridge specification for \({\mu_{n,\theta}}\), setting \({\Sigma_{n, \theta}} \propto {\kappa_{n,p} \sum_{i = 1}^n g_i(\theta)g_i(\theta)^\top}\), where \({\kappa_{n, p}}\) is a scaling factor that depends on the relative values of \({p}\) and \({n}\), ensures the regularization scale is commensurate with the dimension and observed variability of the estimating functions.
This distinction is most relevant when the convex hull constraint is violated, as the specific method of resolving the violation directly shapes the resulting inference.

The structure of the regularization also establishes a conceptual link with the unit information prior \citep{kass1995reference}, which involves a normal distribution containing an amount of information equivalent to that of a single observation.
Within the reference mixture \({\widetilde{P}_n}\), setting \({\tau_n = 1}\) assigns the auxiliary information the weight of one pseudo-observation.
Its impact is manifested through \({\lambda_{RET}}\), which determines the tilting of the empirical weights \({p_i(\theta)}\) and \({p_c(\theta)}\) required to satisfy the moment condition.
In this context, the bridge approach aligns with the concept of a unit information prior, serving as a reference to ensure the regularization remains conservative.
This link is an interpretation of the regularization rather than a claim that RETEL is a Bayesian model with a complete likelihood.
The auxiliary component is a pseudo-observation that enters \({L_{RET}(\theta)}\) only through the tilting, so \({L_{RET}(\theta)}\) is a pseudo-likelihood for \({\theta}\).

Next, we establish that RETEL retains certain desirable asymptotic properties of EL and ETEL.
We consider RETEL obtained from \({\lambda_{RET}}\) in \eqref{eq:RETEL_minimization2}.
The following condition controls \({p_n(\theta, \lambda)}\) in \eqref{eq:penalty} and ensures that the regularization is well defined and asymptotically stable when it depends on \({n}\) and \({\theta}\).
\begin{condition}\label{condition:retel asymptotics}
\({\tau_n} = {O(\log n)}\) and \({\tau_n^{-1}} = {O(n^{C_0})}\) for some constant \({C_0} \geq {0}\).
With probability~\({1}\), \({\Sigma_{n, \theta}}\) is positive definite for every \({\theta} \in {\Theta}\) and any \({n}\).
For some \({\mu} \in {\mathbb{R}^p}\) and \({\Sigma} \in {\mathbb{R}^{p \times p}}\), \({\mu_{n, \theta_0}} = {\mu} + {o_P(1)}\) and \({\Sigma_{n, \theta_0}} = {\Sigma} + {o_P(1)}\).
There exist a neighborhood \({\mathcal{N}_0}\) of \({\theta_0}\) and constants \({0} < {b_1} \leq {b_2} < {\infty}\) such that, with probability approaching \({1}\), \({\sup_{\theta \in \mathcal{N}_0}\left\vert \mu_{n, \theta} \right\vert} \leq {b_2}\) and the eigenvalues of \({\Sigma_{n, \theta}}\) lie in \({[b_1, b_2]}\) for all \({\theta} \in {\mathcal{N}_0}\).
\end{condition}
\begin{theorem}
Under \cref{condition:parameter space,condition:continuity,condition:full rank,condition:uniform boundedness,condition:retel asymptotics},
\begin{equation*}
\log R_{RET}\left(\theta_0\right)
-
\log\widetilde{R}_{RET}\left(\theta_0\right)
= O_P({n^{-1/2}}).
\end{equation*}
Additionally, both \({-2\log R_{RET}(\theta_0)}\) and \({-2\log \widetilde{R}_{RET}(\theta_0)}\) converge in distribution to \({\chi^2_p}\).
\end{theorem}
As a consequence, the logarithms of the regularized methods are identical up to \({O_P(n^{-1/2})}\), and both methods exhibit Wilks' theorem.
For Bayesian inference, we can obtain the Bernstein--von Mises result for both versions of RETEL.
\begin{condition}\label{condition:prior}
The prior measure admits a density with respect to the Lebesgue measure.
The density \({\pi}(\cdot)\) is continuous in \({\Theta}\) and is positive in a neighborhood of \({\theta_0}\).
\end{condition}
\begin{condition}\label{condition:bvm}
For any \({\delta} > {0}\), there exists \({\epsilon} > {0}\) such that
\begin{equation*}
\textnormal{pr}\left(
\sup_{\left\vert \theta - \theta_0 \right\vert > \delta}
\frac{1}{n}\left(\log L_{RET}\left(\theta\right) - \log L_{RET}\left(\theta_0\right)\right) \leq -\epsilon
\right)
\to
1.
\end{equation*}
For the posterior obtained from \({\textnormal{RETEL}_r}\), the same condition
is assumed with \({\widetilde{L}_{RET}}\) in place of \({L_{RET}}\).
\end{condition}
\cref{condition:prior} and \cref{condition:bvm} are regularity conditions used to establish Bernstein--von Mises theorems for EL and ETEL \citep{chib2018bayesian,yu2024variational}.
\begin{theorem}
Under \cref{condition:parameter space,condition:continuity,condition:full rank,condition:uniform boundedness,condition:retel asymptotics,condition:prior,condition:bvm},
\begin{equation*}
\sup_{\mathcal{B}} \left\vert \pi\left(n^{1/2}\left(\theta - \widehat{\theta}\right) \in \mathcal{B} \relmiddle | \mathcal{D}_n\right) - N\left(0, \Omega\right)\left(\mathcal{B}\right) \right\vert \to 0
\end{equation*}
in probability, where \({\widehat{\theta}}\) is the maximum ETEL estimator, \({\pi(n^{1/2}(\theta - \widehat{\theta}) \mid \mathcal{D}_n)}\) is the posterior distribution of \({n^{1/2}(\theta - \widehat{\theta})}\) obtained from RETEL, and \({\mathcal{B}}\) denotes any Borel subset of \({\mathbb{R}^p}\).
\end{theorem}
This result implies that, when the moment constraints are correctly specified, the total variation distance between the posterior distribution of \({n^{1/2}(\theta - \widehat{\theta})}\) and \({N(0, \Omega)}\) tends to zero in probability.
Since \({n^{1/2}(\widehat{\theta} - \theta_0)}\) converges in distribution to \({N(0, \Omega)}\), the theorem implies posterior consistency and asymptotically correct frequentist coverage of credible sets.

Finally, we note that the proposed framework admits several natural extensions.
We have focused on the auxiliary normal distribution as it provides a general mechanism to satisfy the convex hull constraint for any applicable estimating functions.
This choice is further supported as it is the maximum entropy distribution among all distributions with a given mean and covariance \citep[Theorem~8.6.5]{cover2006elements}.
Nevertheless, the framework can accommodate other exponential family distributions without modification to better align with the specific support or tail behavior of the data, especially outside the context of the convex hull constraint.
Additionally, while \({\tau_n}\) (along with \({\mu_{n, \theta}}\) and \(\Sigma_{n, \theta}\)) has been treated primarily as a fixed tuning parameter, the framework allows for a Bayesian treatment by assigning prior distributions to these parameters.
We provide a brief illustration of these extensions in \cref{subsec:egypt}.

\section{Simulation}\label{sec:simulations}
The proposed RETEL method, alongside ETEL and AETEL, is implemented in the \texttt{R} package \texttt{retel} \citep{retel} via the limited-memory Broyden--Fletcher--Goldfarb--Shanno (L-BFGS) algorithm, while EL and its variants are implemented via the \texttt{melt} package \citep{kim2004melt}.
These packages were used for all computations and numerical experiments conducted in Sections~\ref{sec:simulations} and~\ref{sec:application}. 
All code and data required to reproduce the results are available at \url{https://github.com/markean/retel}.

\subsection{Posterior Coverage}
\citet{monahan1992proper} proposed examining the validity of a pseudo-likelihood \({L(\theta)}\) based on the coverage probabilities of posterior intervals. 
For a parameter \({\theta} \in \mathbb{R}\), let \({\pi(\theta \mid x)}\) be the posterior density obtained using \({L(\theta)}\) with an absolutely continuous prior density \({\pi(\theta)}\) and observed data \({x}\).
For this pseudo-likelihood to be valid by coverage, posterior intervals should provide correct coverage probabilities.
In particular, when \({(X, \theta)}\) is generated from the Bayesian model, the random variable \({H} = {\int_{-\infty}^\theta \pi(t \mid X)dt}\) should follow a uniform distribution \({U(0, 1)}\).  
This approach has been adopted for EL by \citet{lazar2003bayesian} and \citet{cheng2019bayesian}.

To investigate the validity of RETEL for Bayesian inference, we begin by simulating a value of \({\theta}\) from a logistic distribution denoted as \({\textnormal{Logistic}(l, s)}\), where \({l}\) is the location parameter and \({s}\) is the scale parameter.
Next, we generate \({n}\) observations from \({N(\theta, 1)}\) and compute \({H}\) for \({\textnormal{RETEL}_f}\) and \({\textnormal{RETEL}_r}\), employing \({p_n(\theta, \lambda)}\) in \eqref{eq:penalty} with \({\mu_{n, \theta}} = {\overline{X} - \theta}\) and \({\Sigma_{n, \theta}} = {1}\).
For comparison purposes, we also compute \({H}\) using ETEL and AETEL.
Keeping \({l}\) fixed at \({0}\), we repeat this procedure \({\num{10000}}\) times for each combination of \({n} \in {\{5, 20, 50, 100\}}\), \({s} \in {\{1, 5\}}\), and \({\tau_n} \in {\{1, \log n\}}\).
We approximate the posterior distributions on a grid of \({\theta}\) values.
Using the computed \({\num{10000}}\) \({H}\) values, we conduct the Kolmogorov--Smirnov test to evaluate the uniformity of the distributions. 

\begin{figure}[!t]
\centering
\includegraphics[width=\linewidth,keepaspectratio]{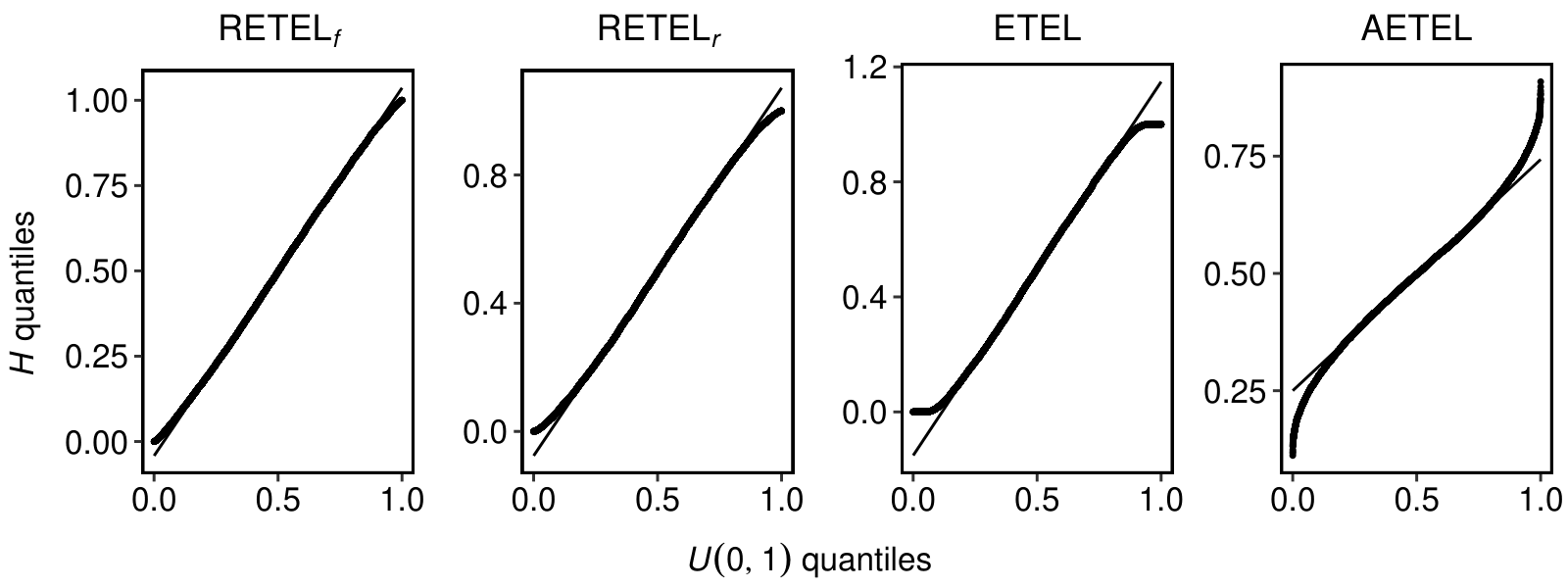}
\caption{
Quantile-quantile plots for the distribution of \({H}\) versus \({U(0, 1)}\) when \({n} = {5}\), \({s} = {5}\), and \({\tau_n} = {1}\).
The light-tailed distribution from ETEL is due to the convex hull constraint.
AETEL produces a heavier-tailed distribution than the others.
}
\label{fig:qq}
\end{figure}

The resulting \({p}\)-values are reported in \cref{tab:mb}, and \cref{fig:qq} displays the quantile-quantile plots for the distribution of \({H}\) versus \({U(0, 1)}\) when \({n} = {5}\), \({s} = {5}\), and \({\tau_n} = {1}\).
The plots highlight the differences in the tails of the distributions that are not apparent from the \({p}\)-values alone.
With a smaller sample size of \({n} = {5}\), RETEL tends to show a closer conformity to \({U(0, 1)}\) compared to ETEL and AETEL. 
The impact of a larger prior variance (\({s} = {5}\)) and a larger \({p_c(\theta)}\) (\({\tau_n} = {\log n}\)) becomes more apparent when \({n} = {50}\).
As the sample size increases, the differences between the posterior distributions of the methods become negligible. 
All of the methods provide an excellent approximation to the null distribution when \({n}\) is \({50}\) or more. 
We emphasize that the Kolmogorov--Smirnov tests are based on a sample of \({\num{10000}}\) replicates and so are able to pick up quite small departures from the uniform distribution.
Additional plots for the full results are provided in the Supplementary Material.
%
\begin{table}[!t]
\centering
\def~{\hphantom{0}}
\caption{\({p}\)-values from the Kolmogorov--Smirnov test for uniformity.}
{\begin{tabular}{cc|ccccccccc}
\toprule
\multicolumn{2}{c}{} &&
\multicolumn{2}{c}{\({\tau_n} = {1}\)} &&
\multicolumn{2}{c}{\({\tau_n} = {\log n}\)} &&
\multicolumn{2}{c}{}\\  
\midrule
\({n}\) & \({s}\) && \({\textnormal{RETEL}_f}\) & \({\textnormal{RETEL}_r}\) && \({\textnormal{RETEL}_f}\) & \({\textnormal{RETEL}_r}\) && ETEL & AETEL\\
\midrule
\multirow{2}{*}{\({5}\)} 
& \({1}\) 
&& \({<0.001}\) & \({0}\) 
&& \({<0.001}\)  & \({<0.001}\) 
&& \({0}\) & \({0}\) \\ 
& \({5}\) 
&& \({<0.001}\) & \({<0.001}\) 
&& \({<0.001}\)  & \({<0.001}\) 
&& \({0}\) & \({0}\) \\[1ex] 
\multirow{2}{*}{\({20}\)} 
& \({1}\) 
&& \({<0.001}\) & \({<0.001}\) 
&& \({<0.001}\)  & \({<0.001}\) 
&& \({<0.001}\) & \({0.007}\) \\ 
& \({5}\) 
&& \({<0.001}\) & \({<0.001}\) 
&& \({0.001}\)  & \({<0.001}\) 
&& \({<0.001}\) & \({0.064}\) \\[1ex]
\multirow{2}{*}{\({50}\)} 
& \({1}\) 
&& \({0.262}\) & \({0.248}\) 
&& \({0.303}\)  & \({0.303}\) 
&& \({0.221}\) & \({0.425}\) \\ 
& \({5}\) 
&& \({0.360}\) & \({0.320}\) 
&& \({0.414}\)  & \({0.423}\) 
&& \({0.303}\) & \({0.466}\) \\[1ex]
\multirow{2}{*}{\({100}\)} 
& \({1}\) 
&& \({0.428}\) & \({0.430}\) 
&& \({0.417}\)  & \({0.418}\) 
&& \({0.364}\) & \({0.714}\) \\ 
& \({5}\) 
&& \({0.363}\) & \({0.367}\) 
&& \({0.389}\)  & \({0.369}\) 
&& \({0.323}\) & \({0.781}\) \\
\bottomrule
\end{tabular}}
\label{tab:mb}
\end{table}

Next, we investigate the frequentist properties of the posterior intervals obtained from RETEL.
We consider a true mean parameter value \({\theta_0} = {0}\) and generate \({n}\) observations from \({N(0, 1)}\).
Using the logistic prior distribution described earlier, we compute a \({95\%}\) posterior credible interval for \({\theta}\) using each of the four methods.
This procedure is repeated \(\num{10000}\) times for different combinations of \({n} \in {\{5, 20, 50, 100\}}\), \({s} \in {\{0.5, 1, 5\}}\), and \({l} \in {\{0, 2\}}\), while fixing \({\tau_n}\) at \({\log n}\).
We then calculate the coverage rate and average length of the central credible intervals.

The results for \({l} = {0}\) are presented in \cref{tab:cr_l0}, where the prior mean matches the true parameter value.
As \({s}\) decreases, indicating stronger prior information on \({\theta}\) at \({0}\), higher coverage rates and shorter intervals are obtained.
The differences between the methods are most pronounced when \({n} = {5}\).
The intervals obtained from ETEL exhibit significantly lower coverage rates compared to those of the other methods.
AETEL produces the widest intervals with coverage rates higher than the nominal level. 
The wider intervals and departure from the nominal coverage rate are related to the boundedness problem of AEL, which arises due to the addition of one pseudo-observation \citep{emerson2009calibration}.
On the other hand, RETEL yields coverage rates closer to the nominal level but features much shorter intervals compared to AETEL.
Within RETEL, the full version produces wider intervals with higher coverage rates than the reduced version, consistent with the findings from the plots in \cref{fig:qq}.
\begin{table}[!ht]
\centering
\def~{\hphantom{0}}
\caption{Coverage rates (CR) in \% and average lengths (AL) of credible intervals when \({l} = {0}\).
The largest standard error of the lengths is \({0.005}\) when \({n} = {5}\) and \({s} = {5}\).
}
{\begin{tabular}{cc|cccccccccccc}
\toprule
\multicolumn{2}{c}{} &&
\multicolumn{2}{c}{\({\textnormal{RETEL}_f}\)} &&
\multicolumn{2}{c}{\({\textnormal{RETEL}_r}\)} &&
\multicolumn{2}{c}{ETEL} &&
\multicolumn{2}{c}{AETEL}\\  
\midrule
\({n}\) & \({s}\) && CR & AL && CR & AL && CR & AL && CR & AL\\
\midrule
\multirow{3}{*}{\({5}\)} 
& \({0.5}\) 
&& \({95.9}\) & \({1.445}\) 
&& \({94.4}\) & \({1.385}\) 
&& \({79.2}\) & \({1.128}\)
&& \({100.0~}\) & \({2.572}\) \\ 
& \({1}\) 
&& \({94.1}\) & \({1.581}\) 
&& \({92.4}\) & \({1.505}\) 
&& \({77.8}\) & \({1.200}\) 
&& \({100.0~}\) & \({5.338}\) \\ 
& \({5}\) 
&& \({93.2}\) & \({1.647}\) 
&& \({91.3}\) & \({1.561}\) 
&& \({77.2}\) & \({1.230}\) 
&& \({100.0~}\) & \({9.367}\) \\[1ex] 
\multirow{3}{*}{\({20}\)} 
& \({0.5}\) 
&& \({94.2}\) & \({0.805}\) 
&& \({94.0}\) & \({0.803}\) 
&& \({93.1}\) & \({0.790}\) 
&& \({96.3}\) & \({0.886}\) \\ 
& \({1}\) 
&& \({93.6}\) & \({0.830}\) 
&& \({93.4}\) & \({0.828}\) 
&& \({92.4}\) & \({0.815}\) 
&& \({96.0}\) & \({0.932}\) \\ 
& \({5}\) 
&& \({93.3}\) & \({0.839}\) 
&& \({93.1}\) & \({0.837}\) 
&& \({92.2}\) & \({0.824}\) 
&& \({96.1}\) & \({0.965}\) \\[1ex] 
\multirow{3}{*}{\({50}\)} 
& \({0.5}\) 
&& \({94.8}\) & \({0.534}\) 
&& \({94.8}\) & \({0.534}\) 
&& \({94.5}\) & \({0.530}\) 
&& \({95.8}\) & \({0.558}\) \\ 
& \({1}\) 
&& \({94.5}\) & \({0.542}\) 
&& \({94.5}\) & \({0.542}\) 
&& \({94.2}\) & \({0.537}\) 
&& \({95.4}\) & \({0.566}\) \\ 
& \({5}\) 
&& \({94.5}\) & \({0.544}\) 
&& \({94.4}\) & \({0.544}\) 
&& \({94.1}\) & \({0.540}\) 
&& \({95.3}\) & \({0.569}\) \\[1ex] 
\multirow{3}{*}{\({100}\)} 
& \({0.5}\) 
&& \({94.8}\) & \({0.385}\) 
&& \({94.8}\) & \({0.385}\) 
&& \({94.6}\) & \({0.380}\) 
&& \({95.2}\) & \({0.394}\) \\ 
& \({1}\) 
&& \({94.7}\) & \({0.387}\) 
&& \({94.7}\) & \({0.387}\) 
&& \({94.4}\) & \({0.383}\) 
&& \({95.1}\) & \({0.397}\) \\ 
& \({5}\) 
&& \({94.6}\) & \({0.388}\) 
&& \({94.6}\) & \({0.388}\) 
&& \({94.3}\) & \({0.384}\) 
&& \({95.1}\) & \({0.398}\) \\
\bottomrule
\end{tabular}}
\label{tab:cr_l0}
\end{table}

\cref{tab:cr_l2} shows the results when \({l} = {2}\), indicating a prior mean that is far from the true parameter value. 
In this case, the credible intervals tend to be wider with lower coverage rates.
ETEL is relatively unaffected due to the convex hull constraint.
However, the effect of different \({l}\) values is noticeable for the other methods.
Particularly when \({n} = {5}\) and \({s} = {0.5}\), the strong prior shifts the intervals toward \({2}\).
AETEL is the most affected, as its coverage rate is considerably lower than that of RETEL, even with wider intervals.
To sum up, RETEL exhibits robust performance across various prior means and variances, demonstrating close-to-nominal posterior coverage rates with small sample sizes. 
\begin{table}[htbp]
\centering
\def~{\hphantom{0}}
\caption{Coverage rates (CR) in \% and average lengths (AL) of credible intervals when \({l} = {2}\).
The largest standard error of the lengths is \({0.006}\) when \({n} = {20}\) and \({s} = {0.5}\).
}
{\begin{tabular}{cc|cccccccccccc}
\toprule
\multicolumn{2}{c}{} &&
\multicolumn{2}{c}{\({\textnormal{RETEL}_f}\)} &&
\multicolumn{2}{c}{\({\textnormal{RETEL}_r}\)} &&
\multicolumn{2}{c}{ETEL} &&
\multicolumn{2}{c}{AETEL}\\  
\midrule
\({n}\) & \({s}\) && CR & AL && CR & AL && CR & AL && CR & AL\\
\midrule
\multirow{3}{*}{\({5}\)} 
& \({0.5}\) 
&& \({86.3}\) & \({1.610}\) 
&& \({85.3}\) & \({1.530}\) 
&& \({73.2}\) & \({1.157}\)
&& \({80.0}\) & \({3.887}\) \\ 
& \({1}\) 
&& \({92.6}\) & \({1.616}\) 
&& \({90.8}\) & \({1.534}\) 
&& \({76.9}\) & \({1.207}\) 
&& \({100.0~}\)  & \({6.174}\) \\ 
& \({5}\) 
&& \({93.1}\) & \({1.647}\) 
&& \({91.3}\) & \({1.561}\) 
&& \({77.2}\) & \({1.230}\) 
&& \({100.0~}\)  & \({10.858}\) \\[1ex] 
\multirow{3}{*}{\({20}\)} 
& \({0.5}\) 
&& \({91.6}\) & \({0.834}\) 
&& \({91.4}\) & \({0.832}\) 
&& \({90.7}\) & \({0.819}\) 
&& \({94.2}\) & \({1.529}\) \\ 
& \({1}\) 
&& \({93.0}\) & \({0.835}\) 
&& \({92.9}\) & \({0.833}\) 
&& \({92.1}\) & \({0.820}\) 
&& \({96.6}\) & \({0.995}\) \\ 
& \({5}\) 
&& \({93.3}\) & \({0.839}\) 
&& \({93.1}\) & \({0.837}\) 
&& \({92.2}\) & \({0.824}\) 
&& \({96.2}\) & \({0.975}\) \\[1ex] 
\multirow{3}{*}{\({50}\)} 
& \({0.5}\) 
&& \({93.5}\) & \({0.543}\) 
&& \({93.4}\) & \({0.543}\) 
&& \({93.3}\) & \({0.539}\) 
&& \({94.4}\) & \({0.569}\) \\ 
& \({1}\) 
&& \({94.3}\) & \({0.543}\) 
&& \({94.3}\) & \({0.543}\) 
&& \({94.1}\) & \({0.538}\) 
&& \({95.2}\) & \({0.568}\) \\ 
& \({5}\) 
&& \({94.5}\) & \({0.554}\) 
&& \({94.5}\) & \({0.544}\) 
&& \({94.1}\) & \({0.540}\) 
&& \({95.3}\) & \({0.569}\) \\[1ex] 
\multirow{3}{*}{\({100}\)} 
& \({0.5}\) 
&& \({94.2}\) & \({0.388}\) 
&& \({94.2}\) & \({0.388}\) 
&& \({94.0}\) & \({0.384}\) 
&& \({94.7}\) & \({0.398}\) \\ 
& \({1}\) 
&& \({94.5}\) & \({0.388}\) 
&& \({94.4}\) & \({0.388}\) 
&& \({94.2}\) & \({0.384}\) 
&& \({95.0}\) & \({0.398}\) \\ 
& \({5}\) 
&& \({94.6}\) & \({0.388}\) 
&& \({94.6}\) & \({0.388}\) 
&& \({94.3}\) & \({0.384}\) 
&& \({95.1}\) & \({0.398}\)  \\
\bottomrule
\end{tabular}}
\label{tab:cr_l2}
\end{table}

\subsection{Expected Kullback--Leibler Divergence}
The restricted posterior domain significantly affects Bayesian inference with EL and ETEL, especially when the sample size is small.
In an extreme example with only two observations, \({X_1}\) and \({X_2}\), where the interest is in the mean parameter \({\theta}\), the posterior domain collapses to a singleton as \({\vert X_1 - X_2 \vert}\) approaches zero.
This example illustrates a problematic aspect of EL and ETEL, where the posterior can appear to provide more definitive information on the parameter when the observations coincide (\({X_1 = X_2}\)).

More generally, consider a parametric model \({\mathcal{M}} = {\{p(x \mid \theta) \mid x \in \mathcal{X}, \theta \in \Theta\}}\) and a prior density \({\pi(\theta)}\).
The expected information obtained from observing \({x}\) from \({\mathcal{M}}\) can be measured using the expected Kullback--Leibler divergence:
\begin{equation*}
 I\left(\pi \relmiddle | \mathcal{M}\right) 
 =
\int_{\mathcal{X}} D_{KL}
\left(
\pi\left(\cdot \relmiddle | x\right)
\relmiddle \Vert 
\pi\left(\cdot\right)
\right)
m\left(x\right)dx,
\end{equation*}
where \({\pi(\theta \mid x)} = {\pi(\theta) p(x \mid \theta) / m(x)}\) and \({m(x)} = {\int_{\Theta}\pi(\theta) p(x \mid \theta)d\theta}\).
Let \({I(\pi\mid\mathcal{M}_n)}\) denote the expected information obtained from the set of observations \({\{x_1, \dots, x_n\}}\).  
It is expected that \({I(\pi\mid\mathcal{M}_n)}\) increases monotonically with \({n}\) \citep{mantovan2006hydrological}.
The following result, based on \citet[Theorem 3]{berger2009formal}, illustrates this monotonicity property.
\begin{proposition}\label{prop:monotone ekl}
Let \({\mathcal{M}} = {\{p(x_1, x_2\mid \theta) \mid x_1 \in \mathcal{X}, x_2 \in \mathcal{X}, \theta \in \Theta\}}\) be a model with a sufficient statistic \({t} = {t(x_1, x_2)} \in {\mathcal{U}}\). 
Suppose \({\pi(\theta)}\) is a strictly positive and continuous prior on \({\Theta}\), where \({\pi(\theta \mid x_1, x_2)} = {\pi(\theta)p(x_1, x_2 \mid \theta) / m(x_1, x_2)}\) and \({m(x_1, x_2)} = {\int_{\Theta}\pi(\theta) p(x_1, x_2 \mid \theta)d\theta} < {\infty}\).
Under \cref{condition:parameter space}, if \(\int_{\mathcal{U}} p(t \mid \theta) \log(p(t \mid \theta) / p(t \mid {\theta}^\prime)) dt < {\infty}\) for any \({\theta} \in {\Theta}\) and \({{\theta}^\prime} \in {\Theta}\), then \({I(\pi\mid\mathcal{M}_1)} \leq {I(\pi\mid\mathcal{M}_2)} < {\infty}\).
\end{proposition}
Based on the above proposition, the approximate validity of a pseudo-likelihood for Bayesian inference can be evaluated by examining whether it preserves the monotonicity property.

To examine the performance of RETEL compared to EL and ETEL, we consider two independent experiments where we obtain independent observations, denoted as \({X_{ij}}\) for \({i} = {1, 2}\) and \({j} = {1, \dots, n}\), from the following hierarchical model:
\begin{align*}
X_{ij} \mid \theta_i &\sim N\left(\theta_i, \sigma^2\right),\\  
\theta_i \mid \mu &\sim
\textnormal{Cauchy}\left(\mu, \gamma\right),\\
\mu &\sim N\left(0, \tau^2\right).
\end{align*}
We assume fixed values of \({\sigma} = {1}\), \({\gamma} = {1}\), and \({\tau} = {10}\).  
We use the four methods in place of the normal density for \(X_{ij}\).  
Our main focus is on the marginal posterior distribution of \({\mu}\), with the density denoted by \({\pi(\mu \mid \mathcal{D}_n)}\).
Given the values of \({\theta_1}\) and \({\theta_2}\) with \({\Delta} = {\vert\theta_1 - \theta_2\vert} > {2}\), the Cauchy distribution for \({\theta_1}\) and \({\theta_2}\) yields two maximum likelihood estimates of \({\mu}\) given by \({(\theta_1 + \theta_2)/2} \pm {\sqrt{\Delta^2 - 1}}\)
\citep{dharmadhikari1985examples}.
Consequently, when combined with the large standard deviation of the prior distribution for \({\mu}\), the restricted posterior domain of \({\theta_1}\) and \({\theta_2}\) from EL and ETEL leads to a bimodal marginal posterior distribution for \({\mu}\).
This bimodality can potentially result in inflated values of \({I(\pi\mid\mathcal{M}_n)}\) for EL and ETEL, particularly when \({n}\) is small.

The marginal likelihood, \({m({x})}\), for the four methods cannot be computed analytically. 
Instead, we can observe that \(I(\pi \mid \mathcal{M})\) can be expressed as follows:
\begin{equation*}
 I\left(\pi \relmiddle | \mathcal{M}\right) 
 =
\int_{\Theta}
\pi\left(\theta\right)
\left\{
\int_{\mathcal{X}} D_{KL}
\left(
\pi\left(\cdot \relmiddle | x\right)
\relmiddle \Vert 
\pi\left(\cdot\right)
\right)
p\left(x\mid \theta\right)dx
\right\}
d\theta,
\end{equation*}
where \({D_{KL}(\pi(\cdot \mid x) \Vert \pi(\cdot))}\) is computed with respect to \({\mu}\).
Since our focus is on comparing \({I(\pi\mid\mathcal{M}_n)}\) for the methods,
we fix \({\theta_1}\) and \({\theta_2}\) at \({-3}\) and \({3}\), respectively.
For each method and \({n} \in {\{2, 4, 6, 8, 10\}}\), we estimate the inner integrand of \({I(\pi\mid\mathcal{M}_n)}\) through simulation using the following steps:
\begin{enumerate}
\item Generate \({X_{1j}}\) from \({N(-3, 1)}\) and \({X_{2j}}\) from \({N(3, 1)}\) for \({j} = {1, \dots, n}\).
\item Generate \({\num{10000}}\) posterior samples of \({\theta_1}\), \({\theta_2}\), and \({\mu}\) with a random-walk Metropolis--Hastings algorithm.
\item Estimate \({\pi(\mu \mid \mathcal{D}_n)}\) from the posterior samples and compute \({D_{KL}(
\pi(\cdot \mid \mathcal{D}_n)\ \Vert \
\pi(\cdot))}\) by numerical integration with adaptive quadrature.
\item Repeat Steps 1--3 \({\num{1000}}\) times and take the average of the estimates from Step 3.
\end{enumerate}

Step 2 is implemented with two chains of length \({\num{5000}}\), ensuring that the potential scale reduction factor \citep{gelman1992inference} remains below \({1.1}\) on average for each method.
For the regularized methods, \({\tau_n} = {1}\) is used when \({n = 2}\), and \({\tau_n} = {\log n}\) is used otherwise.

\begin{figure}[!t]
\centering
\begin{subfigure}{0.495\linewidth}
\includegraphics[width=\linewidth]{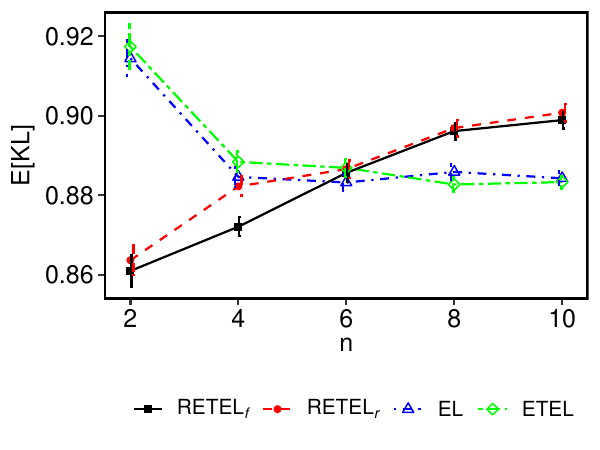}
\caption{}
\label{fig:kl_a}
\end{subfigure}
\hfill
\begin{subfigure}{0.495\linewidth}
\includegraphics[width=\linewidth]{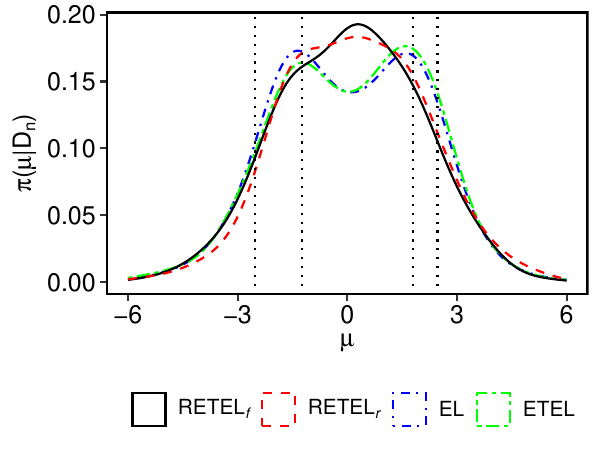}
\caption{}
\label{fig:kl_b}
\end{subfigure}
\caption{
Plots of (a) the expected Kullback--Leibler divergence (inner integrand) and (b) the marginal posterior density \({\pi(\mu \mid \mathcal{D}_n)}\) with \({n} = {2}\).
The error bars in (a) represent plus or minus one standard error. 
The vertical dotted lines in (b) indicate the four realized data points.
}
\label{fig:kl}
\end{figure}

The results are summarized in \cref{fig:kl}.
In \cref{fig:kl_a}, it can be seen that \({I(\pi\mid\mathcal{M}_n)}\) is the smallest when \({n} = {2}\) for RETEL, and it increases monotonically as the sample size grows.
The reduced version tends to produce slightly larger \({I(\pi\mid\mathcal{M}_n)}\) compared to the full version.
On the other hand, EL and ETEL attain the largest \({I(\pi\mid\mathcal{M}_n)}\) when \({n} = {2}\), with values of \({0.914}\) and \({0.917}\), respectively.
The values of \({I(\pi\mid\mathcal{M}_n)}\) decrease as the sample size and the range of the data increase.
EL and ETEL do not exhibit an upward trend in \({I(\pi\mid\mathcal{M}_n)}\) and, even as \({n}\) moves toward \({10}\), do not show a notable improvement.
This discrepancy is caused by the strong bimodality of \({\pi(\mu \mid \mathcal{D}_n)}\), as illustrated in \cref{fig:kl_b}.

\subsection{Confidence Regions}
We close this section with a brief assessment of the frequentist calibration of RETEL for constructing confidence regions, focusing on a comparison with EL and its variants.
The finite-sample undercoverage issue of EL, especially when \({p}\) is large relative to \({n}\), is well documented \citep{tsao2004bounds}.
Nevertheless, EL typically achieves better finite-sample coverage than ETEL, despite their first-order equivalence in yielding \(\chi^2\)-calibrated confidence regions.
This is because higher-order expansions of EL \citep{newey2004higher,schennach2007point} admit some favorable properties, including Bartlett correctability, which ETEL lacks \citep{jing1996exponential}.

We first assess the performance of our proposal for the mean of \({N(0, I_p)}\) with \({p} \in {\{10, 20, 30\}}\), where
\({I_p}\) denotes the \({p \times p}\) identity matrix.
For each \({p}\), we consider relatively small sample sizes, with \({n / p} \in {\{2, 5, 10\}}\), and simulate \({X_i} \sim {N(0, I_p)}\) for \({i} = {1, \dots, n}\).
We then compute the log-likelihood ratio statistic and construct a \({95\%}\) confidence region based on the chi-square calibration.
For comparison with other methods that use pseudo-data, we also compute the statistics from AETEL, AEL, and the balanced adjusted empirical likelihood (BAEL) of \citet{emerson2009calibration}.
The parameters for RETEL are set as 
\begin{equation*}
\tau_n = \log n, \mu_{n, \theta} = \overline{X},\ \textnormal{and}\ \Sigma_{n, \theta} = {p^{1/2}(n - 1)^{-1}\sum_{i = 1}^{n}}(X_i - \overline{X})(X_i - \overline{X})^\top.
\end{equation*}
The scaling factor \({p^{1/2}}\) in \({\Sigma_{n, \theta}}\) accounts for the impact of \({p}\) in the calibration.

\cref{tab:cr_mvnorm} shows the coverage rates of confidence regions for the mean based on \num{1000} simulations for each combination of \({n}\) and \({p}\).
Across different settings, the RETEL methods tend to exhibit coverage rates that are closer to the nominal level than those of other methods.
The higher rates for \({\textnormal{RETEL}_f}\) correspond to the results in the previous sections.
The rates for AETEL are inflated throughout, and those for AEL are inflated when \({n} = {2p}\), which is caused by a known issue where the statistics are bounded above \citep{emerson2009calibration}.
This boundedness issue quickly makes these two calibration methods ineffective as \({p}\) grows.
BAEL, which is specifically designed for the mean, overcomes the issue by meticulously placing two pseudo-observations. 
Still, the performance of regularized methods is more robust to the growing \({p}\) compared to that of BAEL.

Next, we evaluate the methods in a linear regression setting with a response \({Y_i}\) and \({p}\)-variate covariates \({X_i}\), including an intercept.
The estimating functions are given by \({g_i(\theta)} = {X_i(Y_i - X_i^\top\theta)}\).
As in the mean case, we maintain the same \({p} \in {\{10, 20, 30\}}\) with \({n/p} \in {\{5, 10, 20\}}\), conducting \num{1000} simulations for each combination.
The covariates are generated from a multivariate normal distribution with an AR(1) covariance structure, where \({\Sigma_{jk}} = {0.5^{|j-k|}}\) for \({j, k} \in {\{1, \dots, p-1\}}\). 
The true parameter \({\theta_0}\) is drawn uniformly from the unit sphere, and the response is generated by adding independent standard normal noise to \({X_i^\top \theta_0}\).
For RETEL, we employ \({\tau_n} = {\log n}\), \({\mu_{n, \theta}} = {n^{-1}\sum_{i = 1}^n g_i(\theta)}\), and \({\Sigma_{n, \theta}} = {p^{1/2}(n - 1)^{-1}\sum_{i = 1}^n (g_i(\theta) - \mu_{n, \theta})(g_i(\theta) - \mu_{n, \theta})^\top}\).
In this experiment, we consider the extended empirical likelihood (EEL) of \citet{tsao2013empirical, tsao2014extended} alongside the other methods, omitting BAEL.
EEL addresses the convex hull constraint by mapping the likelihood contours onto the full parameter space via a similarity transformation.
The transformation involves an expansion factor that depends on the EL ratio and can be configured to attain the second-order accuracy of Bartlett correction.

\cref{tab:cr_lm} reports the coverage rates for the regression setting at the nominal 95\% level.
The relative sample size \({n/p}\) is kept small to make achieving the nominal level difficult.
Furthermore, the geometry of the estimating functions renders the original parameter space highly nonconvex.
However, as \({p}\) increases relative to \({n}\), the performance gap between the methods becomes more pronounced.
Both \({\textnormal{RETEL}_f}\) and \({\textnormal{RETEL}_r}\) maintain relatively stable coverage rates at or above 80\% across all configurations.
Notably, their performance is often comparable to or exceeds that of EEL when \({n/p} = {5}\).
In contrast, AETEL and AEL suffer significantly as \({p}\) grows. 
AETEL in particular shows severe undercoverage, dropping as low as \({23.1\%}\) for \({p} = {30}\) and \({n}= {150}\), underscoring the lower coverage performance of ETEL compared to EL even with the addition of a pseudo-observation to relieve the convex hull constraint.
Overall, EEL provides the most consistent performance across the full range of settings, approaching the nominal level quickly as \({n}\) increases.
Its competitive results are in part based on the EL-based expansion factor, which accounts for the dimensionality and curvature of the domain via repeated evaluation of the empirical likelihood along a secant line.
This suggests that RETEL's associated parameters could theoretically be further calibrated to the data and estimating functions to achieve coverage that is more stable and closer to the nominal level.
\begin{table}[pt]
\centering
\def~{\hphantom{0}}
\caption{
Coverage rates in \% for the mean of \({p}\)-dimensional multivariate normal distributions.
}
{\begin{tabularx}{0.85\textwidth}{cc|YYYYY}
\toprule
\({p}\) & \({n}\) &
\({\textnormal{RETEL}_f}\) &
\({\textnormal{RETEL}_r}\) &
AETEL &
AEL &
BAEL\\  
\midrule
\multirow{3}{*}{\({10}\)} 
& \({20}\) 
& \({78.1}\) 
& \({72.3}\) 
& \({99.8}\) 
& \({100.0~}\) 
& \({91.1}\) \\ 
& \({50}\) 
& \({91.6}\) 
& \({90.4}\) 
& \({99.7}\) 
& \({88.1}\) 
& \({93.3}\) \\ 
& \({100}\) 
& \({93.6}\) 
& \({92.9}\) 
& \({100.0~}\) 
& \({93.0}\) 
& \({93.6}\) \\[1ex] 
\multirow{3}{*}{\({20}\)} 
& \({40}\) 
& \({84.7}\) 
& \({80.3}\) 
& \({100.0~}\) 
& \({100.0~}\) 
& \({84.2}\) \\ 
& \({100}\) 
& \({92.1}\) 
& \({90.6}\) 
& \({100.0~}\) 
& \({81.1}\) 
& \({86.3}\) \\ 
& \({200}\) 
& \({95.5}\) 
& \({94.5}\) 
& \({100.0~}\) 
& \({92.3}\) 
& \({92.3}\) \\[1ex] 
\multirow{3}{*}{\({30}\)} 
& \({60}\) 
& \({91.3}\) 
& \({88.2}\) 
& \({100.0~}\) 
& \({100.0~}\) 
& \({78.7}\) \\ 
& \({150}\) 
& \({94.1}\) 
& \({93.1}\) 
& \({100.0~}\) 
& \({78.2}\) 
& \({83.1}\) \\ 
& \({300}\) 
& \({94.5}\) 
& \({93.7}\) 
& \({100.0~}\) 
& \({90.4}\) 
& \({90.2}\) \\
\bottomrule
\end{tabularx}}
\label{tab:cr_mvnorm}
\end{table}

\FloatBarrier

\begin{table}[pt]
\centering
\def~{\hphantom{0}}
\caption{Coverage rates in \% for the regression coefficients of the \({p}\)-dimensional linear model.}
{\begin{tabularx}{0.85\textwidth}{cc|YYYYY}
\toprule
\({p}\) & \({n}\) &
\({\textnormal{RETEL}_f}\) &
\({\textnormal{RETEL}_r}\) &
AETEL &
AEL &
EEL\\  
\midrule
\multirow{3}{*}{\({10}\)} 
& \({50}\) 
& \({87.2}\) 
& \({85.1}\) 
& \({57.3}\) 
& \({72.2}\) 
& \({87.2}\) \\ 
& \({100}\) 
& \({87.9}\) 
& \({86.9}\) 
& \({76.5}\) 
& \({85.2}\) 
& \({90.6}\) \\ 
& \({200}\) 
& \({91.8}\) 
& \({91.0}\) 
& \({88.2}\) 
& \({92.1}\) 
& \({93.9}\) \\[1ex]
\multirow{3}{*}{\({20}\)} 
& \({100}\) 
& \({87.9}\) 
& \({84.6}\) 
& \({35.1}\) 
& \({54.7}\) 
& \({82.5}\) \\ 
& \({200}\) 
& \({84.2}\) 
& \({82.8}\) 
& \({66.7}\) 
& \({77.0}\) 
& \({87.2}\) \\ 
& \({400}\) 
& \({86.7}\) 
& \({86.3}\) 
& \({81.3}\) 
& \({87.2}\) 
& \({92.0}\) \\[1ex]
\multirow{3}{*}{\({30}\)} 
& \({150}\) 
& \({88.4}\) 
& \({85.7}\) 
& \({23.1}\) 
& \({41.9}\) 
& \({75.9}\) \\ 
& \({300}\) 
& \({81.8}\) 
& \({80.0}\) 
& \({57.1}\) 
& \({72.6}\) 
& \({85.8}\) \\ 
& \({600}\) 
& \({86.1}\) 
& \({85.5}\) 
& \({78.2}\) 
& \({86.2}\) 
& \({91.1}\) \\
\bottomrule
\end{tabularx}}
\label{tab:cr_lm}
\end{table}

\section{Application}\label{sec:application}
\subsection{Median Income for Four-Person Families}
We present an application of RETEL to the estimation of median 1989 income for four-person families by state in the USA.
In the field of small area estimation \citep{ghosh1994small,rao2003small}, the state-level direct estimates provided by the Census Bureau based on the Current Population Survey data may not be sufficiently accurate for some states due to limited sample sizes.
To address this issue, Bayesian methods have been proposed, which incorporate additional information or related auxiliary variables specific to these small areas \citep{fay1979estimates,datta1996estimation,ghosh1996estimation}.
In particular, EL has been applied to small area estimation in hierarchical Bayesian models \citep{chaudhuri2011empirical,chaudhuri2017hamiltonian}.

Let \({Y_i}\), \({i} = {1, \dots, 51}\), represent the direct estimate of the 1989 median income for four-person families in the \({i}\)th state and the District of Columbia.
We also consider the direct estimate of the 1979 median income, denoted by \({X_{1i}}\), as an auxiliary variable.
Additionally, following \citet{chung2019estimation}, we incorporate the adjusted census median income denoted by \({X_{2i}}\), where \({X_{2i}} = {(\textnormal{PCI}_{i, 1989}/ \textnormal{PCI}_{i, 1979})X_{1i}}\).
Here, \({\textnormal{PCI}_{i, 1979}}\) and \({\textnormal{PCI}_{i, 1989}}\) refer to per capita income from the Bureau of Economic Analysis in 1979 and 1989, respectively.
All variables are standardized to ensure numerical stability and facilitate illustration.

As in the generalized linear model approach of \citet{chaudhuri2011empirical}, we assume that the \({Y_i}\) are conditionally independent given \({\theta_i}\).
Specifically, we assume the following:
\begin{align*}
E\left\{Y_{i} \mid \theta_i\right\} = \theta_i, &\quad \textnormal{Var}\left[Y_{i} \mid \theta_i\right] = V_i,\\  
\theta_i \mid \beta, \sigma^2 &\overset{\textnormal{ind}}{\sim}
N\left(X_i^\top \beta, \sigma^2\right),\\
\beta \mid \sigma^2 &\sim N\left(\beta_0, g\sigma^2 \left(X^\top X\right)^{-1}\right),\\
\sigma^2 &\sim \pi\left(\sigma^2\right) \propto \sigma^{-2}.
\end{align*}
Here, \({\beta} = {(\beta_1, \beta_2)}\), \({X}_i = {(X_{1i}, X_{2i})}\), and \({X}\) is the matrix with the \({i}\)th row given by \({X_i^\top}\).
The sampling variance \({V_i}\) is set to \({1}\).
We adopt the \({g}\)-prior of \citet{zellner1988bayesian} for \({\beta}\) with \({\beta_0} = {(X^\top X)^{-1} X^\top Y}\) and \({g} = {0.1}\), where \({Y} = {(Y_1, \dots, Y_{51})}\).
For the likelihood function, we use RETEL, EL, and ETEL with the bivariate estimating function \({(Y_i - \theta_i, (Y_i - \theta_i)^2 / V_i - 1)}\).

For each method, we use a random-walk Metropolis--Hastings algorithm to draw posterior samples of \({\theta}\), \({\beta}\), and \({\sigma^2}\) from four chains, each of length \({\num{250000}}\).
Both \({\textnormal{RETEL}_f}\) and \({\textnormal{RETEL}_r}\) employ \({\mu_{n, \theta}} = {n^{-1}\sum_{i = 1}^{n}g_i(\theta)}\), \({\Sigma_{n, \theta}} = {(n - 1)^{-1}\sum_{i = 1}^n(g_i(\theta) - \mu_{n, \theta})(g_i(\theta) - \mu_{n, \theta})^\top}\), and \({\tau_n} = {\log n}\), where \({n} = {51}\).
The maximum potential scale reduction factor of all the methods is \({1.0119}\) for \({\theta}\), \({1.0006}\) for \({\beta}\), and \({1.0137}\) for \({\sigma^2}\).
We compute the \({95\%}\) posterior credible interval for each \({\theta_i}\) and use the posterior median \({\widehat{\theta}_i}\) as an estimate for \({Y_i}\).

\cref{tab:theta_metric} provides a summary of the performance of the methods.
The results show that RETEL demonstrates improvement over EL and ETEL, exhibiting smaller deviations in all metrics and providing more accurate estimates.
Although the intervals from RETEL are comparable in length to those from EL and ETEL, the reduced version performs the best among the methods in terms of accuracy.
On the other hand, EL and ETEL exhibit nearly equivalent performances, aligning with the findings in \citet{chaudhuri2011empirical}.
\begin{table}[!t]
\centering
\def~{\hphantom{0}}
\caption{Comparison of the accuracy of estimates of \({\theta}\).
AAD, average absolute deviation; 
AARD, average absolute relative deviation; 
ASD, average squared deviation; 
ASRD, average squared relative deviation;
AL, average length of the \({51}\) credible intervals.
}
{\begin{tabularx}{0.7\textwidth}{c|YYYYY}
\toprule
Method & AAD & AARD & ASD & ASRD & AL\\ 
\midrule
\({\textnormal{RETEL}_f}\) 
& \({0.278}\) & \({0.911}\) & \({0.111}\) & \({3.982}\) & \({3.754}\) \\[1ex]
\({\textnormal{RETEL}_r}\) 
& \({0.272}\) & \({0.900}\) & \({0.109}\) & \({3.831}\) & \({3.781}\) \\[1ex]
EL 
& \({0.280}\) & \({0.947}\) & \({0.112}\) & \({4.587}\) & \({3.755}\) \\[1ex]
ETEL 
& \({0.279}\) & \({0.942}\) & \({0.115}\) & \({4.696}\) & \({3.729}\) \\
\bottomrule
\end{tabularx}}
\label{tab:theta_metric}
\end{table}

\subsection{Roman-Era Egyptian Lifespan}\label{subsec:egypt}
We analyze Roman-era Egyptian lifespan records to illustrate an extension of RETEL, with a focus on using a non-normal auxiliary distribution and hyperpriors for \({\tau_n}\). 
The dataset, notably studied by \citet{pearson1902change}, consists of 141 age-at-death observations from Egyptian mummies. 
\citet{hjort2018hybrid} applied a hybrid EL approach to these data using the estimating equation \({g(X_i, \theta)} = {\mathds{1}\{X_i \in A\} - \theta}\) for \({i} = {1, \dots, 141}\), where \({X_i}\) is the age at death and \({A}\) is a specified interval.
Here, the parameter \({\theta} \in {(0, 1)}\) represents the probability of the age at death falling within \({A}\), with the empirical binomial estimate denoted by \({\widehat{\theta}} = {n^{-1}\sum_{i = 1}^n\mathds{1}\{X_i \in A\}}\), where \({n} = {141}\).

With \({g_i(\theta)} \in {\{-\theta, 1-\theta\}}\), the convex hull constraint is always satisfied for any nontrivial \({A}\); thus, the geometric necessity of an auxiliary normal distribution becomes less relevant.
Instead, we consider an auxiliary random variable \({\widetilde{g}} = {Z - \delta}\), where \({Z} \sim {\textnormal{Bernoulli}(p)}\) and \({\delta} \in {(0, 1)}\).  
This shifted binary structure provides a natural fit for the support of the estimating function.
Following the moment-generating function and Kullback--Leibler divergence arguments for \({\widetilde{g}}\) in \cref{sec:retel}, the RETEL framework applies directly, and the component \({p_n(\theta, \lambda)}\) in \eqref{eq:penalty} is reformulated as \({p_n(\theta, \lambda)} = {\tau_n \exp(-\delta_{n, \theta} \lambda) (1 - p_{n, \theta} + p_{n, \theta} \exp(\lambda))}\).
Consequently, \(p_n(\theta, \lambda)\) remains strictly convex in \({\lambda}\), ensuring that the dual minimization problem has a unique solution.

Regarding the parameterization of \({p_n(\theta, \lambda)}\), we restrict our attention to the case where \({p_{n, \theta}} = {\delta_{n, \theta}}\), which preserves the property \({\lambda_{RET}(\widehat{\theta})} = {0}\) and maintains consistency with the normal case. 
We fix these parameters at \({\check{\theta}}\), the maximum likelihood estimate of \({\theta}\) obtained from a parametric Weibull fit.
In this setup, the regularization strength is governed by \({\tau_n}\) and \({\theta}\). 
This motivates a Bayesian treatment of \({\tau_n}\) to reflect uncertainty in the regularization, obviating the need to choose it manually by instead specifying a prior distribution.
Accordingly, we henceforth write \({\tau}\) instead of \({\tau_n}\) for notational simplicity.
This treatment is a possible extension of the RETEL construction rather than a fully specified Bayesian model.
With \({\tau}\) being a regularization parameter of the pseudo-likelihood, the joint posterior of \({(\theta, \tau)}\) is the pseudo-posterior \({\pi(\theta, \tau \mid \mathcal{D}_n)} \propto {\pi(\theta, \tau)L_{RET}(\theta)}\), where \({L_{RET}(\theta)}\) is computed at \({\tau}\).

Specifically, we assign a \({\textnormal{Gamma}(2, 1)}\) prior to \({\tau}\), which has its mode at \({1}\).
This choice corresponds to a unit-information baseline that weights the auxiliary component as a single pseudo-observation, consistent with an effective sample size interpretation \citep{morita2008determining}.
This specification also mirrors the use of Gamma priors for penalty terms in the Bayesian Lasso \citep{park2008bayesian,casella2010penalized}.
We avoid improper priors for \({\tau}\) to ensure posterior propriety, as the RETEL construction does not admit integrability over an unbounded regularization domain.
For \({\theta}\), we consider two specifications: an unconditional \(\textnormal{Uniform}(0, 1)\) prior and a conditional prior defined as \({\theta \mid \tau} \sim {\textnormal{Beta}(\check{\theta}\tau, (1 - \check{\theta})\tau)}\).
In this formulation, \({\tau}\) serves as the precision parameter (or prior effective sample size), ensuring that as the auxiliary information becomes more dominant, the prior distribution becomes more concentrated around \({\check{\theta}}\).

\begin{figure}[!t]
\centering
\begin{subfigure}{0.495\linewidth}
\includegraphics[width=\linewidth]{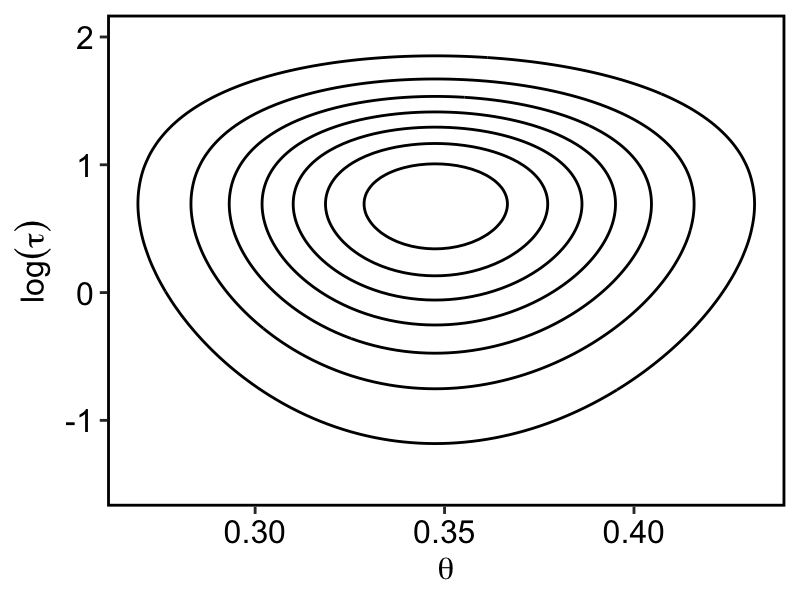}
\caption{}
\label{fig:uniform}
\end{subfigure}
\hfill
\begin{subfigure}{0.495\linewidth}
\includegraphics[width=\linewidth]{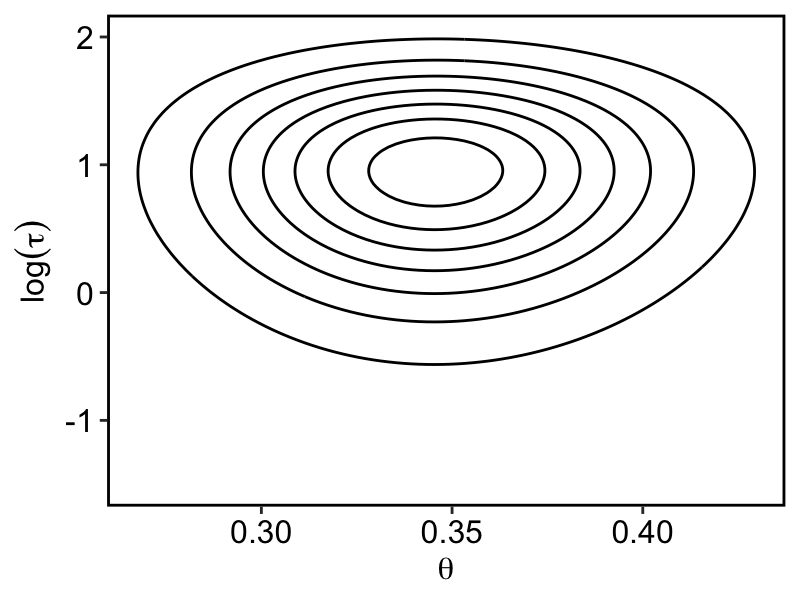}
\caption{}
\label{fig:beta}
\end{subfigure}
\caption{
Contour plots of the joint posterior density of \({(\theta, \log \tau)}\) for the Egyptian lifespan data.
The logarithmic scale for \({\tau}\) is employed to facilitate visualization.
(a) Posterior density under \({\theta \sim \textnormal{Uniform}(0, 1)}\).
(b) Posterior density under \({{\theta} \mid {\tau} \sim \textnormal{Beta}(\check{\theta}\tau, (1 - \check{\theta})\tau)}\), where \({\check{\theta}}\) is the maximum likelihood estimate obtained from a parametric Weibull fit.
Both specifications assume \({\tau \sim \textnormal{Gamma}(2, 1)}\).}
\label{fig:contour}
\end{figure}

For illustration, we set the interval to \({A} = {[10, 25]}\), which yields \({\widehat{\theta}} = {0.348}\) and \({\check{\theta}} = {0.360}\).
We employ \(\textnormal{RETEL}_f\) here, though the difference from 
\(\textnormal{RETEL}_r\) is minimal as the convex hull constraint is not binding.
The joint posterior densities for both prior specifications are presented in \cref{fig:contour}. Under the unconditional prior (\cref{fig:uniform}), the contours exhibit a U-shape in the \((\theta, \log \tau)\) plane. 
When \({\theta}\) is in the vicinity of the empirical and parametric estimates, the auxiliary distribution requires minimal tilting, providing broader support for \({\tau}\) across the vertical plane.
As \({\theta}\) deviates from this central region, the posterior density concentrates toward larger values of \({\tau}\), where the increased regularization dampens the magnitude of the multiplier \({\lambda}\). 
This leaves the lower-left and lower-right corners of the parameter space negligible.
In contrast, under the conditional prior (\cref{fig:beta}), this U-shaped curvature is superseded by a more vertically concentrated density. 
Furthermore, the marginal density of \({\log \tau}\) concentrates at higher values relative to the unconditional case.
This occurs because the prior precision scales with \({\tau}\), which reinforces the likelihood's regularization toward \({\check{\theta}}\). 
This upward shift reflects the stricter penalty imposed on deviations from \({\check{\theta}}\) as \({\tau}\) increases, leading the posterior to concentrate at higher regularization strengths.

\section{Discussion}\label{sec:discussion}
This paper has investigated a suite of methods to deal with the convex hull constraint without the need to invoke parameter-dependent pseudo-data.  
The first step was the development of WETEL as an extension of AEL and AETEL.  
WETEL accommodates fractional observations and reduces the dependence of pseudo-data on the parameter, allowing for a massive expansion of the convex hull while aligning the pseudo-data more closely with the observed data.
As a subsequent step, WETEL leads to the regularization technique of RETEL by passing to a limit where pseudo-data are added in a particular way.  
We also provided a distinct derivation of RETEL as the solution to a Kullback--Leibler divergence optimization problem involving a mixture of the empirical distribution and a continuous exponential family distribution.  

The likelihood ratios from RETEL compare the constrained regularized likelihood to the unconstrained regularized likelihood.
This is implicit in \eqref{eq:RETEL_f} and \eqref{eq:RETEL_r}.  
In essence, RETEL replaces the empirical distribution with a regularized empirical distribution before considering tilts that match constraints.  
This stabilizes the results, particularly for smaller sample sizes.  
It also appears to produce a posterior distribution that is less pathological and more amenable to traditional sampling techniques for model fitting.  
We showed that RETEL retains the desirable properties of EL and ETEL such as Wilks' and Bernstein--von Mises theorems.  
The simulation and data analysis demonstrated that RETEL exhibits improved finite-sample performance compared to EL and ETEL for Bayesian inference.
Overall, our findings highlight the effectiveness of RETEL as a pseudo-likelihood for Bayesian inference in overcoming the convex hull constraint of EL and ETEL. 

There are a number of reasons to replace integration in a Bayesian model with maximization.  
In addition to handling the nuisance parameter, maximization can be much quicker than integration.  
We suspect that an appropriate regularization in RETEL will bring the maximized version of the problem closer to a genuine Bayesian solution.  
This represents a natural direction for future research.
One promising avenue is the development of fully Bayesian extensions of RETEL, including hierarchical formulations, as illustrated in \cref{subsec:egypt}.
Such extensions would include assigning prior distributions to the regularization parameters to formally account for uncertainty in the auxiliary information, as well as exploring non-normal exponential family distributions to better accommodate a broader range of data and model supports.
Another important direction is to investigate whether RETEL retains the robust higher-order asymptotic properties of ETEL.
\citet{schennach2007point} showed that ETEL has robust higher-order asymptotic properties under model misspecification compared to the EL estimator. 
\citet{chib2018bayesian} established Bernstein--von Mises results for ETEL under model misspecification. 
Further research is needed to determine the extent to which these properties hold for RETEL.


\section*{Funding}
This work was supported by the National Science Foundation under Grants Nos.~SES-1921523 and DMS-2015552. 

\section*{Supplementary material}
Supplement to ``Regularized Exponentially Tilted Empirical Likelihood for Bayesian Inference''.
The supplementary material provides technical proofs of the results presented in the main text, additional plots from the simulations, and computational details regarding hardware specifications and empirical runtimes.

\bibliographystyle{plainnat}
\bibliography{bibliography}

\end{document}


\title{Supplementary Material: Regularized Exponentially Tilted Empirical Likelihood for Bayesian Inference}

\author{%
 Eunseop Kim\thanks{\texttt{markean@pm.me}} \quad
 Steven N.~MacEachern\thanks{\texttt{snm@stat.osu.edu}} \quad
 Mario Peruggia\thanks{\texttt{peruggia@stat.osu.edu}} \\[4pt]
 Department of Statistics, The Ohio State University
}
\date{}

\maketitle

\section{Proofs}
All references to theorems and propositions are to the main paper unless otherwise specified.
We restate the conditions of the main paper below.
\begin{condition}\label{condition:parameter space}
The parameter space \({\Theta}\) is compact, with \({\theta_0}\) an interior point of \({\Theta}\) and the unique solution to \({E_P\{g_i(\theta)\}} = {0}\).
\end{condition}
\begin{condition}\label{condition:continuity}
With probability~\({1}\), \({g_i(\theta)}\) is continuous at each \({\theta} \in {\Theta}\) and continuously differentiable in a neighborhood \({\mathcal{N}}\) of \({\theta_0}\), and \({E_P\{\sup_{\theta \in \mathcal{N}} \Vert \partial_{\theta}g_i(\theta) \Vert\}} < {\infty}\).
\end{condition}
\begin{condition}\label{condition:full rank}
\({\textnormal{rank}({G})} = {\textnormal{rank}({V})} = {p}\).
\end{condition}
\begin{condition}\label{condition:uniform boundedness}
For some \({\alpha} > {3}\), \({E_P\{\sup_{\theta \in \Theta} \vert g_i(\theta) \vert^\alpha\}} < {\infty}\).
\end{condition}
\begin{condition}\label{condition:retel asymptotics}
\({\tau_n} = {O(\log n)}\) and \({\tau_n^{-1}} = {O(n^{C_0})}\) for some constant \({C_0} \geq {0}\).
With probability~\({1}\), \({\Sigma_{n, \theta}}\) is positive definite for every \({\theta} \in {\Theta}\) and any \({n}\).
For some \({\mu} \in {\mathbb{R}^p}\) and \({\Sigma} \in {\mathbb{R}^{p \times p}}\), \({\mu_{n, \theta_0}} = {\mu} + {o_P(1)}\) and \({\Sigma_{n, \theta_0}} = {\Sigma} + {o_P(1)}\).
There exist a neighborhood \({\mathcal{N}_0}\) of \({\theta_0}\) and constants \({0} < {b_1} \leq {b_2} < {\infty}\) such that, with probability approaching \({1}\), \({\sup_{\theta \in \mathcal{N}_0}\left\vert \mu_{n, \theta} \right\vert} \leq {b_2}\) and the eigenvalues of \({\Sigma_{n, \theta}}\) lie in \({[b_1, b_2]}\) for all \({\theta} \in {\mathcal{N}_0}\).
\end{condition}
\begin{condition}\label{condition:prior}
The prior measure admits a density with respect to the Lebesgue measure.
The density \({\pi}(\cdot)\) is continuous in \({\Theta}\) and is positive in a neighborhood of \({\theta_0}\).
\end{condition}
\begin{condition}\label{condition:bvm}
For any \({\delta} > {0}\), there exists \({\epsilon} > {0}\) such that
\begin{equation*}
\textnormal{pr}\left(
\sup_{\left\vert \theta - \theta_0 \right\vert > \delta}
\frac{1}{n}\left(\log L_{RET}\left(\theta\right) - \log L_{RET}\left(\theta_0\right)\right) \leq -\epsilon
\right)
\to
1.
\end{equation*}
For the posterior obtained from \({\textnormal{RETEL}_r}\), the same condition
is assumed with \({\widetilde{L}_{RET}}\) in place of \({L_{RET}}\).
\end{condition}

\subsection{Proposition~1}
We employ the same notation as in the main paper and introduce the following quantities:
\begin{align*}
h\left(\theta\right)
&=
E_P\left\{g_i\left(\theta\right)\right\},\quad
h_n\left(\theta\right)
=
\frac{1}{n}\sum_{i = 1}^n g_i\left(\theta\right),\quad
V_n\left(\theta\right)
=
\frac{1}{n}\sum_{i = 1}^n g_i\left(\theta\right)g_i\left(\theta\right)^\top,\\
\widetilde{h}_N\left(\theta\right)
&=
\sum_{i = 1}^{N}w_ig_i\left(\theta\right)
=
\frac{n}{n + 1}h_n\left(\theta\right) + \frac{1}{m\left(n + 1\right)}\sum_{i = n + 1}^{N}g_i\left(\theta\right),\\
\widetilde{V}_N\left(\theta\right)
&=
\sum_{i = 1}^{N}w_ig_i\left(\theta\right){g_i\left(\theta\right)}^\top
=
\frac{n}{n + 1}V_n\left(\theta\right) +
\frac{1}{m\left(n + 1\right)}\sum_{i = n + 1}^{N}g_i\left(\theta\right){g_i\left(\theta\right)}^\top.
\end{align*}
Since the value of \({m}\) is fixed, the terms \({m^{-1}\sum_{i = n + 1}^N g_i(\theta)}\) and \({m^{-1}\sum_{i = n + 1}^N g_i(\theta){g_i(\theta)}^\top}\) are finite for each \({\theta}\).
By the weak law of large numbers and \cref{condition:parameter space}, we have
\begin{equation*}
h_n(\theta_0) = o_P\left({1}\right),\quad \widetilde{h}_N(\theta_0) = o_P\left({1}\right),\quad V_n\left(\theta_0\right) = V + o_P\left({1}\right),\quad \widetilde{V}_N\left({\theta_0}\right) = V + o_P\left({1}\right).
\end{equation*}
Applying the uniform law of large numbers and \cref{condition:continuity}, we obtain
\begin{align*}
\sup_{\theta \in \mathcal{N}}\left\Vert
\partial_{\theta}h_n\left(\theta\right)
-
E_P\left\{\partial_{\theta}g_i\left(\theta\right)\right\}\right\Vert
&=
{o_P\left({1}\right)},\\
\sup_{\theta \in \mathcal{N}}\left\Vert
\partial_{\theta}\widetilde{h}_N\left(\theta\right)
-
E_P\left\{\partial_{\theta}g_i\left(\theta\right)\right\}\right\Vert
&=
{o_P\left({1}\right)}.
\end{align*}
By \cref{condition:full rank} and \citet[Theorem~2.5]{jacod2018review}, there exist consistent estimators \({\widehat{\theta}}\) and \({\widehat{\theta}_w}\) that converge in probability to \({\theta_0}\), and \({h_n(\widehat{\theta})} = {0}\) and \({\widetilde{h}_N(\widehat{\theta}_w)} = {0}\) with probability approaching \({1}\).
Using \cref{condition:full rank} and the central limit theorem, we can show that both \({n^{1/2}{h}_n(\theta_0)}\) and \({n^{1/2}\widetilde{h}_N(\theta_0)}\) are stochastically bounded.
As a result, \({n^{1/2}\vert \widehat{\theta} - \theta_0\vert}\) and \({n^{1/2}\vert \widehat{\theta}_w - \theta_0\vert}\) are also stochastically bounded \citep[Theorem~2.9]{jacod2018review}, which establishes \({\widehat{\theta}_w - \widehat{\theta}} = {o_P(n^{-1/2})}\).

Next, the first-order condition for \({\lambda_{WET}(\theta_0)}\) yields
\begin{equation*}
\sum_{i = 1}^N w_i\exp\left(\lambda^\top g_i\left(\theta_0\right)\right) g_i\left(\theta_0\right)
=
0.
\end{equation*}
Using the weight specification for \({w_i}\) in the main paper, it follows from \citet[Lemma~A2]{newey2004higher}, together with \cref{condition:parameter space,condition:continuity,condition:full rank,condition:uniform boundedness}, that \({\lambda_{WET}(\theta_0)} = {O_P(n^{-1/2})}\) with probability approaching \({1}\).
Expanding the condition around \({\lambda} = {0}\), we obtain
\begin{equation}\label{eq:wet lambda expansion}
0
=
\widetilde{h}_N\left(\theta_0\right) +
\widetilde{V}_N\left(\theta_0\right)
\lambda_{WET}\left(\theta_0\right) +
R_N,
\end{equation}
where
\begin{equation*}
R_N
=
\frac{1}{2}\sum_{i = 1}^N w_i
\exp\left({\xi}^\top{g_i\left(\theta_0\right)}\right)
{\lambda_{WET}\left(\theta_0\right)}^\top {g_i\left(\theta_0\right)}{g_i\left(\theta_0\right)}^\top{\lambda_{WET}\left(\theta_0\right)}{g_i\left(\theta_0\right)}
\end{equation*}
for some \({\xi}\) between \({0}\) and \({\lambda_{WET}(\theta_0)}\).
Let
\begin{equation*}
R_n
=
\frac{1}{2\left(n + 1\right)}\sum_{i = 1}^n
\exp\left({\xi}^\top{g_i\left(\theta_0\right)}\right)
{\lambda_{WET}\left(\theta_0\right)}^\top {g_i\left(\theta_0\right)}{g_i\left(\theta_0\right)}^\top{\lambda_{WET}\left(\theta_0\right)}{g_i\left(\theta_0\right)}
\end{equation*}
and
\begin{equation*}
R_m
=
\frac{1}{2m\left(n + 1\right)}\sum_{i = n + 1}^{N}
\exp\left({\xi}^\top{g_i\left(\theta_0\right)}\right)
{\lambda_{WET}\left(\theta_0\right)}^\top {g_i\left(\theta_0\right)}{g_i\left(\theta_0\right)}^\top{\lambda_{WET}\left(\theta_0\right)}{g_i\left(\theta_0\right)}.
\end{equation*}
Then, combining \({\lambda_{WET}(\theta_0)} = {O_P(n^{-1/2})}\), \cref{condition:uniform boundedness}, and Lemma~A1 in \citet{newey2004higher}, we obtain \({R_N} =
{R_n} + {R_m} = {O_P(n^{-1})} + O_P(n^{-2})\) and
\begin{equation}\label{eq:wet lambda expression}
\lambda_{WET}\left(\theta_0\right)
=
-
{\widetilde{V}_N\left(\theta_0\right)}^{-1}
\widetilde{h}_N\left(\theta_0\right)
+
O_P\left(n^{-1}\right).
\end{equation}
By following the same steps for \({\lambda_{ET}}\), we get \({\lambda_{ET}(\theta_0)} = {-
{V_n(\theta_0)}^{-1}
h_n(\theta_0)}
+
{O_P(n^{-1})}\).
Therefore, we have \({\lambda_{WET}(\theta_0)} - {\lambda_{ET}(\theta_0)} = {O_P(n^{-1})}\).
\(\hfill\blacksquare\)

\subsection{Theorem~1}
The saddle point problem of \({\widehat{\theta}_{w}}\) and \({\lambda_{WET}(\widehat{\theta}_{w})}\) leads to the following first-order conditions:
\begin{align*}
\sum_{i = 1}^N w_i\exp\left(\lambda^\top g_i\left(\theta\right)\right) \partial_{\theta} g_i\left(\theta\right)^\top \lambda
&= 0,\\
\sum_{i = 1}^N w_i\exp\left(\lambda^\top g_i\left(\theta\right)\right) g_i\left(\theta\right)
&= 0.
\end{align*}
With \cref{condition:parameter space,condition:continuity,condition:full rank,condition:uniform boundedness}, we can directly apply the results from \citet[Theorem~3.2]{newey2004higher} and \citet[Theorem~1]{zhu2009adjusted}.
By using \eqref{eq:wet lambda expansion} and expanding the conditions around \({\theta} = {\theta_0}\) and \({\lambda} = {0}\), we obtain
\begin{equation*}
W_N
\begin{pmatrix}
\theta - \theta_0 \\
\lambda
\end{pmatrix}
=
\begin{pmatrix}
0\\
-\widetilde{h}_N\left(\theta_0\right)
\end{pmatrix} + o_P\left(n^{-1/2}\right),
\end{equation*}
where
\begin{equation*}
W_N
=
\begin{pmatrix}
0 & \sum_{i = 1}^N w_i\partial_{\theta} g_i\left(\theta_0\right)^\top\\
\sum_{i = 1}^N w_i\partial_{\theta} g_i\left(\theta_0\right) &
\widetilde{V}_N\left(\theta_0\right)
\end{pmatrix}
\to
W
=
\begin{pmatrix}
0 & G^\top \\
G & V
\end{pmatrix}
\end{equation*}
in probability.
Consequently, we have
\begin{equation*}
W^{-1}
=
\begin{pmatrix}
-\Omega & H\\
H^\top & P
\end{pmatrix},
\end{equation*}
where \({H} = {\Omega{G}^\top {V}^{-1}}\) and \({P} = {{V}^{-1} - {V}^{-1}G\Omega{G}^\top {V}^{-1}}\).
Hence, we obtain
\begin{equation*}
\widehat{\theta}_{w} - \theta_0 = -H\widetilde{h}_N\left(\theta_0\right) + o_P\left(n^{-1/2}\right),
\end{equation*}
and the first result follows by noting that \({n^{1/2}\widetilde{h}_N(\theta_0)}\) converges in distribution to \({N(0, V)}\) and \({HVH^\top} = {\Omega}\).
For the second result, observe that
\begin{align}\label{eq:wet lr}
\begin{split}
-2&\log R_{WET}\left(\theta_0\right)\\
&= -2N\sum_{i = 1}^N w_i
\left\{
{\lambda_{WET}\left(\theta_0\right)}^\top g_i\left(\theta_0\right)
-
\log\left(
\sum_{j = 1}^N w_j \exp\left({\lambda_{WET}\left(\theta_0\right)}^\top g_j\left(\theta_0\right)\right)
\right)
\right\} \\
&= -2N {\lambda_{WET}\left(\theta_0\right)}^\top \widetilde{h}_N\left(\theta_0\right)
+
2N \log\left(
\sum_{i = 1}^N w_i \exp\left({\lambda_{WET}\left(\theta_0\right)}^\top g_i\left(\theta_0\right)\right)
\right)
\end{split}
\end{align}
and
\begin{align}\label{eq:wet log expansion}
\begin{split}
\log&\left(
\sum_{i = 1}^N w_i \exp\left({\lambda_{WET}\left(\theta_0\right)}^\top g_i\left(\theta_0\right)\right)\right)\\
&= \log\left(1 + {\lambda_{WET}\left(\theta_0\right)}^\top \widetilde{h}_N\left(\theta_0\right)
+
{\lambda_{WET}\left(\theta_0\right)}^\top \widetilde{V}_N\left(\theta_0\right) \lambda_{WET}\left(\theta_0\right) / 2
+ o_P\left(n^{-1}\right)
\right) \\
&= {\lambda_{WET}\left(\theta_0\right)}^\top \widetilde{h}_N\left(\theta_0\right)
+
{\lambda_{WET}\left(\theta_0\right)}^\top \widetilde{V}_N\left(\theta_0\right) \lambda_{WET}\left(\theta_0\right) / 2
+ o_P\left(n^{-1}\right).
\end{split}
\end{align}
Substituting the expressions in \eqref{eq:wet log expansion} and \eqref{eq:wet lambda expression} into \eqref{eq:wet lr}, we obtain
\begin{align*}
-2\log R_{WET}\left(\theta_0\right)
&=
N {\lambda_{WET}\left(\theta_0\right)}^\top \widetilde{V}_N\left(\theta_0\right) {\lambda_{WET}\left(\theta_0\right)}
+
o_P\left(1\right)\\
&=
N {\widetilde{h}_N\left(\theta_0\right)}^\top{\widetilde{V}_N\left(\theta_0\right)}^{-1}\widetilde{h}_N\left(\theta_0\right) +
o_P\left(1\right),
\end{align*}
and the result follows.
\(\hfill\blacksquare\)

\subsection{Proposition~2}
Fix \({\theta} \in {\Theta}\).
The function \({d_n(\theta, \cdot)}\) is finite, nonnegative, convex, and continuous on \({\mathbb{R}^p}\), with \({d_n(\theta, 0)} = {n}\).
The function \({p}\) is finite and continuous on \({\mathbb{R}^p}\).
Since the covariance matrix \({\Sigma}\) has full rank, \({p}\) is strictly convex and \({p(\lambda)} \to {\infty}\) as \({\vert\lambda\vert} \to {\infty}\).
Hence, the function \({c}\) is finite, continuous, strictly convex, and level-bounded.
The level set \({\{\lambda \in \mathbb{R}^p : c(\lambda) \leq n + 1\}}\) is therefore compact, and it is nonempty since \({c(0)} = {n + 1}\).
The minimum of \({c}\) over this set is attained at some \({\lambda_{RET}}\), which is the unique global minimizer of \({c}\) since \({c} > {n + 1} \geq {c(\lambda_{RET})}\) outside the set and \({c}\) is strictly convex.

For the second claim, \({p_m}\) epi-converges to \({p}\) as \({m} \to {\infty}\) with probability~\({1}\) \citep[Theorem~2.3]{artstein1995consistency}.
The event on which this holds does not depend on \({\theta}\).
Since \({d_n(\theta, \cdot)}\) is continuous, \({c_m}\) epi-converges to \({c}\) on the same event.
Fix an outcome in this event.
Each \({c_m}\) is finite, continuous, and convex, with \({c_m(0)} = {n + 1}\).

Choose \({R} > {0}\) such that \({c(\lambda)} \geq {2n + 4}\) whenever \({\vert\lambda\vert} = {R}\), which is possible since \({c} \geq {p}\) and \({p(\lambda)} \to {\infty}\) as \({\vert\lambda\vert} \to {\infty}\).
Let \({\lambda_{R, m}}\) attain \({\min_{\vert\lambda\vert = R}c_m(\lambda)}\).
Choose a subsequence \({m_k}\) along which these minimum values converge to their limit inferior.
By compactness of the sphere \({\{\lambda \in \mathbb{R}^p : \vert\lambda\vert = R\}}\), pass to a further subsequence, still indexed by \({m_k}\), such that \({\lambda_{R, m_k}} \to {\lambda_R}\) with \({\vert\lambda_R\vert} = {R}\).
By the construction of the subsequence, the epi-convergence of \({c_{m_k}}\) to \({c}\), and the choice of \({R}\),
\begin{equation*}
\liminf_{m \to \infty}\min_{\vert\lambda\vert = R}c_m(\lambda)
=
\lim_{k \to \infty}c_{m_k}(\lambda_{R, m_k})
\geq
c(\lambda_R)
\geq
2n + 4.
\end{equation*}
Let \({m_*}\) be such that \({\min_{\vert\lambda\vert = R}c_m(\lambda)} \geq {2n + 3}\) for all \({m} \geq {m_*}\).
For \({m} \geq {m_*}\) and \({\vert\lambda\vert} > {R}\), write \({t} = {\vert\lambda\vert/R} > {1}\), so that \({\vert\lambda/t\vert} = {R}\).
The convexity of \({c_m}\) gives \({c_m(\lambda/t)} \leq {(1/t)c_m(\lambda) + (1 - 1/t)c_m(0)}\), and thus
\begin{equation*}
c_m(\lambda)
\geq
c_m(0) + t\left(c_m\left(\frac{\lambda}{t}\right) - c_m(0)\right)
\geq
n + 1 + \left(n + 2\right)\frac{\left\vert\lambda\right\vert}{R}.
\end{equation*}
Let \({\alpha} \in {\mathbb{R}}\), \({m} \geq {m_*}\), and \({\lambda}\) satisfy \({c_m(\lambda)} \leq {\alpha}\).
Either \({\vert\lambda\vert} \leq {R}\), or \({\vert\lambda\vert} > {R}\) and
\begin{equation*}
n + 1 + \left(n + 2\right)\frac{\left\vert\lambda\right\vert}{R} \leq c_m\left(\lambda\right) \leq \alpha,
\end{equation*}
so that \({\vert\lambda\vert} \leq {R(\alpha - n - 1)/(n + 2)} \leq {R\alpha/(n + 2)}\).
In either case,
\begin{equation*}
\left\vert\lambda\right\vert \leq R\max\left\{1, \frac{\alpha}{n + 2}\right\}.
\end{equation*}
The level set \({\{\lambda \in \mathbb{R}^p : c_m(\lambda) \leq \alpha\}}\) is therefore contained in the ball of radius \({R\max\{1, \alpha/(n + 2)\}}\); that is, \({\{c_m\}}\) is eventually level-bounded.

By the epi-convergence of \({c_m}\) to \({c}\), there is a sequence \({\lambda_m^\prime} \to {\lambda_{RET}}\) with
\begin{equation*}
\limsup_{m \to \infty}\inf_{\lambda \in \mathbb{R}^p}c_m(\lambda)
\leq
\limsup_{m \to \infty}c_m(\lambda_m^\prime)
\leq
c(\lambda_{RET}).
\end{equation*}
Let \({\epsilon_m} \downarrow {0}\) and write \({S_m} = {\{\epsilon_m\textnormal{-}\argmin_{\lambda \in \mathbb{R}^p} c_m(\lambda)\}}\).
For \({m} \geq {m_*}\), \({c_m}\) is continuous and level-bounded, so it attains its minimum and \({S_m}\) is nonempty.
Let \({\lambda_m} \in {S_m}\) for \({m} \geq {m_*}\).
Since \({c_m(\lambda_m)} \leq {\inf_{\lambda \in \mathbb{R}^p} c_m(\lambda) + \epsilon_m} \leq {n + 1 + \epsilon_1}\),
\begin{equation*}
\left\vert\lambda_m\right\vert
\leq
R\max\left\{1, \frac{n + 1 + \epsilon_1}{n + 2}\right\}.
\end{equation*}
If \({\lambda^*}\) is a cluster point of \({\{\lambda_m\}}\), with \({\lambda_{m_k}} \to {\lambda^*}\), the epi-convergence of \({c_{m_k}}\) to \({c}\) gives
\begin{align*}
c(\lambda^*)
&\leq
\liminf_{k \to \infty}c_{m_k}(\lambda_{m_k})\\
&\leq
\liminf_{k \to \infty}\left(\inf_{\lambda \in \mathbb{R}^p}c_{m_k}(\lambda) + \epsilon_{m_k}\right)\\
&\leq
\limsup_{k \to \infty}\left(\inf_{\lambda \in \mathbb{R}^p}c_{m_k}(\lambda) + \epsilon_{m_k}\right)\\
&=
\limsup_{k \to \infty}\inf_{\lambda \in \mathbb{R}^p}c_{m_k}(\lambda)\\
&\leq
\limsup_{m \to \infty}\inf_{\lambda \in \mathbb{R}^p}c_m(\lambda)\\
&\leq
c(\lambda_{RET}),
\end{align*}
and thus \({\lambda^*} = {\lambda_{RET}}\).
Every cluster point of the bounded sequence \({\{\lambda_m\}}\) is therefore \({\lambda_{RET}}\), and \({\lambda_m} \to {\lambda_{RET}}\).
Since \({S_m}\) is nonempty for \({m} \geq {m_*}\) and every sequence of points \({\lambda_m} \in {S_m}\) converges to \({\lambda_{RET}}\), both the inner and outer limits of \({\{S_m\}}\) equal \({\{\lambda_{RET}\}}\).
The set convergence \({\lim_{m \to \infty}S_m} = {\{\lambda_{RET}\}}\) follows.
\(\hfill\blacksquare\)

\subsection{Proposition~3}
We fix \({\theta}\) and consider maximizing
\begin{align*}
-D_{KL}\left(\widetilde{P}_{\lambda} \relmiddle \Vert \widetilde{P}_n\right)
&=
-\int_{\mathcal{X}}
\log\left(
\frac{\widetilde{P}_{\lambda}\left(d\omega\right)}{\widetilde{P}_n\left(d\omega\right)}
\right)
\widetilde{P}_{\lambda}\left(d\omega\right)\\
&=
-\sum_{i = 1}^n p_i \log\left(\left(n + \tau_n\right)p_i\right) -
p_c\log\left(
\frac{n + \tau_n}{\tau_n}p_c\right) -
\frac{p_c}{2}\lambda^\top \Sigma_{n, \theta} \lambda,
\end{align*}
subject to the moment constraint
\begin{equation*}
\sum_{i = 1}^n p_i g_i\left(\theta\right) + p_c E_{\widetilde{P}_{\lambda}}\left\{{\widetilde{g}_{\lambda}}\right\}
=
\sum_{i = 1}^n p_i g_i\left(\theta\right) + p_c\left(\mu_{n, \theta} + \Sigma_{n, \theta}\lambda\right)
=
0.
\end{equation*}
The Lagrangian associated with the constrained maximization problem is
\begin{align*}
L
=&
-\sum_{i = 1}^n p_i \log p_i -
p_c\log p_c + p_c \log \tau_n -
\frac{p_c}{2}\lambda^\top \Sigma_{n, \theta} \lambda\\
&+\kappa^\top\left(\sum_{i = 1}^n p_ig_i\left(\theta\right) + p_c\left(\mu_{n, \theta} + \Sigma_{n, \theta}\lambda\right)\right) + \nu \left(\sum_{i = 1}^n p_i
+ p_c - 1\right),
\end{align*}
where \({\kappa} \in {\mathbb{R}^p}\) and \({\nu} \in {\mathbb{R}}\) are Lagrange multipliers.
Differentiating the Lagrangian expression with respect to each \({p_i}\) and \({p_c}\), and equating the derivatives to zero, we have \({\kappa} = {\lambda}\) and
\begin{equation*}
\nu =
\sum_{i = 1}^n p_i \log p_i + p_c \log p_c + \frac{p_c}{2}\lambda^\top \Sigma_{n, \theta}\lambda - p_c \log \tau_n + 1.
\end{equation*}
After some algebra, it can be shown that
\begin{equation*}
p_i\left(\theta\right) = \frac{\exp\left({\lambda_{RET}}^\top g_i\left(\theta\right)\right)}{c_n\left(\theta, \lambda_{RET}\right)}\quad
\left(i = 1, \dots, n\right),\quad
p_c\left(\theta\right) = \frac{p_n\left(\theta, \lambda_{RET}\right)}{c_n\left(\theta, \lambda_{RET}\right)},
\end{equation*}
where \({c_n(\theta, \lambda_{RET})}\) is the normalizing constant.
This leads to solving the dual problem, and the result follows.
\(\hfill\blacksquare\)

\subsection{Theorem~2}
We begin by establishing that \({\lambda_{RET}(\theta_0)} = {O_P(n^{-1/2})}\).
Observe that
\begin{align*}
\frac{1}{n}c_n\left(\theta_0, 0\right)
&= \frac{1}{n}d_n\left(\theta_0, 0\right)
+
\frac{1}{n}p_n\left(\theta_0, 0\right)\\
&= 1 + \frac{\tau_n}{n}\\
&\geq
\frac{1}{n}c_n\left(\theta_0, \lambda_{RET}\left(\theta_0\right)\right),
\end{align*}
where the last inequality follows from the definition of \({\lambda_{RET}(\theta_0)}\).
We perform a Taylor expansion of \({c_n(\theta_0, \lambda_{RET}(\theta_0))}\) around \({\lambda_{RET}(\theta_0)} = {0}\), yielding
\begin{align*}
\frac{1}{n}c_n&\left(\theta_0, \lambda_{RET}\left(\theta_0\right)\right)
=
1 + \frac{\tau_n}{n} + {\lambda_{RET}\left(\theta_0\right)}^\top \left(h_n\left(\theta_0\right) + \frac{\tau_n}{n}\mu_{n, \theta_0}\right) \\
&+ \frac{1}{2n}{\lambda_{RET}\left(\theta_0\right)}^\top \left(
\sum_{i = 1}^n\exp\left({\widetilde{\lambda}}^\top g_i\left(\theta_0\right)\right)g_i\left(\theta_0\right){g_i\left(\theta_0\right)}^\top
\right)
\lambda_{RET}\left(\theta_0\right)\\
&+ \frac{\tau_n}{2n}{\lambda_{RET}\left(\theta_0\right)}^\top \left(
\mu_{n, \theta_0}{\mu_{n, \theta_0}}^\top + \Sigma_{n, \theta_0}\right)
\lambda_{RET}\left(\theta_0\right),
\end{align*}
where \({\widetilde{\lambda}}\) lies between \({0}\) and \({\lambda_{RET}}(\theta_0)\).
Using the above expansion, we find
\begin{align*}
0
\geq
&{\lambda_{RET}\left(\theta_0\right)}^\top
\left(h_n\left(\theta_0\right) + \frac{\tau_n}{n}\mu_{n, \theta_0}\right) \\
&-
\frac{1}{2n}{\lambda_{RET}\left(\theta_0\right)}^\top \left(
\sum_{i = 1}^n\left(-\exp\left({\widetilde{\lambda}}^\top g_i\left(\theta_0\right)\right)\right)g_i\left(\theta_0\right){g_i\left(\theta_0\right)}^\top
\right)
\lambda_{RET}\left(\theta_0\right)\\
&+
\frac{\tau_n}{2n}{\lambda_{RET}\left(\theta_0\right)}^\top
\left(\mu_{n, \theta_0}{\mu_{n, \theta_0}}^\top + \Sigma_{n, \theta_0}\right)
\lambda_{RET}\left(\theta_0\right)\\
\geq&
{\lambda_{RET}\left(\theta_0\right)}^\top
\left(h_n\left(\theta_0\right) + \frac{\tau_n}{n}\mu_{n, \theta_0}\right)\\
&-
\max_{1 \leq i \leq n}\left\{
-\exp\left({\widetilde{\lambda}}^\top g_i\left(\theta_0\right)\right)
\right\}\frac{1}{2}{\lambda_{RET}\left(\theta_0\right)}^\top V_n\left(\theta_0\right)\lambda_{RET}\left(\theta_0\right)\\
&+ \frac{\tau_n}{2n}{\lambda_{RET}\left(\theta_0\right)}^\top
\left(\mu_{n, \theta_0}{\mu_{n, \theta_0}}^\top + \Sigma_{n, \theta_0}\right)
\lambda_{RET}\left(\theta_0\right)\\
\geq&
-\left\vert \lambda_{RET}\left(\theta_0\right)\right\vert
\left\vert h_n\left(\theta_0\right) + \frac{\tau_n}{n}\mu_{n, \theta_0}\right\vert \\
&-
\max_{1 \leq i \leq n}\left\{
-\exp\left({\widetilde{\lambda}}^\top g_i\left(\theta_0\right)\right)
\right\}\frac{1}{2}{\lambda_{RET}\left(\theta_0\right)}^\top V_n\left(\theta_0\right)\lambda_{RET}\left(\theta_0\right)\\
&+\frac{\tau_n}{4n}{\lambda_{RET}\left(\theta_0\right)}^\top
\left(\mu_{n, \theta_0}{\mu_{n, \theta_0}}^\top + \Sigma_{n, \theta_0}\right)
\lambda_{RET}\left(\theta_0\right)\\
\geq&
-\left\vert \lambda_{RET}\left(\theta_0\right)\right\vert
\left\vert h_n\left(\theta_0\right) + \frac{\tau_n}{n}\mu_{n, \theta_0}\right\vert \\
&+
\frac{1}{4}{\lambda_{RET}\left(\theta_0\right)}^\top V_n\left(\theta_0\right)\lambda_{RET}\left(\theta_0\right)\\
&+
\frac{\tau_n}{4n}
{\lambda_{RET}\left(\theta_0\right)}^\top
\left(\mu_{n, \theta_0}{\mu_{n, \theta_0}}^\top + \Sigma_{n, \theta_0}\right)
\lambda_{RET}\left(\theta_0\right).
\end{align*}
The last inequality holds since \({\max_{1 \leq i \leq n}\{-\exp({\widetilde{\lambda}}^\top g_i(\theta_0))}\} < {-1/2}\) with probability approaching \({1}\) \citep[Lemma~A1]{newey2004higher}.
Let \({\lambda_{RET}(\theta_0)} = {\vert \lambda_{RET}(\theta_0)\vert \xi}\) with \({\vert\xi\vert} = {1}\).
Rearranging the terms in the last inequality gives
\begin{equation*}
\frac{1}{4}\left\vert \lambda_{RET}\left(\theta_0\right)\right\vert
{\xi}^\top
\left(V_n\left(\theta_0\right)
+
\frac{\tau_n}{n}\left(\mu_{n, \theta_0}{\mu_{n, \theta_0}}^\top + \Sigma_{n, \theta_0}\right)\right)
\xi
\leq
\left\vert h_n\left(\theta_0\right) + \frac{\tau_n}{n}\mu_{n, \theta_0}\right\vert.
\end{equation*}
Since \({\tau_n{n}^{-1}(\mu_{n, \theta_0}{\mu_{n, \theta_0}}^\top + \Sigma_{n, \theta_0})} = {O_P(n^{-1}\log n)}\) by \cref{condition:retel asymptotics}, \cref{condition:full rank} implies that
\begin{equation*}
C\left\vert \lambda_{RET}\left(\theta_0\right)\right\vert
\leq
\left\vert h_n\left(\theta_0\right) + \frac{\tau_n}{n}\mu_{n, \theta_0}\right\vert
\end{equation*}
for some constant \({C} > {0}\) with probability approaching \({1}\).
Thus, we have
\begin{equation}\label{eq:retel lambda bound}
\lambda_{RET}\left(\theta_0\right)
=
O_P(n^{-1/2}).
\end{equation}

Next, we rewrite the first-order condition for \({\lambda_{RET}(\theta_0)}\) as follows:
\begin{equation}\label{eq:retel foc}
\frac{1}{n}\sum_{i = 1}^n
\exp\left({\lambda}^\top g_i\left(\theta_0\right)\right)g_i\left(\theta_0\right)
+
\frac{1}{n}p_n\left(\theta_0, \lambda\right)
\left(\mu_{n, \theta_0} + \Sigma_{n,\theta_0} \lambda\right)
=
0.
\end{equation}
By considering \cref{condition:retel asymptotics} and \eqref{eq:retel lambda bound}, we find that
\begin{equation*}
\frac{1}{n}p_n\left(\theta_0, \lambda_{RET}\left(\theta_0\right)\right)
\left(\mu_{n, \theta_0} + \Sigma_{n,\theta_0} \lambda_{RET}\left(\theta_0\right)\right)
=
O_P\left(n^{-1}\log n\right).
\end{equation*}
Expanding the left-hand side of \eqref{eq:retel foc} for \({\lambda_{RET}(\theta_0)}\) around \({\lambda} = {0}\), we obtain
\begin{equation*}
0
=
h_n\left(\theta_0\right) + V_n\left(\theta_0\right)\lambda_{RET}\left(\theta_0\right) + R_1 + O_P\left(n^{-1}\log n\right),
\end{equation*}
where
\begin{equation*}
\left\vert
R_1
\right\vert
\leq
\frac{C}{n}\sum_{i = 1}^n {\left\vert
g_i\left(\theta_0\right)
\right\vert}^3 {\left\vert
\lambda_{RET}\left(\theta_0\right)
\right\vert}^2
\end{equation*}
for some constant \({C} > {0}\) with probability approaching \({1}\).
From \cref{condition:uniform boundedness}, it follows that \({R_1} = {O_P(n^{-1})}\) and
\begin{equation}\label{eq:retel lambda expansion}
\lambda_{RET}\left(\theta_0\right)
=
-{V_n\left(\theta_0\right)}^{-1}h_n\left(\theta_0\right) + O_P\left(n^{-1}\log n\right).
\end{equation}
Next, we evaluate the expressions:
\begin{align*}
\frac{d_n\left(\theta_0, \lambda_{RET}\left(\theta_0\right)\right)}{n + \tau_n}
=
\frac{1}{n + \tau_n}&\sum_{i = 1}^n \exp\left({\lambda_{RET}\left(\theta_0\right)}^\top g_i\left(\theta_0\right)\right)\\
=
\frac{n}{n + \tau_n}& + \frac{n}{n + \tau_n} {\lambda_{RET}\left(\theta_0\right)}^\top
h_n\left(\theta_0\right)\\
&+
\frac{n}{2\left(n + \tau_n\right)} {\lambda_{RET}\left(\theta_0\right)}^\top
V_n\left(\theta_0\right)\lambda_{RET}\left(\theta_0\right)
+
R_2,
\end{align*}
where \({R_2} = {O_P(n^{-3/2})}\).
Similarly,
\begin{align*}
&\frac{p_n\left(\theta_0, \lambda_{RET}\left(\theta_0\right)\right)}{n + \tau_n}
=
\frac{\tau_n}{n + \tau_n}\exp\left({\lambda_{RET}\left(\theta_0\right)}^\top \mu_{n, \theta_0} + \frac{1}{2}{\lambda_{RET}\left(\theta_0\right)}^\top \Sigma_{n, \theta_0} \lambda_{RET}\left(\theta_0\right)\right)\\
&=
\frac{\tau_n}{n + \tau_n}\left(1 + {\lambda_{RET}\left(\theta_0\right)}^\top \mu_{n, \theta_0} + \frac{1}{2}{\lambda_{RET}\left(\theta_0\right)}^\top \Sigma_{n, \theta_0} \lambda_{RET}\left(\theta_0\right) + R_3\right)
\end{align*}
with \({R_3} = {O_P(n^{-1})}\).
From \eqref{eq:retel lambda bound}, we get
\begin{equation*}
\frac{p_n\left(\theta_0, \lambda_{RET}\left(\theta_0\right)\right)}{n + \tau_n}
=
\frac{\tau_n}{n + \tau_n} + O_P\left(n^{-3/2}\right).
\end{equation*}
Putting the above expressions together, we have
\begin{align*}
\frac{c_n\left(\theta_0, \lambda_{RET}\left(\theta_0\right)\right)}{n + \tau_n}
=
1 &+
\frac{n}{n + \tau_n} {\lambda_{RET}\left(\theta_0\right)}^\top
h_n\left(\theta_0\right)\\
&+
\frac{n}{2\left(n + \tau_n\right)} {\lambda_{RET}\left(\theta_0\right)}^\top
V_n\left(\theta_0\right)\lambda_{RET}\left(\theta_0\right)
+
O_P\left(n^{-3/2}\right),
\end{align*}
and
\begin{align}\label{eq:retel difference}
\begin{split}
\log&\left(\frac{c_n\left(\theta_0, \lambda_{RET}\left(\theta_0\right)\right)}{n + \tau_n}\right)\\
&=
\frac{n}{n + \tau_n} {\lambda_{RET}\left(\theta_0\right)}^\top
h_n\left(\theta_0\right)\\
&\hspace{1.5em} +
\frac{n}{2\left(n + \tau_n\right)} {\lambda_{RET}\left(\theta_0\right)}^\top
V_n\left(\theta_0\right)\lambda_{RET}\left(\theta_0\right)
+
O_P\left(n^{-3/2}\right)
+
O_P\left(n^{-2}\right)\\
&=
\frac{n}{n + \tau_n} {\lambda_{RET}\left(\theta_0\right)}^\top
h_n\left(\theta_0\right)\\
&\hspace{1.5em} +
\frac{n}{2\left(n + \tau_n\right)} {\lambda_{RET}\left(\theta_0\right)}^\top
V_n\left(\theta_0\right)\lambda_{RET}\left(\theta_0\right)
+
O_P\left(n^{-3/2}\right)\\
&=
O_P\left(n^{-1/2}\right).
\end{split}
\end{align}

From \eqref{eq:retel difference}, it follows that
\begin{align*}
\log& \left( \frac{R_{RET}\left(\theta_0\right)}{\widetilde{R}_{RET}\left(\theta_0\right)}\right)\\
&=
\log\left(
\frac{n + \tau_n}{\tau_n}p_c\left(\theta_0, \lambda_{RET}\left(\theta_0\right)\right)
\right)\\
&=
\log\left(
\frac{n + \tau_n}{c_n\left(\theta_0, \lambda_{RET}\left(\theta_0\right)\right)}\exp\left(
{\lambda_{RET}\left(\theta_0\right)}^\top \mu_{n, \theta_0} + \frac{1}{2}{\lambda_{RET}\left(\theta_0\right)}^\top \Sigma_{n, \theta_0} \lambda_{RET}\left(\theta_0\right)
\right)
\right)\\
&=
{\lambda_{RET}\left(\theta_0\right)}^\top \mu_{n, \theta_0} + \frac{1}{2}{\lambda_{RET}\left(\theta_0\right)}^\top \Sigma_{n, \theta_0} \lambda_{RET}\left(\theta_0\right) - \log\left(\frac{c_n\left(\theta_0, \lambda_{RET}\left(\theta_0\right)\right)}{n + \tau_n}\right)\\
&=
O_P\left(n^{-1/2}\right),
\end{align*}
establishing the first result.
For the second result, it suffices to show that \({-2\log \widetilde{R}_{RET}(\theta_0)}\) converges in distribution to \({\chi^2_p}\).
We have
\begin{align*}
-2\log \widetilde{R}_{RET}\left(\theta_0\right)
&=
-2\sum_{i = 1}^n\log\left(
\left(n + \tau_n\right)p_i\left(\theta_0\right)
\right)\\
&=
-2\sum_{i = 1}^n\left(
{\lambda_{RET}\left(\theta_0\right)}^\top g_i\left(\theta_0\right)
-
\log\left(\frac{c_n\left(\theta_0, \lambda_{RET}\left(\theta_0\right)\right)}{n + \tau_n}\right)
\right)\\
&=
-2n{\lambda_{RET}\left(\theta_0\right)}^\top h_n\left(\theta_0\right)
+
2n\log\left(\frac{c_n\left(\theta_0, \lambda_{RET}\left(\theta_0\right)\right)}{n + \tau_n}\right).
\end{align*}
Applying \eqref{eq:retel difference} and rearranging the terms with \eqref{eq:retel lambda expansion}, we obtain
\begin{align*}
-2\log \widetilde{R}_{RET}\left(\theta_0\right)
&=
-2\left(\frac{n\tau_n}{n + \tau_n}\right){\lambda_{RET}\left(\theta_0\right)}^\top h_n\left(\theta_0\right)\\
&\hspace{1.5em} +
\left(\frac{n}{n + \tau_n}\right) n {\lambda_{RET}\left(\theta_0\right)}^\top V_n\left(\theta_0\right) \lambda_{RET}\left(\theta_0\right)
+
o_P\left(1\right)\\
&=
\left(\frac{n}{n + \tau_n}\right) n {\lambda_{RET}\left(\theta_0\right)}^\top V_n\left(\theta_0\right)\lambda_{RET}\left(\theta_0\right)
+
o_P\left(1\right)\\
&=
n {h_n\left(\theta_0\right)}^\top {V_n\left(\theta_0\right)}^{-1} h_n\left(\theta_0\right) + o_P\left(1\right),
\end{align*}
which converges in distribution to \({\chi^2_p}\).
This establishes the second result.
\(\hfill\blacksquare\)

\subsection{Theorem~3}
Throughout this subsection we consider the posterior obtained from \({\textnormal{RETEL}_f}\), that is, from \({L_{RET}}\).
We write \({\ell(\theta)} = {\log L_{RET}(\theta)}\) and \({\widehat{\theta}}\) for the maximum ETEL estimator that solves \({h_n(\theta)} = {0}\).
When no root exists, we set \({\widehat{\theta}} = {\theta_0}\), which does not affect any statement below.
By \cref{condition:retel asymptotics}, \({\lambda_{RET}(\theta)}\) exists and is unique for every \({\theta} \in {\Theta}\), so \({L_{RET}(\theta)}\) is well defined and, being a product of strictly positive probabilities, lies in \({(0, 1]}\).
Since \({\pi}\) is a probability density on \({\Theta}\) by \cref{condition:prior}, we have \({0} < {\int_\Theta \pi(\theta)L_{RET}(\theta)d\theta} \leq {1}\), where the strict inequality holds because \({L_{RET}(\theta)} > {0}\) for every \({\theta}\) and \({\pi}\) is positive on a neighborhood of the interior point \({\theta_0}\) by \cref{condition:parameter space,condition:prior}.
Thus the posterior distribution is well defined.

In addition to \({h(\theta)}\) and \({V_n(\theta)}\) from the proof of Proposition~1 and \({G}\), \({V}\), and \({\Omega}\) from the main paper, let \({V(\theta)} = {E_P\{g_i(\theta)g_i(\theta)^\top\}}\) and, on the neighborhood \({\mathcal{N}}\) of \({\theta_0}\) in \cref{condition:continuity}, \({G(\theta)} = {E_P\{\partial_\theta g_i(\theta)\}}\), with \({V(\theta_0)} = {V}\) and \({G(\theta_0)} = {G}\).
Let \({\sigma_V}\) be the smallest eigenvalue of \({V}\) and \({\sigma_G}\) the smallest singular value of \({G}\).
We write \({\bar{V}_n(\theta)} = {V_n(\theta) - h_n(\theta)h_n(\theta)^\top}\) and \({\bar{V}(\theta)} = {V(\theta) - h(\theta)h(\theta)^\top}\).
Let
\begin{equation*}
m_n\left(\theta, \lambda\right)
=
\frac{1}{n}\sum_{i = 1}^n \exp\left(\lambda^\top g_i\left(\theta\right)\right),
\quad
q_n\left(\theta, \lambda\right)
=
\exp\left(\lambda^\top \mu_{n, \theta} + \frac{1}{2}\lambda^\top \Sigma_{n, \theta}\lambda\right).
\end{equation*}
In the notation of the main paper, \({d_n(\theta, \lambda)} = {n m_n(\theta, \lambda)}\), \({p_n(\theta, \lambda)} = {\tau_n q_n(\theta, \lambda)}\), and
\begin{equation*}
c_n\left(\theta, \lambda\right) = n m_n\left(\theta, \lambda\right) + \tau_n q_n\left(\theta, \lambda\right).
\end{equation*}
Write \({a_n} = {\tau_n / (n + \tau_n)}\), \({t_n} = {\tau_n / n}\), and \({K_n(\theta, \lambda)} = {\log(c_n(\theta, \lambda) / (n + \tau_n))}\).
Since \({\Sigma_{n, \theta}}\) is positive definite by \cref{condition:retel asymptotics}, the map \({\lambda} \mapsto {c_n(\theta, \lambda)}\) is strictly convex and coercive.
We write \({\lambda(\theta)} = {\lambda_{RET}(\theta)}\) for its minimizer, so that \({K_n(\theta, \lambda(\theta))} \leq {K_n(\theta, 0)} = {0}\).
Define the convex function
\begin{equation*}
\psi_\theta\left(\lambda\right)
=
\log m_n\left(\theta, \lambda\right) - \lambda^\top h_n\left(\theta\right)
=
\log\left(\frac{1}{n}\sum_{i = 1}^n \exp\left(\lambda^\top \left(g_i\left(\theta\right) - h_n\left(\theta\right)\right)\right)\right),
\end{equation*}
which satisfies \({\psi_\theta(0)} = {0}\) and, by Jensen's inequality, \({\psi_\theta(\lambda)} \geq {0}\) for all \({\lambda}\), and define
\begin{equation*}
A\left(\theta\right)
=
n\lambda\left(\theta\right)^\top h_n\left(\theta\right)
-
n\log\left(n m_n\left(\theta, \lambda\left(\theta\right)\right)\right).
\end{equation*}
Then \({A(\theta) + n \log n} = {-n\psi_\theta(\lambda(\theta))} \leq {0}\).
Since \({p_i(\theta)} = {\exp(\lambda(\theta)^\top g_i(\theta)) / c_n(\theta, \lambda(\theta))}\) and \({c_n} = {n m_n(1 + t_n q_n / m_n)}\) at \({\lambda} = {\lambda(\theta)}\), we have
\begin{equation*}
\sum_{i = 1}^n \log p_i\left(\theta\right) = A\left(\theta\right) - n \log\left(1 + \frac{t_n q_n\left(\theta, \lambda\left(\theta\right)\right)}{m_n\left(\theta, \lambda\left(\theta\right)\right)}\right).
\end{equation*}
Let
\begin{equation}\label{eq:bvm pen}
\textnormal{pen}\left(\theta\right)
=
\log p_c\left(\theta\right)
-
n\log\left(1 + \frac{t_n q_n\left(\theta, \lambda\left(\theta\right)\right)}{m_n\left(\theta, \lambda\left(\theta\right)\right)}\right).
\end{equation}
By the definition \({L_{RET}(\theta)} = {p_c(\theta)\prod_{i = 1}^n p_i(\theta)}\),
\begin{equation}\label{eq:bvm decomposition}
\ell\left(\theta\right)
=
A\left(\theta\right) + \textnormal{pen}\left(\theta\right).
\end{equation}
Since \({p_c(\theta)} = {\tau_n q_n / c_n} \in {(0, 1)}\) and \({t_n q_n / m_n} \geq {0}\) at \({\lambda} = {\lambda(\theta)}\), \({\textnormal{pen}(\theta)} \leq {0}\) for every \({\theta}\).
Moreover,
\begin{equation}\label{eq:bvm pc}
\log p_c\left(\theta\right)
=
\log a_n + \log q_n\left(\theta, \lambda\left(\theta\right)\right) - K_n\left(\theta, \lambda\left(\theta\right)\right).
\end{equation}
Finally, let \({\zeta_n} = {\max_{1 \leq i \leq n}\sup_{\theta \in \Theta}\vert g_i(\theta)\vert}\), \({B_n^{(3)}} = {n^{-1}\sum_{i = 1}^n \sup_{\theta \in \Theta}\vert g_i(\theta)\vert^3}\),
\begin{equation*}
B_n^{\left(\alpha\right)} = \frac{1}{n}\sum_{i = 1}^n \sup_{\theta \in \Theta}\left\vert g_i\left(\theta\right)\right\vert^\alpha, 
\end{equation*}
and \({B_\delta} = {\{\theta \in \Theta : \vert \theta - \theta_0\vert \leq \delta\}}\), and fix \({b} = {(1/3 + 1/\alpha)/2}\) and \({r_n} = {n^{-b}}\), so that \({1/\alpha} < {b} < {1/3}\) by \cref{condition:uniform boundedness}.
Throughout, \({C}\) denotes a generic positive constant, possibly different at each occurrence, that may depend on \({\delta}\), \({p}\), \({\alpha}\), \({b_2}\), \({C_0}\), and \({P}\), but not on \({n}\), \({\theta}\), \({\lambda}\), or \({u}\).
Numbered constants retain their values once introduced.

The proof is organized around four lemmas.
\cref{lemma:bvm preliminaries} collects consequences of \cref{condition:parameter space,condition:continuity,condition:full rank,condition:uniform boundedness}.
\cref{lemma:bvm working event} fixes a localization radius and constructs the working event on which the remaining lemmas operate.
\cref{lemma:bvm multiplier} controls the multiplier \({\lambda_{RET}(\theta)}\) uniformly over a fixed neighborhood of \({\theta_0}\).
\cref{lemma:bvm expansions} converts these bounds into local asymptotic normality of \({\ell}\) at \({\widehat{\theta}}\), with a self-normalized remainder, and exponential decay away from \({\widehat{\theta}}\).
The theorem then follows from a total variation argument over three zones.
Two remarks follow the proof.
\cref{rmk:bvm centering} records the equivalent formulation for \({n^{1/2}(\theta - \theta_0)}\) with a random centering, and \cref{rmk:retel reduced} gives the modification for \({\textnormal{RETEL}_r}\).

\begin{lemma}\label{lemma:bvm preliminaries}
Assume \cref{condition:parameter space,condition:continuity,condition:full rank,condition:uniform boundedness}.
\begin{enumerate}
\item[(i)]
The suprema satisfy \({\sup_{\theta \in \Theta}\vert h_n(\theta) - h(\theta)\vert} = {o_P(1)}\) and \({\sup_{\theta \in \Theta}\Vert V_n(\theta) - V(\theta)\Vert} = {o_P(1)}\), and \({h(\cdot)}\) and \({V(\cdot)}\) are continuous on \({\Theta}\).
There exists a closed ball \({\mathcal{N}^\prime} \subseteq {\mathcal{N}}\) centered at \({\theta_0}\) such that \({\sup_{\theta \in \mathcal{N}^\prime}\Vert (\partial h_n/\partial\theta^\top)(\theta) - G(\theta)\Vert} = {o_P(1)}\) and \({G(\cdot)}\) is continuous on \({\mathcal{N}^\prime}\).
\item[(ii)]
With probability approaching \({1}\), \({\widehat{\theta}}\) exists with \({h_n(\widehat{\theta})} = {0}\).
The estimator \({\widehat{\theta}}\) converges in probability to \({\theta_0}\), \({n^{1/2}(\widehat{\theta} - \theta_0)} = {-G^{-1}n^{1/2}h_n(\theta_0)} + {o_P(1)}\), and \({n^{1/2}(\widehat{\theta} - \theta_0)}\) converges in distribution to \({N(0, \Omega)}\).
\end{enumerate}
\end{lemma}

\noindent\textit{Proof.}
(i).
All claims follow from one uniform law of large numbers over a compact index set, which we prove once and then apply entrywise.
Let \({K}\) be a compact set, and let \({w(x, \theta)}\) be real-valued, continuous in \({\theta} \in {K}\) for \({P}\)-almost every \({x}\), with \({E_P\{\sup_{\theta \in K}\vert w(X, \theta)\vert\}} < {\infty}\).
Set \({m_n(\theta)} = {n^{-1}\sum_{i = 1}^n w(X_i, \theta)}\) and \({m(\theta)} = {E_P\{w(X, \theta)\}}\).
For \({\theta^\prime} \in {K}\) and \({r} > {0}\), let \({\omega(x, \theta^\prime, r)} = {\sup_{\theta \in K, \vert\theta - \theta^\prime\vert \leq r}\vert w(x, \theta) - w(x, \theta^\prime)\vert}\).
By the continuity of \({w(x, \theta)}\) at \({\theta^\prime}\), \({\omega(x, \theta^\prime, r)} \downarrow {0}\) as \({r} \downarrow {0}\) for \({P}\)-almost every \({x}\), and \({\omega(x, \theta^\prime, r)} \leq {2\sup_{\theta \in K}\vert w(x, \theta) \vert}\), so \({E_P\{\omega(X, \theta^\prime, r)\}} \downarrow {0}\) as \({r} \downarrow {0}\) by dominated convergence.
Given \({\eta} > {0}\), cover the compact \({K}\) by finitely many balls with centers \({\theta_j} \in {K}\) and radii \({r_j}\) such that \({E_P\{\omega(X, \theta_j, r_j)\}} \leq {\eta}\).
On each ball,
\begin{align*}
&\sup_{\theta \in K, \vert\theta - \theta_j\vert \leq r_j}\left\vert m_n\left(\theta\right) - m\left(\theta\right)\right\vert\\
&\leq
\sup_{\theta \in K, \vert\theta - \theta_j\vert \leq r_j}\left\vert m_n\left(\theta_j\right) - m_n\left(\theta\right)\right\vert
+
\left\vert m_n\left(\theta_j\right) - m\left(\theta_j\right)\right\vert
+
\sup_{\theta \in K, \vert\theta - \theta_j\vert \leq r_j}\left\vert m\left(\theta_j\right) - m\left(\theta\right)\right\vert\\
&\leq
\frac{1}{n}\sum_{i = 1}^n \omega\left(X_i, \theta_j, r_j\right)
+
\left\vert m_n\left(\theta_j\right) - m\left(\theta_j\right)\right\vert
+
E_P\left\{\omega\left(X, \theta_j, r_j\right)\right\},
\end{align*}
the first term converging almost surely to \({E_P\{\omega(X, \theta_j, r_j)\}} \leq {\eta}\) and the second to \({0}\) by the strong law of large numbers.
Taking the maximum over the finite cover gives
\begin{equation*}
\limsup_{n \to \infty}\sup_{\theta \in K}\left\vert m_n\left(\theta\right) - m\left(\theta\right)\right\vert
\leq
2\eta
\end{equation*}
almost surely.
Letting \({\eta} \downarrow {0}\) yields \({\sup_{\theta \in K}\vert m_n(\theta) - m(\theta)\vert} \to {0}\) almost surely, so
\begin{equation*}
\sup_{\theta \in K}\left\vert m_n\left(\theta\right) - m\left(\theta\right)\right\vert = o_P\left(1\right).
\end{equation*}
For \({\theta_k} \to {\theta^\prime}\) in \({K}\), \({w(x, \theta_k)} \to {w(x, \theta^\prime)}\) for \({P}\)-almost every \({x}\) by continuity, and \({\vert w(x, \theta_k)\vert} \leq {\sup_{\theta \in K}\vert w(x, \theta)\vert}\) with \({E_P\{\sup_{\theta \in K}\vert w(X, \theta)\vert\}} < {\infty}\).
By dominated convergence,
\begin{equation*}
m\left(\theta_k\right) = E_P\left\{w\left(X, \theta_k\right)\right\} \to E_P\left\{w\left(X, \theta^\prime\right)\right\} = m\left(\theta^\prime\right),
\end{equation*}
and thus \({m}\) is continuous on \({K}\).
Now apply the uniform law three times.
For \({h_n}\), take \({w}\) to be each entry of \({g}\), with \({K} = {\Theta}\) and \({\vert w \vert} \leq {\vert g \vert}\).
For \({V_n}\), take each entry of \({gg^\top}\), with \({K} = {\Theta}\) and \({\vert w \vert} \leq {\vert g \vert^2}\).
For \({\partial h_n/\partial\theta^\top}\), take each entry of \({\partial g/\partial\theta^\top}\), with \({K} = {\mathcal{N}^\prime}\) and \({\vert w \vert} \leq {\Vert \partial_\theta g \Vert}\), where the closed ball \({\mathcal{N}^\prime} \subseteq {\mathcal{N}}\) centered at \({\theta_0}\) is fixed arbitrarily because the covering argument requires a compact index set.
The continuity in \({\theta}\) of \({g}\) and \({\partial_\theta g}\) holds by \cref{condition:continuity}.
In each case, \({E_P\{\sup_{\theta \in K}\vert w(X, \theta)\vert\}} < {\infty}\), since \({E_P\{\sup_{\theta \in \Theta}\vert g \vert^s\}} \leq {1 + E_P\{\sup_{\theta \in \Theta}\vert g \vert^\alpha\}} < {\infty}\) for \({s} \in {\{1, 2\}}\) by \cref{condition:uniform boundedness} and \({E_P\{\sup_{\theta \in \mathcal{N}}\Vert \partial_\theta g \Vert\}} < {\infty}\) by \cref{condition:continuity}.
Each array has finitely many entries, so the entrywise uniform convergence yields the three supremum statements, and the continuity of \({h(\cdot)}\), \({V(\cdot)}\), and \({G(\cdot)}\) on their domains is the continuity of \({m}\) in the three instances.

\noindent (ii).
For \({\delta} > {0}\), let \({\iota(\delta)} = {\inf_{\vert\theta - \theta_0\vert \geq \delta}\vert h(\theta)\vert}\).
By \cref{condition:parameter space}, \({h(\theta_0)} = {0}\) and \({\theta_0}\) is the unique root of \({h}\) on the compact \({\Theta}\), and \({h}\) is continuous on \({\Theta}\) by (i), so \({\iota(\delta)} > {0}\) for every \({\delta} > {0}\).
Since \({\theta_0}\) is interior to \({\Theta}\) by \cref{condition:parameter space}, \({G(\cdot)}\) is continuous on \({\mathcal{N}^\prime}\) by (i), and \({G}\) is invertible with \({\sigma_G} > {0}\) by \cref{condition:full rank}, choose \({\delta^\prime} > {0}\) with \({\{\theta : \vert\theta - \theta_0\vert \leq \delta^\prime\}} \subseteq {\Theta \cap \mathcal{N}^\prime}\) and \({\Vert G(\theta) - G \Vert} \leq {\sigma_G/2}\) for all \({\theta} \in {B_{\delta^\prime}}\).
Define
\begin{equation*}
\mathcal{A}_n
=
\left\{\sup_{\theta \in B_{\delta^\prime}}\left\Vert \frac{\partial h_n}{\partial\theta^\top}\left(\theta\right) - G\left(\theta\right)\right\Vert \leq \frac{\sigma_G}{4}\right\}
\cap
\left\{\left\vert h_n\left(\theta_0\right)\right\vert \leq \frac{\sigma_G\delta^\prime}{8}\right\},
\end{equation*}
whose probability tends to \({1}\) by (i), directly for the first event, and for the second because \({\vert h_n(\theta_0)\vert} = {\vert h_n(\theta_0) - h(\theta_0)\vert} \leq {\sup_{\theta \in \Theta}\vert h_n(\theta) - h(\theta)\vert}\).
For \({\theta, \theta^{\prime\prime}} \in {B_{\delta^\prime}}\), write \({\gamma(t)} = {\theta^{\prime\prime} + t(\theta - \theta^{\prime\prime})}\) for \({t} \in {[0, 1]}\).
Then
\begin{align*}
\left\vert \gamma\left(t\right) - \theta_0 \right\vert
&=
\left\vert \left(1 - t\right)\left(\theta^{\prime\prime} - \theta_0\right) + t\left(\theta - \theta_0\right) \right\vert\\
&\leq
\left(1 - t\right)\left\vert \theta^{\prime\prime} - \theta_0 \right\vert + t\left\vert \theta - \theta_0 \right\vert\\
&\leq
\left(1 - t\right)\delta^\prime + t\delta^\prime\\
&=
\delta^\prime,
\end{align*}
so \({\gamma(t)} \in {\Theta \cap \mathcal{N}^\prime}\), and hence \({\gamma(t)} \in {B_{\delta^\prime}}\), by the choice of \({\delta^\prime}\).
With probability~\({1}\), every \({g_i(\cdot)}\) is continuously differentiable on \({\mathcal{N}}\) by \cref{condition:continuity}, so each component of \({t} \mapsto {h_n(\gamma(t))}\) is continuously differentiable on \({[0, 1]}\), and integrating each component's derivative over \({[0, 1]}\) gives
\begin{equation*}
h_n\left(\theta\right) - h_n\left(\theta^{\prime\prime}\right) - G\left(\theta - \theta^{\prime\prime}\right)
=
\left(\int_0^1 \left(\frac{\partial h_n}{\partial\theta^\top}\left(\gamma\left(t\right)\right) - G\right)dt\right)\left(\theta - \theta^{\prime\prime}\right).
\end{equation*}
On \({\mathcal{A}_n}\), for every \({t} \in {[0, 1]}\),
\begin{equation*}
\left\Vert \frac{\partial h_n}{\partial\theta^\top}\left(\gamma\left(t\right)\right) - G \right\Vert
\leq
\left\Vert \frac{\partial h_n}{\partial\theta^\top}\left(\gamma\left(t\right)\right) - G\left(\gamma\left(t\right)\right) \right\Vert
+
\left\Vert G\left(\gamma\left(t\right)\right) - G \right\Vert
\leq
\frac{\sigma_G}{4} + \frac{\sigma_G}{2}
=
\frac{3\sigma_G}{4},
\end{equation*}
so \({\vert h_n(\theta) - h_n(\theta^{\prime\prime}) - G(\theta - \theta^{\prime\prime})\vert} \leq {(3\sigma_G/4)\vert\theta - \theta^{\prime\prime}\vert}\).
For \({w} \in {\mathbb{R}^p}\), \({\vert w \vert} = {\vert GG^{-1}w \vert} \geq {\sigma_G\vert G^{-1}w \vert}\), that is, \({\vert G^{-1}w\vert} \leq {\sigma_G^{-1}\vert w\vert}\).
Let \({F(\theta)} = {\theta - G^{-1}h_n(\theta)}\).
On \({\mathcal{A}_n}\),
\begin{align*}
\left\vert F\left(\theta\right) - F\left(\theta^{\prime\prime}\right)\right\vert
&=
\left\vert G^{-1}\left(G\left(\theta - \theta^{\prime\prime}\right) - \left(h_n\left(\theta\right) - h_n\left(\theta^{\prime\prime}\right)\right)\right)\right\vert\\
&\leq
\sigma_G^{-1}\frac{3\sigma_G}{4}\left\vert\theta - \theta^{\prime\prime}\right\vert\\
&=
\frac{3}{4}\left\vert\theta - \theta^{\prime\prime}\right\vert,
\end{align*}
and \({\vert F(\theta_0) - \theta_0\vert} = {\vert G^{-1}h_n(\theta_0)\vert} \leq {\delta^\prime/8}\).
On \({\mathcal{A}_n}\), for every \({\theta} \in {B_{\delta^\prime}}\),
\begin{equation*}
\left\vert F\left(\theta\right) - \theta_0\right\vert
\leq
\left\vert F\left(\theta\right) - F\left(\theta_0\right)\right\vert + \left\vert F\left(\theta_0\right) - \theta_0\right\vert
\leq
\frac{3\delta^\prime}{4} + \frac{\delta^\prime}{8}
=
\frac{7\delta^\prime}{8}
<
\delta^\prime.
\end{equation*}
On \({\mathcal{A}_n}\), since \({F}\) maps \({B_{\delta^\prime}}\) into itself and is a contraction, the Banach fixed point theorem gives a root of \({h_n}\) in \({B_{\delta^\prime}}\), and hence \({h_n(\widehat{\theta})} = {0}\) by the definition of \({\widehat{\theta}}\).
For every \({\delta} \in {(0, \delta^\prime]}\), on \({\mathcal{A}_n} \cap {\{\sup_{\theta \in \Theta}\vert h_n(\theta) - h(\theta)\vert \leq \iota(\delta)/2\}}\), whose probability tends to \({1}\) by (i), \({\vert h(\widehat{\theta})\vert} = {\vert h(\widehat{\theta}) - h_n(\widehat{\theta})\vert} \leq {\sup_{\theta \in \Theta}\vert h_n(\theta) - h(\theta)\vert} \leq {\iota(\delta)/2} < {\iota(\delta)}\), and thus \({\vert\widehat{\theta} - \theta_0\vert} < {\delta}\).
Therefore \({\widehat{\theta}}\) converges in probability to \({\theta_0}\).

Since \({V}\) is finite by \cref{condition:uniform boundedness}, \({n^{1/2}h_n(\theta_0)}\) converges in distribution to \({N(0, V)}\) by the central limit theorem.
On \({\mathcal{A}_n} \cap {\{\sup_{\theta \in \Theta}\vert h_n(\theta) - h(\theta)\vert \leq \iota(\delta^\prime)/2\}}\), where \({\widehat{\theta}} \in {B_{\delta^\prime}}\), write \({0} = {h_n(\widehat{\theta})} = {h_n(\theta_0) + \bar{G}_n(\widehat{\theta} - \theta_0)}\) with \({\bar{G}_n} = {\int_0^1 (\partial h_n/\partial\theta^\top)(\theta_0 + t(\widehat{\theta} - \theta_0))dt}\).
On the same event, for \({w} \in {\mathbb{R}^p}\),
\begin{align*}
\left\vert \bar{G}_nw\right\vert
&=
\left\vert Gw - \left(G - \bar{G}_n\right)w\right\vert\\
&\geq
\left\vert Gw\right\vert - \left\vert \left(G - \bar{G}_n\right)w \right\vert\\
&\geq
\sigma_G \left\vert w\right\vert - \left\Vert G - \bar{G}_n\right\Vert \left\vert w\right\vert\\
&\geq
\sigma_G\left\vert w\right\vert - \frac{3\sigma_G}{4}\left\vert w\right\vert\\
&=
\frac{\sigma_G}{4}\left\vert w\right\vert.
\end{align*}
Then the smallest singular value of \({\bar{G}_n}\) is at least \({\sigma_G/4}\), and thus \({\bar{G}_n}\) is invertible with \({\vert\bar{G}_n^{-1}w\vert} \leq {(4/\sigma_G)\vert w\vert}\) for every \({w}\).
For \({\epsilon} > {0}\), since \({G(\cdot)}\) is continuous at \({\theta_0}\) by (i), there is \({r} \in {(0, \delta^\prime]}\) with \({\Vert G(\theta) - G \Vert} \leq {\epsilon/2}\) for all \({\theta} \in {B_r}\).
On \({\{\vert\widehat{\theta} - \theta_0\vert \leq r\}}\), the segment \({\theta_0 + t(\widehat{\theta} - \theta_0)}\), \({t} \in {[0, 1]}\), lies in \({B_r}\), and
\begin{align*}
\left\Vert \bar{G}_n - G \right\Vert
&=
\left\Vert \int_0^1 \left(\frac{\partial h_n}{\partial\theta^\top}\left(\theta_0 + t\left(\widehat{\theta} - \theta_0\right)\right) - G\right)dt \right\Vert\\
&\leq
\sup_{\theta \in B_r}\left\Vert \frac{\partial h_n}{\partial\theta^\top}\left(\theta\right) - G\left(\theta\right) \right\Vert
+
\sup_{\theta \in B_r}\left\Vert G\left(\theta\right) - G \right\Vert\\
&\leq
\sup_{\theta \in \mathcal{N}^\prime}\left\Vert \frac{\partial h_n}{\partial\theta^\top}\left(\theta\right) - G\left(\theta\right) \right\Vert
+
\frac{\epsilon}{2}.
\end{align*}
Then
\begin{equation*}
\textnormal{pr}\left(\left\Vert\bar{G}_n - G\right\Vert > \epsilon\right)
\leq
\textnormal{pr}\left(\left\vert\widehat{\theta} - \theta_0\right\vert > r\right) + \textnormal{pr}\left(\sup_{\theta \in \mathcal{N}^\prime}\left\Vert \frac{\partial h_n}{\partial\theta^\top}\left(\theta\right) - G\left(\theta\right)\right\Vert > \frac{\epsilon}{2}\right),
\end{equation*}
the first probability tending to zero by the consistency of \({\widehat{\theta}}\) and the second by (i).
Hence \({\bar{G}_n} = {G + o_P(1)}\).
On \({\mathcal{A}_n} \cap {\{\sup_{\theta \in \Theta}\vert h_n(\theta) - h(\theta)\vert \leq \iota(\delta^\prime)/2\}}\), \({n^{1/2}(\widehat{\theta} - \theta_0)} = {-\bar{G}_n^{-1}n^{1/2}h_n(\theta_0)}\) and
\begin{align*}
\left\vert n^{1/2}\left(\widehat{\theta} - \theta_0\right) + G^{-1}n^{1/2}h_n\left(\theta_0\right)\right\vert
&=
\left\vert \bar{G}_n^{-1}\left(\bar{G}_n - G\right)G^{-1}n^{1/2}h_n\left(\theta_0\right)\right\vert\\
&\leq
\frac{4}{\sigma_G}\left\Vert \bar{G}_n - G \right\Vert \left\vert G^{-1}n^{1/2}h_n\left(\theta_0\right)\right\vert\\
&\leq
\frac{4}{\sigma_G^2}\left\Vert \bar{G}_n - G \right\Vert n^{1/2}\left\vert h_n\left(\theta_0\right)\right\vert\\
&=
o_P(1),
\end{align*}
and thus \({n^{1/2}(\widehat{\theta} - \theta_0)} = {-G^{-1}n^{1/2}h_n(\theta_0) + o_P(1)}\), on the event of probability tending to \({1}\).
Then \({n^{1/2}(\widehat{\theta} - \theta_0)}\) converges in distribution to \({N(0, G^{-1}VG^{-\top})}\), and \({G^{-1}VG^{-\top}} = {(G^\top V^{-1}G)^{-1}} = {\Omega}\).
Therefore \({n^{1/2}(\widehat{\theta} - \theta_0)}\) converges in distribution to \({N(0, \Omega)}\).
\(\hfill\blacksquare\)

\begin{lemma}\label{lemma:bvm working event}
Assume \cref{condition:parameter space,condition:continuity,condition:full rank,condition:uniform boundedness,condition:retel asymptotics}.
There exists \({\delta_0} > {0}\) such that, for every \({\delta} \in {(0, \delta_0]}\),
\begin{equation*}
\left\{\theta : \left\vert\theta - \theta_0\right\vert \leq \delta\right\} \subseteq \Theta \cap \mathcal{N}^\prime \cap \mathcal{N}_0,
\end{equation*}
where \({\mathcal{N}^\prime}\) is the closed ball in \cref{lemma:bvm preliminaries}(i) and \({\mathcal{N}_0}\) is the neighborhood in \cref{condition:retel asymptotics}.
Let \({\delta} \in {(0, \delta_0]}\) and define \({E_n}\) as the event on which the following hold simultaneously.
\begin{enumerate}
\item[(a)]
\({r_n\zeta_n} \leq {1}\).
\item[(b)] \({B_n^{(3)} \vee B_n^{(\alpha)}} \leq {C_B}\), where \({C_B} = {E_P\{\sup_{\theta \in \Theta}\vert g_i(\theta) \vert^3\} + E_P\{\sup_{\theta \in \Theta}\vert g_i(\theta) \vert^\alpha\} + 1}\).
\item[(c)] \({h_n(\widehat{\theta})} = {0}\).
\item[(d)] The eigenvalues of \({V_n(\theta)}\) lie in \({[\sigma_V/4, 2C_V]}\) and the eigenvalues of \({\bar{V}_n(\theta)}\) are at least \({\sigma_V/4}\), for all \({\theta} \in {B_\delta}\), where \({C_V} = {\Vert V \Vert + 1}\).
\item[(e)] \({\sup_{\theta \in B_\delta}\vert h_n(\theta)\vert} \leq {C_h}\), where \({C_h} = {\sup_{\theta \in \Theta}\vert h(\theta)\vert + 1}\), and, for all \({\theta^\prime, \theta^{\prime\prime}} \in {B_\delta}\),
\begin{equation}\label{eq:bvm lipschitz}
c_G\left\vert \theta^\prime - \theta^{\prime\prime}\right\vert
\leq
\left\vert h_n\left(\theta^\prime\right) - h_n\left(\theta^{\prime\prime}\right)\right\vert
\leq
C_G\left\vert \theta^\prime - \theta^{\prime\prime}\right\vert,
\end{equation}
where \({c_G} = {\sigma_G/4}\) and \({C_G} = {\Vert G \Vert + \sigma_G}\).
\item[(f)] \({\sup_{\theta \in \mathcal{N}_0}\vert \mu_{n, \theta}\vert} \leq {b_2}\) and the eigenvalues of \({\Sigma_{n, \theta}}\) lie in \({[b_1, b_2]}\) for all \({\theta} \in {\mathcal{N}_0}\).
\end{enumerate}
Then \({\textnormal{pr}(E_n)} \to {1}\).
\end{lemma}

\noindent\textit{Proof.}
Since \({\theta_0}\) is interior to \({\Theta}\) by \cref{condition:parameter space}, there is \({\delta_1} > {0}\) with \({\{\theta : \vert\theta - \theta_0\vert \leq \delta_1\}} \subseteq {\Theta \cap \mathcal{N}^\prime \cap \mathcal{N}_0}\).
For \({\delta} \leq {\delta_1}\), \({B_\delta} = {\{\theta : \vert\theta - \theta_0\vert \leq \delta\}}\), and \({B_\delta}\) is convex.
Since \({h(\theta_0)} = {0}\) by \cref{condition:parameter space}, \({\bar{V}(\theta_0)} = {V(\theta_0)} = {V}\), and since \({V}\) is positive definite by \cref{condition:full rank,condition:uniform boundedness}, \({\sigma_V} > {0}\).
Since \({V(\cdot)}\), \({\bar{V}(\cdot)} = {V(\cdot) - h(\cdot)h(\cdot)^\top}\), and \({G(\cdot)}\) are continuous at \({\theta_0}\) by \cref{lemma:bvm preliminaries}(i) and \({\sigma_G} > {0}\) by \cref{condition:full rank}, there is \({\delta_0} \in {(0, \delta_1]}\) such that, for all \({\theta} \in {B_{\delta_0}}\), \({\Vert V(\theta) - V \Vert} \leq {\min\{\sigma_V/2, 1\}}\), \({\Vert \bar{V}(\theta) - V \Vert} \leq {\sigma_V/2}\), and \({\Vert G(\theta) - G \Vert} \leq {\sigma_G/2}\).
Fix \({\delta} \in {(0, \delta_0]}\).
For (a), since \({\{r_n\zeta_n \leq 1\}} = {\{\zeta_n \leq n^b\}}\),
\begin{align*}
\textnormal{pr}\left(\zeta_n > n^b\right)
&\leq
n\textnormal{pr}\left(\sup_{\theta \in \Theta}\left\vert g_1\left(\theta\right)\right\vert > n^b\right)\\
&=
n\textnormal{pr}\left(\sup_{\theta \in \Theta}\left\vert g_1\left(\theta\right)\right\vert^\alpha > n^{\alpha b}\right)\\
&\leq
n^{1 - \alpha b}
E_P\left\{\sup_{\theta \in \Theta}\left\vert g_1\left(\theta\right)\right\vert^\alpha\right\}.
\end{align*}
With \({\alpha b} > {1}\) and \({E_P\{\sup_{\theta \in \Theta}\vert g_1(\theta)\vert^\alpha\}} < {\infty}\) by \cref{condition:uniform boundedness}, \({\textnormal{pr}(r_n\zeta_n \leq 1)} \to {1}\).
For (b), by Jensen's inequality applied to the concave map \({t} \mapsto {t^{3/\alpha}}\) on \({[0, \infty)}\) with \({\alpha} > {3}\), \({E_P\{\sup_{\theta \in \Theta}\vert g_i(\theta)\vert^3\}} \leq {(E_P\{\sup_{\theta \in \Theta}\vert g_i(\theta)\vert^\alpha\})^{3/\alpha}} < {\infty}\), and the weak law of large numbers gives
\begin{equation*}
B_n^{\left(3\right)} = E_P\left\{\sup_{\theta \in \Theta}\left\vert g_i\left(\theta\right)\right\vert^3\right\} + o_P\left(1\right)\
\end{equation*}
and \({B_n^{(\alpha)}} = {E_P\{\sup_{\theta \in \Theta}\vert g_i(\theta)\vert^\alpha\} + o_P(1)}\).
Then \({B_n^{(3)} \vee B_n^{(\alpha)}} \leq {C_B}\) with probability approaching \({1}\).
For (c), \({h_n(\widehat{\theta})} = {0}\) with probability approaching \({1}\) by \cref{lemma:bvm preliminaries}(ii).
For (d), let \({Q_n} = {\sup_{\theta \in \Theta}\Vert V_n(\theta) - V(\theta)\Vert}\), \({Q_n^\prime} = {\sup_{\theta \in \Theta}\vert h_n(\theta) - h(\theta)\vert}\), and \({\bar{Q}_n} = {Q_n + (2\sup_{\theta \in \Theta}\vert h(\theta) \vert + 1)Q_n^\prime}\).
By \cref{lemma:bvm preliminaries}(i), \({Q_n} = {o_P(1)}\) and \({Q_n^\prime} = {o_P(1)}\).
Since \({h(\cdot)}\) is continuous by \cref{lemma:bvm preliminaries}(i) and \({\Theta}\) is compact by \cref{condition:parameter space}, \({\sup_{\theta \in \Theta}\vert h(\theta) \vert} < {\infty}\).
For every \({\theta} \in {\Theta}\),
\begin{align*}
\left\Vert \bar{V}_n\left(\theta\right) - \bar{V}\left(\theta\right) \right\Vert
&=
\left\Vert V_n\left(\theta\right) - V\left(\theta\right) - \left(h_n\left(\theta\right)h_n\left(\theta\right)^\top - h\left(\theta\right)h\left(\theta\right)^\top\right)\right\Vert\\
&=
\left\Vert V_n\left(\theta\right) - V\left(\theta\right) - h_n\left(\theta\right)\left(h_n\left(\theta\right) - h\left(\theta\right)\right)^\top - \left(h_n\left(\theta\right) - h\left(\theta\right)\right)h\left(\theta\right)^\top\right\Vert\\
&\leq
\left\Vert V_n\left(\theta\right) - V\left(\theta\right)\right\Vert + \left\vert h_n\left(\theta\right)\right\vert\left\vert h_n\left(\theta\right) - h\left(\theta\right)\right\vert + \left\vert h_n\left(\theta\right) - h\left(\theta\right)\right\vert\left\vert h\left(\theta\right)\right\vert\\
&\leq
Q_n + \left(\left\vert h_n\left(\theta\right)\right\vert + \left\vert h\left(\theta\right)\right\vert\right)Q_n^\prime.
\end{align*}
On \({\{Q_n^\prime \leq 1\}}\), \({\vert h_n(\theta)\vert} \leq {\vert h(\theta)\vert + Q_n^\prime} \leq {\sup_{\theta \in \Theta}\vert h(\theta)\vert + 1}\) for every \({\theta} \in {\Theta}\), and hence
\begin{equation*}
\sup_{\theta \in \Theta}\left\Vert \bar{V}_n\left(\theta\right) - \bar{V}\left(\theta\right)\right\Vert \leq \bar{Q}_n.
\end{equation*}
On \({\{Q_n \leq \sigma_V/4\}} \cap {\{Q_n^\prime \leq 1\}} \cap {\{\bar{Q}_n \leq \sigma_V/4\}}\), whose probability tends to \({1}\), for every \({\theta} \in {B_\delta}\) and every unit vector \({u}\),
\begin{align*}
u^\top V_n\left(\theta\right)u
&=
u^\top V u + u^\top\left(V\left(\theta\right) - V\right)u + u^\top\left(V_n\left(\theta\right) - V\left(\theta\right)\right)u\\
&\geq
\sigma_V - \left\vert u^\top\left(V\left(\theta\right) - V\right)u\right\vert - \left\vert u^\top\left(V_n\left(\theta\right) - V\left(\theta\right)\right)u\right\vert\\
&\geq
\sigma_V - \left\Vert V\left(\theta\right) - V\right\Vert - \left\Vert V_n\left(\theta\right) - V\left(\theta\right)\right\Vert\\
&\geq
\sigma_V - \frac{\sigma_V}{2} - \frac{\sigma_V}{4}\\
&=
\frac{\sigma_V}{4},\\
u^\top V_n\left(\theta\right)u
&\leq
\left\vert u^\top V u \right\vert + \left\vert u^\top\left(V\left(\theta\right) - V\right)u\right\vert + \left\vert u^\top\left(V_n\left(\theta\right) - V\left(\theta\right)\right)u\right\vert\\
&\leq
\left\Vert V \right\Vert + \left\Vert V\left(\theta\right) - V\right\Vert + \left\Vert V_n\left(\theta\right) - V\left(\theta\right)\right\Vert\\
&\leq
\left\Vert V \right\Vert + 1 + \frac{\sigma_V}{4}\\
&=
C_V + \frac{\sigma_V}{4}\\
&\leq
2C_V,\\
u^\top \bar{V}_n\left(\theta\right)u
&=
u^\top V u + u^\top\left(\bar{V}\left(\theta\right) - V\right)u + u^\top\left(\bar{V}_n\left(\theta\right) - \bar{V}\left(\theta\right)\right)u\\
&\geq
\sigma_V - \left\vert u^\top\left(\bar{V}\left(\theta\right) - V\right)u\right\vert - \left\vert u^\top\left(\bar{V}_n\left(\theta\right) - \bar{V}\left(\theta\right)\right)u\right\vert\\
&\geq
\sigma_V - \left\Vert \bar{V}\left(\theta\right) - V\right\Vert - \left\Vert \bar{V}_n\left(\theta\right) - \bar{V}\left(\theta\right)\right\Vert\\
&\geq
\sigma_V - \frac{\sigma_V}{2} - \frac{\sigma_V}{4}\\
&=
\frac{\sigma_V}{4}.
\end{align*}
Then the eigenvalues of \({V_n(\theta)}\) lie in \({[\sigma_V/4, 2C_V]}\) and the eigenvalues of \({\bar{V}_n(\theta)}\) are at least \({\sigma_V/4}\) for all \({\theta} \in {B_\delta}\).
For (e), on \({\{Q_n^\prime \leq 1\}}\), \({\sup_{\theta \in B_\delta}\vert h_n(\theta)\vert} \leq {\sup_{\theta \in \Theta}\vert h(\theta) \vert + 1} = {C_h}\).
For \eqref{eq:bvm lipschitz} and \({\theta^\prime, \theta^{\prime\prime}} \in {B_\delta}\), since \({B_\delta}\) is convex and contained in \({\mathcal{N}^\prime}\) and \({h_n(\cdot)}\) is continuously differentiable on \({\mathcal{N}^\prime}\) by \cref{condition:continuity}, write \({h_n(\theta^\prime) - h_n(\theta^{\prime\prime})} = {M_n(\theta^\prime - \theta^{\prime\prime})}\) with
\begin{equation*}
M_n
=
\int_0^1\frac{\partial h_n}{\partial\theta^\top}\left(\theta^{\prime\prime} + t\left(\theta^\prime - \theta^{\prime\prime}\right)\right)dt.
\end{equation*}
On \({\{\sup_{\theta \in \mathcal{N}^\prime}\Vert (\partial h_n/\partial\theta^\top)(\theta) - G(\theta)\Vert \leq \sigma_G/4\}}\), whose probability tends to \({1}\) by \cref{lemma:bvm preliminaries}(i),
\begin{align*}
\left\Vert M_n - G \right\Vert
&\leq
\int_0^1\left\Vert \frac{\partial h_n}{\partial\theta^\top}\left(\theta^{\prime\prime} + t\left(\theta^\prime - \theta^{\prime\prime}\right)\right) - G \right\Vert dt\\
&\leq
\int_0^1\left\Vert \frac{\partial h_n}{\partial\theta^\top}\left(\theta^{\prime\prime} + t\left(\theta^\prime - \theta^{\prime\prime}\right)\right) - G\left(\theta^{\prime\prime} + t\left(\theta^\prime - \theta^{\prime\prime}\right)\right)\right\Vert dt\\
&\quad+
\int_0^1\left\Vert G\left(\theta^{\prime\prime} + t\left(\theta^\prime - \theta^{\prime\prime}\right)\right) - G\right\Vert dt\\
&\leq
\frac{\sigma_G}{4} + \frac{\sigma_G}{2}\\
&=
\frac{3\sigma_G}{4}.
\end{align*}
Then
\begin{align*}
\left\vert h_n\left(\theta^\prime\right) - h_n\left(\theta^{\prime\prime}\right) \right\vert
&=
\left\vert M_n\left(\theta^{\prime} - \theta^{\prime\prime}\right) \right\vert\\
&\geq
\left\vert G\left(\theta^{\prime} - \theta^{\prime\prime}\right) \right\vert - \left\vert \left(M_n - G\right)\left(\theta^{\prime} - \theta^{\prime\prime}\right) \right\vert\\
&\geq
\sigma_G\left\vert\theta^{\prime} - \theta^{\prime\prime}\right\vert - \left\Vert M_n - G\right\Vert\left\vert\theta^{\prime} - \theta^{\prime\prime}\right\vert\\
&\geq
\sigma_G\left\vert\theta^{\prime} - \theta^{\prime\prime}\right\vert - \frac{3\sigma_G}{4}\left\vert\theta^{\prime} - \theta^{\prime\prime}\right\vert\\
&=
c_G\left\vert\theta^{\prime} - \theta^{\prime\prime}\right\vert,\\
\left\vert h_n\left(\theta^\prime\right) - h_n\left(\theta^{\prime\prime}\right) \right\vert
&=
\left\vert M_n\left(\theta^{\prime} - \theta^{\prime\prime}\right) \right\vert\\
&\leq
\left\Vert M_n - G\right\Vert\left\vert\theta^{\prime} - \theta^{\prime\prime}\right\vert + \left\Vert G\right\Vert\left\vert\theta^{\prime} - \theta^{\prime\prime}\right\vert\\
&\leq
\frac{3\sigma_G}{4}\left\vert\theta^{\prime} - \theta^{\prime\prime}\right\vert + \left\Vert G\right\Vert\left\vert\theta^{\prime} - \theta^{\prime\prime}\right\vert\\
&\leq
C_G\left\vert\theta^{\prime} - \theta^{\prime\prime}\right\vert,
\end{align*}
and thus \eqref{eq:bvm lipschitz} holds for all \({\theta^\prime, \theta^{\prime\prime}} \in {B_\delta}\).
For (f), the bounds on \({\mu_{n, \theta}}\) and \({\Sigma_{n, \theta}}\) hold with probability approaching \({1}\) by \cref{condition:retel asymptotics}.
Since each of (a)--(f) holds with probability approaching \({1}\), \({\textnormal{pr}(E_n)} \to {1}\).
\(\hfill\blacksquare\)

In what follows we fix \({\delta} \in {(0, \delta_0]}\), so that \({E_n}\) denotes the event of \cref{lemma:bvm working event} at this \({\delta}\).
Write \({C_1} = {C_G + 1}\) and define
\begin{equation*}
\Theta_{n, 1}
=
\left\{\theta \in B_\delta : \left\vert h_n\left(\theta\right)\right\vert \leq C_1 n^{-1/3}\right\},
\quad
\Theta_{n, 2}
=
\left\{\theta \in B_\delta : \left\vert h_n\left(\theta\right)\right\vert > C_1 n^{-1/3}\right\}.
\end{equation*}

\begin{lemma}\label{lemma:bvm multiplier}
Assume \cref{condition:parameter space,condition:continuity,condition:full rank,condition:uniform boundedness,condition:retel asymptotics}.
Then there exist constants \({c_2} > {0}\), \({C_3}\), \({C_4}\), \({C_5}\), and a deterministic \({n_*}\) such that, on \({E_n}\) and for \({n} \geq {n_*}\), the following hold simultaneously for all \({\theta} \in {B_\delta}\).
\begin{enumerate}
\item[(i)] For all \({\lambda} \in {\mathbb{R}^p}\),
\begin{equation}\label{eq:bvm psi bound}
\psi_\theta\left(\lambda\right)
\geq
\log\left(1 + \frac{\sigma_V}{24}\min\left\{\left\vert\lambda\right\vert, \bar{\lambda}\right\}^2\right),
\end{equation}
where \({\bar{\lambda}} = {(\sigma_V/(8B^\prime))^{1/(\alpha - 2)}}\) and \({B^\prime} = {2^{\alpha - 1}(C_B + C_h^\alpha)}\).
\item[(ii)] If \({\theta} \in {\Theta_{n, 2}}\), then \({\vert\lambda(\theta)\vert} \geq {c_2 n^{-1/3}}\).
\item[(iii)] If \({\theta} \in {\Theta_{n, 1}}\), then \({\vert\lambda(\theta)\vert} \leq {C_3(\vert h_n(\theta)\vert + t_n)}\), \({\max_{1 \leq i \leq n}\vert\lambda(\theta)^\top g_i(\theta)\vert} \leq {C_4}\), and
\begin{align}\label{eq:bvm lambda fine}
\begin{split}
\lambda\left(\theta\right)
&=
-V_n\left(\theta\right)^{-1}\left(h_n\left(\theta\right) + t_n \mu_{n, \theta}\right)
+
\rho\left(\theta\right),\\
\left\vert\rho\left(\theta\right)\right\vert
&\leq
C_5\left(\left\vert h_n\left(\theta\right)\right\vert^2 + t_n\left\vert h_n\left(\theta\right)\right\vert + t_n^2\right).
\end{split}
\end{align}
\end{enumerate}
\end{lemma}

\noindent\textit{Proof.}
(i).
Fix \({\theta} \in {B_\delta}\) and \({\lambda} \in {\mathbb{R}^p}\), and set \({x_i} = {\lambda^\top(g_i(\theta) - h_n(\theta))}\), so that \({\sum_{i = 1}^n x_i} = {0}\) and \({\psi_\theta(\lambda)} = {\log(n^{-1}\sum_{i = 1}^n \exp(x_i))}\).
Applying \({\exp(x_i)} \geq {1 + x_i + x_i^2/3}\) for \({\vert x_i\vert} \leq {1}\) and \({\exp(x_i)} \geq {1 + x_i}\) otherwise,
\begin{align*}
\frac{1}{n}\sum_{i = 1}^n \exp\left(x_i\right)
&\geq
1 + \frac{1}{n}\sum_{i = 1}^n x_i + \frac{1}{3n}\sum_{i : \vert x_i\vert \leq 1} x_i^2\\
&=
1 + \frac{1}{3n}\sum_{i = 1}^n x_i^2 - \frac{1}{3n}\sum_{i : \vert x_i\vert > 1} x_i^2\\
&=
1 + \frac{1}{3}\left(\lambda^\top \bar{V}_n\left(\theta\right)\lambda - \frac{1}{n}\sum_{i : \vert x_i\vert > 1} x_i^2\right).
\end{align*}
For \({\vert x_i\vert} > {1}\),
\begin{align*}
x_i^2
&\leq
\left\vert x_i \right\vert^\alpha\\
&\leq
\left\vert \lambda \right\vert^\alpha
\left\vert g_i\left(\theta\right) - h_n\left(\theta\right) \right\vert^\alpha\\
&\leq
\left\vert \lambda \right\vert^\alpha
\left(\left\vert g_i\left(\theta\right) \right\vert + \left\vert h_n\left(\theta\right) \right\vert\right)^\alpha\\
&=
\left\vert \lambda \right\vert^\alpha
2^\alpha
\left(\frac{\left\vert g_i\left(\theta\right) \right\vert + \left\vert h_n\left(\theta\right) \right\vert}{2}\right)^\alpha\\
&\leq
\left\vert \lambda \right\vert^\alpha
2^{\alpha - 1}
\left(\left\vert g_i\left(\theta\right) \right\vert^\alpha + \left\vert h_n\left(\theta\right) \right\vert^\alpha\right)\\
&\leq
\left\vert \lambda \right\vert^\alpha
2^{\alpha - 1}
\left(\sup_{\vartheta \in \Theta}\left\vert g_i\left(\vartheta\right)\right\vert^\alpha + C_h^\alpha\right).
\end{align*}
Then,
\begin{align*}
\frac{1}{n}\sum_{i : \vert x_i\vert > 1} x_i^2
&\leq
\left\vert \lambda \right\vert^\alpha
2^{\alpha - 1}\frac{1}{n}\sum_{i = 1}^n
\left(\sup_{\vartheta \in \Theta}\left\vert g_i\left(\vartheta\right)\right\vert^\alpha + C_h^\alpha\right)\\
&=
\left\vert \lambda \right\vert^\alpha
2^{\alpha - 1} \left(B_n^{(\alpha)} + C_h^\alpha\right)\\
&\leq
\left\vert \lambda \right\vert^\alpha
2^{\alpha - 1} \left(C_B + C_h^\alpha\right)\\
&=
\left\vert \lambda \right\vert^\alpha B^\prime.
\end{align*}
With \({\lambda^\top\bar{V}_n(\theta)\lambda} \geq {(\sigma_V/4)\vert\lambda\vert^2}\),
\begin{align*}
\frac{1}{n}\sum_{i = 1}^n \exp\left(x_i\right)
&\geq
1 + \frac{1}{3}\left(\frac{\sigma_V}{4}\left\vert\lambda\right\vert^2 - \left\vert \lambda \right\vert^\alpha B^\prime\right)\\
&=
1 + \frac{\sigma_V}{12}\left\vert\lambda\right\vert^2 - \frac{B^\prime}{3}\left\vert\lambda\right\vert^\alpha.
\end{align*}
For \({\vert\lambda\vert} \leq {\bar{\lambda}}\),
\begin{equation*}
\frac{B^\prime}{3}\left\vert\lambda\right\vert^\alpha \leq \frac{B^\prime}{3}\left\vert\lambda\right\vert^2 \bar{\lambda}^{\alpha - 2} = \frac{\sigma_V}{24}\left\vert\lambda\right\vert^2,
\end{equation*}
so
\({\psi_\theta(\lambda)} \geq {\log(1 + (\sigma_V/24)\vert\lambda\vert^2)}\).
For \({\vert\lambda\vert} > {\bar{\lambda}}\), set \({\lambda^\prime} = {(\bar{\lambda}/\vert\lambda\vert)\lambda}\).
The convexity of \({\psi_\theta}\), \({\psi_\theta(0)} = {0}\), and \({\psi_\theta(\lambda^\prime)} \leq {(\bar{\lambda}/\vert\lambda\vert)\psi_\theta(\lambda)}\) give \({\psi_\theta(\lambda)} \geq {(\vert\lambda\vert/\bar{\lambda})\psi_\theta(\lambda^\prime)} \geq {\psi_\theta(\lambda^\prime)} \geq {\log(1 + (\sigma_V/24)\bar{\lambda}^2)}\).

\noindent (ii).
Fix \({\theta} \in {\Theta_{n, 2}}\) and set \({\lambda^*} = {-s h_n(\theta)/\vert h_n(\theta)\vert}\), where \({s} = {\min\{\kappa\vert h_n(\theta)\vert, (2\zeta_n)^{-1}\}}\) and \({\kappa} = {\min\{1/C_h, 2/(3b_2C_h), 1/(16C_V)\}}\).
Since \({{\lambda^*}^\top g_i(\theta)} \leq {s\zeta_n} \leq {1/2}\), applying termwise the inequality \({\exp({\lambda^*}^\top g_i(\theta))} \leq {1 + {\lambda^*}^\top g_i(\theta) + ({\lambda^*}^\top g_i(\theta))^2}\),
\begin{align*}
m_n\left(\theta, \lambda^*\right)
&\leq
1 + {\lambda^*}^\top h_n\left(\theta\right) + {\lambda^*}^\top V_n\left(\theta\right)\lambda^*\\
&=
1 - s\left\vert h_n\left(\theta\right)\right\vert + {\lambda^*}^\top V_n\left(\theta\right)\lambda^*\\
&\leq
1 - s\left\vert h_n\left(\theta\right)\right\vert + 2C_V s^2.
\end{align*}
Since \({s} \leq {\kappa\vert h_n(\theta)\vert} \leq {\kappa C_h} \leq {1}\), so that \({b_2 s + b_2 s^2/2} \leq {3b_2 s/2} \leq {3b_2 C_h\kappa/2} \leq {1}\),
\begin{align*}
q_n\left(\theta, \lambda^*\right)
&=
\exp\left({\lambda^*}^\top \mu_{n, \theta} + \frac{1}{2}{\lambda^*}^\top \Sigma_{n, \theta}\lambda^*\right)\\
&\leq
\exp\left(b_2 s + \frac{b_2 s^2}{2}\right)\\
&\leq
\exp\left(\frac{3b_2 s}{2}\right)\\
&\leq
1 + 3b_2 s.
\end{align*}
Since \({\lambda(\theta)}\) minimizes \({c_n(\theta, \cdot)}\), \({c_n(\theta, \lambda)/(n + \tau_n)} = {(1 - a_n)m_n(\theta, \lambda) + a_n q_n(\theta, \lambda)}\), and \({1 - a_n} \geq {1/2}\) for \({n} \geq {\tau_n}\),
\begin{align*}
K_n\left(\theta, \lambda\left(\theta\right)\right)
&\leq
K_n\left(\theta, \lambda^*\right)\\
&=
\log\left(\left(1 - a_n\right)m_n\left(\theta, \lambda^*\right) + a_n q_n\left(\theta, \lambda^*\right)\right)\\
&\leq
\left(1 - a_n\right)m_n\left(\theta, \lambda^*\right) + a_n q_n\left(\theta, \lambda^*\right) - 1\\
&\leq
\left(1 - a_n\right)\left(1 - s\left\vert h_n\left(\theta\right)\right\vert + 2C_V s^2\right) + a_n\left(1 + 3b_2 s\right) - 1\\
&=
-\left(1 - a_n\right)s\left\vert h_n\left(\theta\right)\right\vert + 2C_V\left(1 - a_n\right)s^2 + 3b_2 a_n s\\
&\leq
-\frac{1}{2}s\left\vert h_n\left(\theta\right)\right\vert + 2C_V s^2 + 3b_2 a_n s.
\end{align*}
With \({\log m_n(\theta, \lambda)} \geq {\lambda^\top h_n(\theta)}\) by Jensen's inequality and \({\log(1 - a_n)} \geq {-2a_n}\) for \({a_n} \leq {1/2}\),
\begin{align*}
K_n\left(\theta, \lambda\left(\theta\right)\right)
&=
\log\left(\left(1 - a_n\right)m_n\left(\theta, \lambda\left(\theta\right)\right) + a_n q_n\left(\theta, \lambda\left(\theta\right)\right)\right)\\
&\geq
\log\left(\left(1 - a_n\right)m_n\left(\theta, \lambda\left(\theta\right)\right)\right)\\
&=
\log\left(1 - a_n\right) + \log m_n\left(\theta, \lambda\left(\theta\right)\right)\\
&\geq
-2a_n + \lambda\left(\theta\right)^\top h_n\left(\theta\right)\\
&\geq
-2a_n - \left\vert\lambda\left(\theta\right)\right\vert\left\vert h_n\left(\theta\right)\right\vert.
\end{align*}
Combining the two bounds on \({K_n(\theta, \lambda(\theta))}\),
\begin{equation*}
\left\vert\lambda\left(\theta\right)\right\vert\left\vert h_n\left(\theta\right)\right\vert
\geq
\frac{1}{2}s\left\vert h_n\left(\theta\right)\right\vert - 2C_V s^2 - \left(3b_2 s + 2\right)a_n.
\end{equation*}
The bounds \({\kappa\vert h_n(\theta)\vert} > {\kappa C_1 n^{-1/3}}\) and \({(2\zeta_n)^{-1}} \geq {(2n^b)^{-1}} \geq {(2n^{1/3})^{-1}}\) give \({s} \geq {c_6 n^{-1/3}}\) with \({c_6} = {\min\{\kappa C_1, 1/2\}}\) and \({s\vert h_n(\theta)\vert} \geq {c_6 C_1 n^{-2/3}}\).
The choice of \({\kappa}\) gives \({2C_V s^2} \leq {2C_V\kappa s\vert h_n(\theta)\vert} \leq {(1/8)s\vert h_n(\theta)\vert}\).
Since \({\tau_n} = {O(\log n)}\) by \cref{condition:retel asymptotics}, there is a deterministic \({n_*}\) such that \({\tau_n} \leq {n}\) and \({(3b_2 + 2)t_n} \leq {(1/8)c_6 C_1 n^{-2/3}}\) for all \({n} \geq {n_*}\).
With \({a_n} \leq {t_n}\) and \({s} \leq {1}\), for \({n} \geq {n_*}\), \({(3b_2 s + 2)a_n} \leq {(3b_2 + 2)t_n} \leq {(1/8)c_6 C_1 n^{-2/3}} \leq {(1/8)s\vert h_n(\theta)\vert}\).
Then,
\begin{equation*}
\left\vert\lambda\left(\theta\right)\right\vert\left\vert h_n\left(\theta\right)\right\vert
\geq
\frac{1}{2}s\left\vert h_n\left(\theta\right)\right\vert - \frac{1}{8}s\left\vert h_n\left(\theta\right)\right\vert - \frac{1}{8}s\left\vert h_n\left(\theta\right)\right\vert
=
\frac{1}{4}s\left\vert h_n\left(\theta\right)\right\vert,
\end{equation*}
so \({\vert\lambda(\theta)\vert} \geq {s/4} \geq {c_2 n^{-1/3}}\) with \({c_2} = {c_6/4}\).

\noindent (iii).
Fix \({\theta} \in {\Theta_{n, 1}}\).
If \({\lambda(\theta)} = {0}\), then \({\partial c_n(\theta, 0)/\partial\lambda} = {n h_n(\theta) + \tau_n\mu_{n, \theta}} = {0}\), and all three claims hold with \({\rho(\theta)} = {0}\).
Suppose \({\lambda(\theta)} \neq {0}\) and set \({s} = {\min\{\vert\lambda(\theta)\vert, r_n/2\}}\) and \({\lambda^*} = {s\lambda(\theta)/\vert\lambda(\theta)\vert}\).
Since \({\lambda^*}\) lies on the segment from \({0}\) to \({\lambda(\theta)}\) and \({c_n(\theta, \cdot)}\) is convex with \({c_n(\theta, \lambda(\theta))} \leq {c_n(\theta, 0)}\),
\begin{equation*}
c_n\left(\theta, \lambda^*\right)
\leq
\frac{s}{\left\vert\lambda\left(\theta\right)\right\vert}c_n\left(\theta, \lambda\left(\theta\right)\right)
+
\left(1 - \frac{s}{\left\vert\lambda\left(\theta\right)\right\vert}\right)c_n\left(\theta, 0\right)
\leq
c_n\left(\theta, 0\right)
=
n + \tau_n.
\end{equation*}
Let \({\varphi(t)} = {c_n(\theta, t\lambda^*)}\).
Then \({\varphi(1)} = {\varphi(0) + \varphi^\prime(0) + \varphi^{\prime\prime}(\tilde{t})/2}\) for some \({\tilde{t}} \in {(0, 1)}\), with \({\varphi^\prime(0)} = {{\lambda^*}^\top(n h_n(\theta) + \tau_n\mu_{n, \theta})}\).
Since \({\vert \tilde{t}{\lambda^*}^\top g_i(\theta)\vert} \leq {\vert\lambda^*\vert\vert g_i(\theta)\vert} \leq {s\zeta_n} \leq {r_n\zeta_n/2} \leq {1/2}\),
\begin{align*}
\varphi^{\prime\prime}\left(\tilde{t}\right)
&=
\sum_{i = 1}^n \exp\left(\tilde{t}{\lambda^*}^\top g_i\left(\theta\right)\right)\left({\lambda^*}^\top g_i\left(\theta\right)\right)^2\\
&\quad
+
\tau_n q_n\left(\theta, \tilde{t}\lambda^*\right)\left(\left({\lambda^*}^\top\mu_{n, \theta} + \tilde{t}{\lambda^*}^\top\Sigma_{n, \theta}\lambda^*\right)^2 + {\lambda^*}^\top\Sigma_{n, \theta}\lambda^*\right)\\
&\geq
\sum_{i = 1}^n \exp\left(\tilde{t}{\lambda^*}^\top g_i\left(\theta\right)\right)\left({\lambda^*}^\top g_i\left(\theta\right)\right)^2\\
&\geq
\sum_{i = 1}^n \left(1 + \tilde{t}{\lambda^*}^\top g_i\left(\theta\right)\right)\left({\lambda^*}^\top g_i\left(\theta\right)\right)^2\\
&\geq
\frac{1}{2}\sum_{i = 1}^n \left({\lambda^*}^\top g_i\left(\theta\right)\right)^2\\
&=
\frac{1}{2}{\lambda^*}^\top\left(\sum_{i = 1}^n g_i\left(\theta\right)g_i\left(\theta\right)^\top\right)\lambda^*\\
&=
\frac{n}{2}{\lambda^*}^\top V_n\left(\theta\right)\lambda^*\\
&\geq
\frac{n}{2}\frac{\sigma_V}{4}\left\vert\lambda^*\right\vert^2\\
&=
\frac{\sigma_V}{8} n s^2.
\end{align*}
Since \({\varphi^\prime(0) + \varphi^{\prime\prime}(\tilde{t})/2} = {\varphi(1) - \varphi(0)} \leq {0}\) and \({\vert\mu_{n, \theta}\vert} \leq {b_2}\),
\begin{align*}
\frac{\sigma_V}{16} n s^2
&\leq
\frac{1}{2}\varphi^{\prime\prime}\left(\tilde{t}\right)\\
&\leq
-\varphi^\prime(0)\\
&\leq
\left\vert\varphi^\prime(0)\right\vert\\
&\leq
\left\vert\lambda^*\right\vert\left\vert nh_n\left(\theta\right) + \tau_n\mu_{n, \theta}\right\vert\\
&\leq
s\left(n\left\vert h_n\left(\theta\right)\right\vert + \tau_nb_2\right),
\end{align*}
so \({s} \leq {C_3(\vert h_n(\theta)\vert + t_n)}\) with \({C_3} = {(16/\sigma_V)\max\{1, b_2\}}\).
If \({\vert\lambda(\theta)\vert} > {r_n/2}\), then \({s} = {r_n/2}\) and \({r_n/2} \leq {C_3(C_1 n^{-1/3} + t_n)}\), which contradicts \({C_3(C_1 n^{-1/3} + t_n)} < {n^{-b}/2} = {r_n/2}\) for \({n} \geq {n_*}\), enlarging \({n_*}\) if necessary, since \({b} < {1/3}\) and \({t_n} = {O(n^{-1}\log n)}\) by \cref{condition:retel asymptotics}.
Hence \({\vert\lambda(\theta)\vert} \leq {r_n/2}\) and \({s} = {\vert\lambda(\theta)\vert}\), so \({\vert\lambda(\theta)\vert} \leq {C_3(\vert h_n(\theta)\vert + t_n)}\).
Write \({\lambda} = {\lambda(\theta)}\).
Since \({\tau_n} \leq {n^{2/3}}\) for \({n} \geq {n_*}\), enlarging \({n_*}\) if necessary, \({\vert\lambda\vert} \leq {C_3(C_1 n^{-1/3} + t_n)} = {C_3(C_1 + \tau_n n^{-2/3})n^{-1/3}} \leq {C_4 n^{-1/3}}\) with \({C_4} = {C_3(C_1 + 1)}\).
With \({\zeta_n} \leq {n^b} \leq {n^{1/3}}\), \({\vert\lambda^\top g_i(\theta)\vert} \leq {\vert\lambda\vert\zeta_n} \leq {C_4 n^{-1/3}n^{1/3}} = {C_4}\).
For \eqref{eq:bvm lambda fine}, dividing \({\partial c_n(\theta, \lambda)/\partial\lambda} = {0}\) by \({n}\),
\begin{equation*}
h_n\left(\theta\right) + V_n\left(\theta\right)\lambda + \rho_1 + t_n q_n\left(\theta, \lambda\right)\left(\mu_{n, \theta} + \Sigma_{n, \theta}\lambda\right) = 0,
\end{equation*}
where \({\rho_1} = {n^{-1}\sum_{i = 1}^n (\exp(\lambda^\top g_i(\theta)) - 1 - \lambda^\top g_i(\theta))g_i(\theta)}\), so
\begin{equation*}
V_n\left(\theta\right)\lambda
=
-\left(h_n\left(\theta\right) + t_n\mu_{n, \theta}\right)
-
\rho_1
-
t_n\Sigma_{n, \theta}\lambda
-
t_n\left(q_n\left(\theta, \lambda\right) - 1\right)\left(\mu_{n, \theta} + \Sigma_{n, \theta}\lambda\right).
\end{equation*}
Since \({\vert\lambda^\top g_i(\theta)\vert} \leq {C_4}\),
\begin{equation*}
\left\vert\exp\left(\lambda^\top g_i\left(\theta\right)\right) - 1 - \lambda^\top g_i\left(\theta\right)\right\vert
\leq
\frac{\exp\left(C_4\right)}{2}\left(\lambda^\top g_i\left(\theta\right)\right)^2,
\end{equation*}
so
\begin{align*}
\left\vert\rho_1\right\vert
&\leq
\frac{1}{n}\sum_{i = 1}^n \left\vert\exp\left(\lambda^\top g_i\left(\theta\right)\right) - 1 - \lambda^\top g_i\left(\theta\right)\right\vert\left\vert g_i\left(\theta\right)\right\vert\\
&\leq
\frac{\exp\left(C_4\right)}{2}\frac{1}{n}\sum_{i = 1}^n \left(\lambda^\top g_i\left(\theta\right)\right)^2\left\vert g_i\left(\theta\right)\right\vert\\
&\leq
\frac{\exp\left(C_4\right)}{2}\left\vert\lambda\right\vert^2\frac{1}{n}\sum_{i = 1}^n \left\vert g_i\left(\theta\right)\right\vert^3\\
&\leq
\frac{\exp\left(C_4\right)}{2}\left\vert\lambda\right\vert^2\frac{1}{n}\sum_{i = 1}^n \sup_{\vartheta \in \Theta}\left\vert g_i\left(\vartheta\right)\right\vert^3\\
&=
\frac{\exp\left(C_4\right)}{2}B_n^{(3)}\left\vert\lambda\right\vert^2\\
&\leq
C\left\vert\lambda\right\vert^2.
\end{align*}
Since \({\vert\lambda\vert} \leq {C_4 n^{-1/3}} \leq {\min\{1, 1/(2b_2)\}}\) for \({n} \geq {n_*}\), enlarging \({n_*}\) if necessary,
\begin{equation*}
\left\vert\lambda^\top\mu_{n, \theta} + \frac{1}{2}\lambda^\top\Sigma_{n, \theta}\lambda\right\vert
\leq
b_2\left\vert\lambda\right\vert + \frac{1}{2}b_2\left\vert\lambda\right\vert^2
\leq
2b_2\left\vert\lambda\right\vert
\leq
1,
\end{equation*}
so
\begin{align*}
\left\vert q_n\left(\theta, \lambda\right) - 1\right\vert
&=
\left\vert\exp\left(\lambda^\top\mu_{n, \theta} + \frac{1}{2}\lambda^\top\Sigma_{n, \theta}\lambda\right) - 1\right\vert\\
&\leq
2\left\vert\lambda^\top\mu_{n, \theta} + \frac{1}{2}\lambda^\top\Sigma_{n, \theta}\lambda\right\vert\\
&\leq
4b_2\left\vert\lambda\right\vert.
\end{align*}
Multiplying by \({V_n(\theta)^{-1}}\),
\begin{equation*}
\lambda
=
-V_n\left(\theta\right)^{-1}\left(h_n\left(\theta\right) + t_n\mu_{n, \theta}\right) + \rho\left(\theta\right),
\end{equation*}
where
\begin{equation*}
\rho\left(\theta\right) = -V_n\left(\theta\right)^{-1}\left(\rho_1 + t_n\Sigma_{n, \theta}\lambda + t_n\left(q_n\left(\theta, \lambda\right) - 1\right)\left(\mu_{n, \theta} + \Sigma_{n, \theta}\lambda\right)\right).
\end{equation*}
With the eigenvalue floor \({\sigma_V/4}\) of \({V_n(\theta)}\) and \({\vert\lambda\vert} \leq {1}\),
\begin{align*}
\left\vert\rho\left(\theta\right)\right\vert
&\leq
\frac{4}{\sigma_V}\left\vert\rho_1 + t_n\Sigma_{n, \theta}\lambda + t_n\left(q_n\left(\theta, \lambda\right) - 1\right)\left(\mu_{n, \theta} + \Sigma_{n, \theta}\lambda\right)\right\vert\\
&\leq
\frac{4}{\sigma_V}\left(\left\vert\rho_1\right\vert + t_n\left\vert\Sigma_{n, \theta}\lambda\right\vert + t_n\left\vert q_n\left(\theta, \lambda\right) - 1\right\vert\left\vert\mu_{n, \theta} + \Sigma_{n, \theta}\lambda\right\vert\right)\\
&\leq
\frac{4}{\sigma_V}\left(C\left\vert\lambda\right\vert^2 + b_2 t_n\left\vert\lambda\right\vert + 4b_2 t_n\left\vert\lambda\right\vert\left(b_2 + b_2\left\vert\lambda\right\vert\right)\right)\\
&=
\frac{4}{\sigma_V}\left(C\left\vert\lambda\right\vert^2 + \left(b_2 + 4b_2^2\right)t_n\left\vert\lambda\right\vert + 4b_2^2 t_n\left\vert\lambda\right\vert^2\right)\\
&\leq
C\left(\left\vert\lambda\right\vert^2 + t_n\left\vert\lambda\right\vert\right).
\end{align*}
The bound \({\vert\lambda\vert} \leq {C_3(\vert h_n(\theta)\vert + t_n)}\) yields \eqref{eq:bvm lambda fine} with a suitable \({C_5}\).
\(\hfill\blacksquare\)

\begin{lemma}\label{lemma:bvm expansions}
Assume \cref{condition:parameter space,condition:continuity,condition:full rank,condition:uniform boundedness,condition:retel asymptotics,condition:prior,condition:bvm}.
Write \({\theta_u} = {\widehat{\theta} + n^{-1/2}u}\).
Then, enlarging \({n_*}\) if necessary, the following hold.
\begin{enumerate}
\item[(i)] On \({E_n \cap \{\widehat{\theta} \in B_{\delta/4}\}}\), for \({n} \geq {n_*}\), \({A(\widehat{\theta}) + n\log n} \geq {-\tau_n}\) and
\begin{equation}\label{eq:bvm A fine}
\sup_{\theta \in \Theta_{n, 1}}
\frac{\left\vert A\left(\theta\right) + n\log n + \frac{n}{2} h_n\left(\theta\right)^\top V_n\left(\theta\right)^{-1}h_n\left(\theta\right)\right\vert}
{1 + n\left\vert h_n\left(\theta\right)\right\vert^2}
\leq
\varepsilon_n
\end{equation}
with deterministic \({\varepsilon_n} \to {0}\).
\item[(ii)] On \({E_n \cap \{\widehat{\theta} \in B_{\delta/4}\}}\), for \({n} \geq {n_*}\), \({\sup_{\theta \in \Theta_{n, 1}}\vert \textnormal{pen}(\theta) - \textnormal{pen}(\widehat{\theta})\vert} \leq {\varepsilon_n^\prime}\) with deterministic \({\varepsilon_n^\prime} \to {0}\), and \({\textnormal{pen}(\widehat{\theta})} \geq {-(1 + C_0)\log n - 4\tau_n - C}\).
\item[(iii)] With probability approaching \({1}\), \({\theta_u} \in {\Theta_{n, 1}}\) for all \({\vert u\vert} \leq {n^{1/6}}\), and
\begin{equation}\label{eq:bvm lan}
\ell\left(\theta_u\right) - \ell\left(\widehat{\theta}\right) = -\frac{1}{2}u^\top\Omega^{-1}u + R_n\left(u\right),\quad \sup_{\vert u\vert \leq n^{1/6}}\frac{\left\vert R_n\left(u\right)\right\vert}{1 + \left\vert u\right\vert^2} = o_P\left(1\right).
\end{equation}
\item[(iv)] There exists \({c_7} > {0}\) such that
\begin{equation}\label{eq:bvm coarse}
\textnormal{pr}\left(
\sup_{\theta \in B_\delta : \vert\theta - \widehat{\theta}\vert > n^{-1/3}}
\left(\ell\left(\theta\right) - \ell\left(\widehat{\theta}\right)\right)
\leq
-c_7 n^{1/3}
\right)
\to
1.
\end{equation}
\end{enumerate}
\end{lemma}

\noindent\textit{Proof.}
On \({E_n \cap \{\widehat{\theta} \in B_{\delta/4}\}}\), \({\widehat{\theta}} \in {\Theta_{n, 1}}\) since \({h_n(\widehat{\theta})} = {0}\).
Set \({\varrho_n} = {C_3(C_1 n^{-1/3} + t_n)}\), so that \({\sup_{\theta \in \Theta_{n, 1}}\vert\lambda(\theta)\vert} \leq {\varrho_n}\) by \cref{lemma:bvm multiplier}(iii) for \({n} \geq {n_*}\), and \({\varrho_n} \to {0}\) since \({\tau_n} = {O(\log n)}\) by \cref{condition:retel asymptotics}.

\noindent (i).
Fix \({\theta} \in {\Theta_{n, 1}}\) and write \({\lambda} = {\lambda(\theta)}\), \({h} = {h_n(\theta)}\), \({\mu} = {\mu_{n, \theta}}\), \({\rho} = {\rho(\theta)}\), and \({w} = {h + t_n\mu}\).
With \({A(\theta) + n\log n} = {n(\lambda^\top h - \log m_n(\theta, \lambda))}\), write \({z} = {m_n(\theta, \lambda) - 1}\), so that
\begin{equation*}
z
=
\lambda^\top h + \frac{1}{2}\lambda^\top V_n\left(\theta\right)\lambda + \rho_2,
\end{equation*}
where \({\rho_2} = {n^{-1}\sum_{i = 1}^n (\exp(\lambda^\top g_i(\theta)) - 1 - \lambda^\top g_i(\theta) - (\lambda^\top g_i(\theta))^2/2)}\).
Since we have \({\vert\lambda^\top g_i(\theta)\vert} \leq {C_4}\) by \cref{lemma:bvm multiplier}(iii),
\begin{equation*}
\left\vert\exp\left(\lambda^\top g_i\left(\theta\right)\right) - 1 - \lambda^\top g_i\left(\theta\right) - \frac{1}{2}\left(\lambda^\top g_i\left(\theta\right)\right)^2\right\vert
\leq
\frac{\exp\left(C_4\right)}{6}\left\vert\lambda^\top g_i\left(\theta\right)\right\vert^3,
\end{equation*}
so
\begin{align*}
\left\vert\rho_2\right\vert
&\leq
\frac{\exp\left(C_4\right)}{6}\frac{1}{n}\sum_{i = 1}^n \left\vert\lambda^\top g_i\left(\theta\right)\right\vert^3\\
&\leq
\frac{\exp\left(C_4\right)}{6}\left\vert\lambda\right\vert^3\frac{1}{n}\sum_{i = 1}^n \sup_{\vartheta \in \Theta}\left\vert g_i\left(\vartheta\right)\right\vert^3\\
&=
\frac{\exp\left(C_4\right)}{6}B_n^{(3)}\left\vert\lambda\right\vert^3\\
&\leq
C\left\vert\lambda\right\vert^3.
\end{align*}
With \({\vert\lambda\vert} \leq {C_3(\vert h\vert + t_n)}\) by \cref{lemma:bvm multiplier}(iii), \({\vert\lambda\vert} \leq {\varrho_n} \leq {1}\) for \({n} \geq {n_*}\), enlarging \({n_*}\) if necessary, and the eigenvalue ceiling \({2C_V}\) of \({V_n(\theta)}\),
\begin{align*}
\left\vert z\right\vert
&\leq
\left\vert\lambda\right\vert\left\vert h\right\vert + C_V\left\vert\lambda\right\vert^2 + C\left\vert\lambda\right\vert^3\\
&\leq
\left\vert\lambda\right\vert\left\vert h\right\vert + C\left\vert\lambda\right\vert^2\\
&\leq
C_3\left(\left\vert h\right\vert + t_n\right)\left\vert h\right\vert + C C_3^2\left(\left\vert h\right\vert + t_n\right)^2\\
&\leq
C\left(\left\vert h\right\vert + t_n\right)^2.
\end{align*}
Substituting \({\vert\lambda\vert} \leq {\varrho_n}\) and \({\vert h\vert} \leq {C_1 n^{-1/3}}\) instead,
\begin{equation*}
\left\vert z\right\vert
\leq
\left\vert\lambda\right\vert\left\vert h\right\vert + C\left\vert\lambda\right\vert^2
\leq
C_1\varrho_n n^{-1/3} + C\varrho_n^2
\leq
\frac{1}{2}
\end{equation*}
for \({n} \geq {n_*}\), enlarging \({n_*}\) if necessary.
Write \({r_3} = {\log(1 + z) - z + z^2/2}\), so that \({\vert r_3\vert} \leq {3\vert z\vert^3}\) and
\begin{equation*}
\lambda^\top h - \log m_n\left(\theta, \lambda\right)
=
\lambda^\top h - z + \frac{1}{2}z^2 - r_3
=
-\frac{1}{2}\lambda^\top V_n\left(\theta\right)\lambda - \rho_2 + \frac{1}{2}z^2 - r_3.
\end{equation*}
Substituting \eqref{eq:bvm lambda fine}, \({\lambda^\top V_n(\theta)\lambda} = {w^\top V_n(\theta)^{-1}w - 2\rho^\top w + \rho^\top V_n(\theta)\rho}\), and thus
\begin{align*}
&A\left(\theta\right) + n\log n + \frac{n}{2}h^\top V_n\left(\theta\right)^{-1}h\\
&=
-\frac{n}{2}\lambda^\top V_n\left(\theta\right)\lambda - n\rho_2 + \frac{n}{2}z^2 - n r_3 + \frac{n}{2}h^\top V_n\left(\theta\right)^{-1}h\\
&=
-\frac{n}{2}w^\top V_n\left(\theta\right)^{-1}w + n\rho^\top w - \frac{n}{2}\rho^\top V_n\left(\theta\right)\rho - n\rho_2 + \frac{n}{2}z^2 - n r_3 + \frac{n}{2}h^\top V_n\left(\theta\right)^{-1}h\\
&=
-\tau_n\mu^\top V_n\left(\theta\right)^{-1}h - \frac{\tau_n^2}{2n}\mu^\top V_n\left(\theta\right)^{-1}\mu + n\rho^\top w - \frac{n}{2}\rho^\top V_n\left(\theta\right)\rho - n\rho_2 + \frac{n}{2}z^2 - n r_3.
\end{align*}
With \({\vert\mu\vert} \leq {b_2}\), the eigenvalues of \({V_n(\theta)}\) contained in \({[\sigma_V/4, 2C_V]}\), \({\vert\rho\vert} \leq {C(\vert h\vert + t_n)^2}\) by \eqref{eq:bvm lambda fine}, \({\vert\lambda\vert} \leq {C_3(\vert h\vert + t_n)}\), \({\vert z\vert} \leq {C(\vert h\vert + t_n)^2}\), \({\vert h\vert} \leq {C_1 n^{-1/3}}\), and \({t_n} \leq {n^{-1/3}}\) for \({n} \geq {n_*}\),
\begin{align*}
\tau_n\left\vert\mu^\top V_n\left(\theta\right)^{-1}h\right\vert
&\leq
\frac{4b_2}{\sigma_V}\tau_n\left\vert h\right\vert
\leq
C\tau_n n^{-1/3},\\
\frac{\tau_n^2}{2n}\left\vert\mu^\top V_n\left(\theta\right)^{-1}\mu\right\vert
&\leq
\frac{2b_2^2}{\sigma_V}\frac{\tau_n^2}{n}
\leq
C\frac{\tau_n^2}{n},\\
n\left\vert\rho\right\vert\left\vert w\right\vert + n\left\vert\rho_2\right\vert
&\leq
Cn\left(\left\vert h\right\vert + t_n\right)^3\\
&\leq
Cn\left(\left\vert h\right\vert^3 + t_n^3\right)\\
&=
C\left(\left\vert h\right\vert\left(n\left\vert h\right\vert^2\right) + \tau_n t_n^2\right)\\
&\leq
C\left(n^{-1/3}\left(n\left\vert h\right\vert^2\right) + \tau_n t_n^2\right),\\
\frac{n}{2}\left\vert\rho^\top V_n\left(\theta\right)\rho\right\vert + \frac{n}{2}z^2
&\leq
Cn\left(\left\vert h\right\vert + t_n\right)^4\\
&\leq
Cn\left(\left\vert h\right\vert^4 + t_n^4\right)\\
&=
C\left(\left\vert h\right\vert^2\left(n\left\vert h\right\vert^2\right) + \tau_n t_n^3\right)\\
&\leq
C\left(n^{-2/3}\left(n\left\vert h\right\vert^2\right) + \tau_n t_n^3\right),\\
n\left\vert r_3\right\vert
&\leq
Cn\left(\left\vert h\right\vert + t_n\right)^6\\
&\leq
Cn\left(\left\vert h\right\vert^6 + t_n^6\right)\\
&=
C\left(\left\vert h\right\vert^4\left(n\left\vert h\right\vert^2\right) + \tau_n t_n^5\right)\\
&\leq
C\left(n^{-4/3}\left(n\left\vert h\right\vert^2\right) + n^{-4/3}\tau_n t_n\right)\\
&=
Cn^{-4/3}\left(n\left\vert h\right\vert^2 + \tau_n t_n\right).
\end{align*}
Each of the five bounds above is at most \({C((1 + \tau_n)n^{-1/3} + \tau_n^2 n^{-1})(1 + n\vert h\vert^2)}\) with \({\tau_n} \leq {n^{2/3}}\), and thus
\begin{equation*}
\left\vert A\left(\theta\right) + n\log n + \frac{n}{2}h^\top V_n\left(\theta\right)^{-1}h\right\vert \leq C\left(\left(1 + \tau_n\right)n^{-1/3} + \frac{\tau_n^2}{n}\right)\left(1 + n\left\vert h\right\vert^2\right).
\end{equation*}
Write \({\varepsilon_n} = {C((1 + \tau_n)n^{-1/3} + \tau_n^2 n^{-1})}\).
Then \({\varepsilon_n} \to {0}\), and dividing by \({1 + n\vert h\vert^2}\) and taking the supremum over \({\Theta_{n, 1}}\) gives \eqref{eq:bvm A fine}.
At \({\widehat{\theta}}\), \({h_n(\widehat{\theta})} = {0}\) gives \({A(\widehat{\theta}) + n\log n} = {-n\log m_n(\widehat{\theta}, \lambda(\widehat{\theta}))}\), and it follows from \({n m_n(\widehat{\theta}, \lambda(\widehat{\theta}))} \leq {c_n(\widehat{\theta}, \lambda(\widehat{\theta}))} \leq {c_n(\widehat{\theta}, 0)} = {n + \tau_n}\) that \({\log m_n(\widehat{\theta}, \lambda(\widehat{\theta}))} \leq {\log(1 + t_n)} \leq {t_n}\).
Hence \({A(\widehat{\theta}) + n\log n} \geq {-\tau_n}\).

\noindent (ii).
For \({\theta} \in {\Theta_{n, 1}}\), write \({x_\theta} = {t_n q_n(\theta, \lambda(\theta))/m_n(\theta, \lambda(\theta))}\).
By \eqref{eq:bvm pen} and \eqref{eq:bvm pc},
\begin{equation*}
\textnormal{pen}\left(\theta\right)
=
\log a_n + \log q_n\left(\theta, \lambda\left(\theta\right)\right) - K_n\left(\theta, \lambda\left(\theta\right)\right) - n\log\left(1 + x_\theta\right).
\end{equation*}
Since \({\widehat{\theta}} \in {\Theta_{n, 1}}\),
\begin{align*}
\textnormal{pen}\left(\theta\right) - \textnormal{pen}\left(\widehat{\theta}\right)
&=
\left(\log q_n\left(\theta, \lambda\left(\theta\right)\right) - \log q_n\left(\widehat{\theta}, \lambda\left(\widehat{\theta}\right)\right)\right)\\
&\quad
-
\left(K_n\left(\theta, \lambda\left(\theta\right)\right) - K_n\left(\widehat{\theta}, \lambda\left(\widehat{\theta}\right)\right)\right)\\
&\quad
-
n\left(\log\left(1 + x_\theta\right) - \log\left(1 + x_{\widehat{\theta}}\right)\right).
\end{align*}
For the first difference, since \({\varrho_n} \to {0}\), \({8b_2\varrho_n} \leq {1}\) for \({n} \geq {n_*}\), enlarging \({n_*}\) if necessary.
With \({\vert\lambda(\theta)\vert} \leq {\varrho_n} \leq {1}\),
\begin{align*}
\left\vert\log q_n\left(\theta, \lambda\left(\theta\right)\right)\right\vert
&=
\left\vert\lambda\left(\theta\right)^\top\mu_{n, \theta} + \frac{1}{2}\lambda\left(\theta\right)^\top\Sigma_{n, \theta}\lambda\left(\theta\right)\right\vert\\
&\leq
b_2\varrho_n + \frac{1}{2}b_2\varrho_n^2\\
&\leq
b_2\varrho_n + \frac{1}{2}b_2\varrho_n\\
&=
\frac{3}{2}b_2\varrho_n\\
&\leq
2b_2\varrho_n\\
&\leq
\frac{1}{4}.
\end{align*}
Then \({\vert q_n(\theta, \lambda(\theta)) - 1\vert} \leq {2\vert\log q_n(\theta, \lambda(\theta))\vert} \leq {4b_2\varrho_n} \leq {1/2}\) and \({q_n(\theta, \lambda(\theta))} \in {[1/2, 3/2]} \subseteq {[1/2, 2]}\).
For the second difference, with \({z} = {m_n(\theta, \lambda(\theta)) - 1}\),
\begin{equation*}
K_n\left(\theta, \lambda\left(\theta\right)\right)
=
\log\left(1 + \left(1 - a_n\right)z + a_n\left(q_n\left(\theta, \lambda\left(\theta\right)\right) - 1\right)\right),
\end{equation*}
and, with \({\vert z\vert} \leq {C_1\varrho_n n^{-1/3} + C\varrho_n^2}\) from (i) and \({a_n} \leq {t_n}\),
\begin{align*}
\left\vert\left(1 - a_n\right)z + a_n\left(q_n\left(\theta, \lambda\left(\theta\right)\right) - 1\right)\right\vert
&\leq
\left(1 - a_n\right)\left\vert z\right\vert + a_n\left\vert q_n\left(\theta, \lambda\left(\theta\right)\right) - 1\right\vert\\
&\leq
\left\vert z\right\vert + 4b_2 t_n\varrho_n\\
&\leq
C\left(\varrho_n n^{-1/3} + \varrho_n^2 + t_n\varrho_n\right)\\
&\leq
\frac{1}{2}
\end{align*}
for \({n} \geq {n_*}\), enlarging \({n_*}\) if necessary.
Then \({\vert K_n(\theta, \lambda(\theta))\vert} \leq {C(\varrho_n n^{-1/3} + \varrho_n^2 + t_n\varrho_n)}\).
For the third difference, with \({\vert z\vert} \leq {1/2}\) from (i), \({m_n(\theta, \lambda(\theta))} \geq {1/2}\) and \({q_n(\theta, \lambda(\theta))} \leq {2}\) give \({0} \leq {x_\theta} \leq {4t_n}\).
Then,
\begin{align*}
\left\vert n\log\left(1 + x_\theta\right) - n x_\theta\right\vert
&\leq
\frac{n}{2} x_\theta^2\\
&\leq
8\tau_n t_n,\\
\left\vert n x_\theta - \tau_n\right\vert
&=
\tau_n\left\vert\frac{q_n\left(\theta, \lambda\left(\theta\right)\right)}{m_n\left(\theta, \lambda\left(\theta\right)\right)} - 1\right\vert\\
&=
\tau_n\left\vert\frac{q_n\left(\theta, \lambda\left(\theta\right)\right) - 1 - z}{m_n\left(\theta, \lambda\left(\theta\right)\right)}\right\vert\\
&\leq
2\tau_n\left\vert q_n\left(\theta, \lambda\left(\theta\right)\right) - 1 - z\right\vert\\
&\leq
2\tau_n\left(\left\vert q_n\left(\theta, \lambda\left(\theta\right)\right) - 1\right\vert + \left\vert z\right\vert\right)\\
&\leq
2\tau_n\left(4b_2\varrho_n + C_1\varrho_n n^{-1/3} + C\varrho_n^2\right)\\
&\leq
2\tau_n\left(4b_2\varrho_n + C_1\varrho_n + C\varrho_n\right)\\
&\leq
C\tau_n\varrho_n,
\end{align*}
and \({\vert n\log(1 + x_\theta) - \tau_n\vert} \leq {\vert n\log(1 + x_\theta) - n x_\theta\vert + \vert n x_\theta - \tau_n\vert} \leq {8\tau_n t_n + C\tau_n\varrho_n}\).
Combining the three bounds at \({\theta}\) and \({\widehat{\theta}}\),
\begin{align*}
&\left\vert\textnormal{pen}\left(\theta\right) - \textnormal{pen}\left(\widehat{\theta}\right)\right\vert\\
&\leq
\left\vert\log q_n\left(\theta, \lambda\left(\theta\right)\right) - \log q_n\left(\widehat{\theta}, \lambda\left(\widehat{\theta}\right)\right)\right\vert
+
\left\vert K_n\left(\theta, \lambda\left(\theta\right)\right) - K_n\left(\widehat{\theta}, \lambda\left(\widehat{\theta}\right)\right)\right\vert\\
&\quad
+
n\left\vert\log\left(1 + x_\theta\right) - \log\left(1 + x_{\widehat{\theta}}\right)\right\vert\\
&\leq
\left\vert\log q_n\left(\theta, \lambda\left(\theta\right)\right)\right\vert + \left\vert\log q_n\left(\widehat{\theta}, \lambda\left(\widehat{\theta}\right)\right)\right\vert
+
\left\vert K_n\left(\theta, \lambda\left(\theta\right)\right)\right\vert + \left\vert K_n\left(\widehat{\theta}, \lambda\left(\widehat{\theta}\right)\right)\right\vert\\
&\quad
+
\left\vert n\log\left(1 + x_\theta\right) - \tau_n\right\vert + \left\vert n\log\left(1 + x_{\widehat{\theta}}\right) - \tau_n\right\vert\\
&\leq
2b_2\varrho_n + 2b_2\varrho_n + C\left(\varrho_n n^{-1/3} + \varrho_n^2 + t_n\varrho_n\right) + C\left(\varrho_n n^{-1/3} + \varrho_n^2 + t_n\varrho_n\right)\\
&\quad
+
\left(8\tau_n t_n + C\tau_n\varrho_n\right) + \left(8\tau_n t_n + C\tau_n\varrho_n\right)\\
&=
4b_2\varrho_n + 2C\left(\varrho_n n^{-1/3} + \varrho_n^2 + t_n\varrho_n\right) + 16\tau_n t_n + 2C\tau_n\varrho_n\\
&\leq
4b_2\varrho_n + 6C\varrho_n + 16\frac{\tau_n^2}{n} + 2C\tau_n\varrho_n\\
&\leq
C\left(\varrho_n + \tau_n\varrho_n + \frac{\tau_n^2}{n}\right).
\end{align*}
Hence, \({\sup_{\theta \in \Theta_{n, 1}}\vert\textnormal{pen}(\theta) - \textnormal{pen}(\widehat{\theta})\vert} \leq {\varepsilon_n^\prime}\) with \({\varepsilon_n^\prime} = {C(\varrho_n + \tau_n\varrho_n + \tau_n^2 n^{-1})}\), and \({\varepsilon_n^\prime} \to {0}\).
At \({\widehat{\theta}}\), \({K_n(\widehat{\theta}, \lambda(\widehat{\theta}))} \leq {K_n(\widehat{\theta}, 0)} = {0}\), \({n\log(1 + x_{\widehat{\theta}})} \leq {n x_{\widehat{\theta}}} \leq {4\tau_n}\), \({\log q_n(\widehat{\theta}, \lambda(\widehat{\theta}))} \geq {-2b_2\varrho_n} \geq {-1}\), and, since \({\tau_n} \leq {n}\) and \({\tau_n^{-1}} = {O(n^{C_0})}\) by \cref{condition:retel asymptotics},
\begin{equation*}
\log a_n
=
\log\frac{\tau_n}{n + \tau_n}
\geq
\log\frac{\tau_n}{2n}
\geq
-\left(1 + C_0\right)\log n - C,
\end{equation*}
and thus \({\textnormal{pen}(\widehat{\theta})} \geq {-(1 + C_0)\log n - 4\tau_n - C}\).

\noindent (iii).
The event that \({E_n}\) holds, \({\widehat{\theta}} \in {B_{\delta/4}}\), and \({n} \geq {n_*}\) has probability tending to \({1}\) by \cref{lemma:bvm working event} and \cref{lemma:bvm preliminaries}(ii).
On this event, for \({\vert u\vert} \leq {n^{1/6}}\), \({\vert\theta_u - \theta_0\vert} \leq {\delta/4 + n^{-1/3}} \leq {\delta}\), enlarging \({n_*}\) if necessary.
Then \({\theta_u} \in {\Theta_{n, 1}}\), since by \eqref{eq:bvm lipschitz} \({\vert h_n(\theta_u)\vert} = {\vert h_n(\theta_u) - h_n(\widehat{\theta})\vert} \leq {C_G\vert\theta_u - \widehat{\theta}\vert} \leq {C_G n^{-1/3}} < {C_1 n^{-1/3}}\).
For \eqref{eq:bvm lan}, by \eqref{eq:bvm decomposition},
\begin{equation*}
\ell\left(\theta_u\right) - \ell\left(\widehat{\theta}\right)
=
-\frac{n}{2}h_n\left(\theta_u\right)^\top V_n\left(\theta_u\right)^{-1}h_n\left(\theta_u\right) + R_n^{(1)}\left(u\right),
\end{equation*}
where
\begin{align*}
R_n^{(1)}\left(u\right)
&=
\left(A\left(\theta_u\right) + n\log n + \frac{n}{2}h_n\left(\theta_u\right)^\top V_n\left(\theta_u\right)^{-1}h_n\left(\theta_u\right)\right)\\
&\quad
-
\left(A\left(\widehat{\theta}\right) + n\log n\right)\\
&\quad
+
\left(\textnormal{pen}\left(\theta_u\right) - \textnormal{pen}\left(\widehat{\theta}\right)\right).
\end{align*}
By (i) at \({\theta_u}\) and \({\widehat{\theta}}\), (ii), and \eqref{eq:bvm lipschitz} with \({n\vert h_n(\theta_u)\vert^2} \leq {C_G^2\vert u\vert^2}\),
\begin{align*}
\left\vert R_n^{(1)}\left(u\right)\right\vert
&\leq
\varepsilon_n\left(1 + n\left\vert h_n\left(\theta_u\right)\right\vert^2\right) + \varepsilon_n + \varepsilon_n^\prime\\
&\leq
\varepsilon_n\left(1 + C_G^2\left\vert u\right\vert^2\right) + \varepsilon_n + \varepsilon_n^\prime\\
&\leq
\left(2\varepsilon_n + \varepsilon_n^\prime\right)\max\{1, C_G^2\}\left(1 + \left\vert u\right\vert^2\right)
\end{align*}
uniformly over \({\vert u\vert} \leq {n^{1/6}}\), with \({\varepsilon_n^{\prime\prime}} = {(2\varepsilon_n + \varepsilon_n^\prime)\max\{1, C_G^2\}}\), and \({\varepsilon_n^{\prime\prime}} \to {0}\).
Let
\begin{equation*}
\bar{G}_u = \int_0^1 \frac{\partial h_n}{\partial\theta^\top}\left(\widehat{\theta} + t n^{-1/2}u\right)dt.
\end{equation*}
Then \({h_n(\theta_u)} = {\bar{G}_u n^{-1/2}u}\) and
\begin{equation*}
n h_n\left(\theta_u\right)^\top V_n\left(\theta_u\right)^{-1}h_n\left(\theta_u\right)
=
u^\top\bar{G}_u^\top V_n\left(\theta_u\right)^{-1}\bar{G}_u u.
\end{equation*}
Write \({\rho_n^\prime} = {\vert\widehat{\theta} - \theta_0\vert + n^{-1/3}}\), so that \({\rho_n^\prime} = {o_P(1)}\) by \cref{lemma:bvm preliminaries}(ii) and the segment from \({\widehat{\theta}}\) to \({\theta_u}\) lies in \({\{\theta : \vert\theta - \theta_0\vert \leq \rho_n^\prime\}}\).
By \cref{lemma:bvm preliminaries}(i), uniformly over \({\vert u\vert} \leq {n^{1/6}}\),
\begin{align*}
\left\Vert\bar{G}_u - G\right\Vert
&=
\left\Vert\int_0^1 \left(\frac{\partial h_n}{\partial\theta^\top}\left(\widehat{\theta} + t n^{-1/2}u\right) - G\right)dt\right\Vert\\
&\leq
\int_0^1 \left\Vert\frac{\partial h_n}{\partial\theta^\top}\left(\widehat{\theta} + t n^{-1/2}u\right) - G\left(\widehat{\theta} + t n^{-1/2}u\right)\right\Vert dt\\
&\quad
+
\int_0^1 \left\Vert G\left(\widehat{\theta} + t n^{-1/2}u\right) - G\right\Vert dt\\
&\leq
\sup_{\theta \in \mathcal{N}^\prime}\left\Vert\frac{\partial h_n}{\partial\theta^\top}\left(\theta\right) - G\left(\theta\right)\right\Vert
+
\sup_{\vert\theta - \theta_0\vert \leq \rho_n^\prime}\left\Vert G\left(\theta\right) - G\right\Vert\\
&=
o_P\left(1\right),
\end{align*}
and
\begin{align*}
\sup_{\vert u\vert \leq n^{1/6}}\left\Vert V_n\left(\theta_u\right) - V\right\Vert
&\leq
\sup_{\vert u\vert \leq n^{1/6}}\left\Vert V_n\left(\theta_u\right) - V\left(\theta_u\right)\right\Vert
+
\sup_{\vert u\vert \leq n^{1/6}}\left\Vert V\left(\theta_u\right) - V\right\Vert\\
&\leq
\sup_{\theta \in \Theta}\left\Vert V_n\left(\theta\right) - V\left(\theta\right)\right\Vert
+
\sup_{\vert\theta - \theta_0\vert \leq \rho_n^\prime}\left\Vert V\left(\theta\right) - V\right\Vert\\
&=
o_P\left(1\right).
\end{align*}
Then,
\begin{align*}
\sup_{\vert u\vert \leq n^{1/6}}\left\Vert V_n\left(\theta_u\right)^{-1} - V^{-1}\right\Vert
&=
\sup_{\vert u\vert \leq n^{1/6}}\left\Vert V_n\left(\theta_u\right)^{-1}\left(V - V_n\left(\theta_u\right)\right)V^{-1}\right\Vert\\
&\leq
\sup_{\vert u\vert \leq n^{1/6}}\left\Vert V_n\left(\theta_u\right)^{-1}\right\Vert\left\Vert V - V_n\left(\theta_u\right)\right\Vert\left\Vert V^{-1}\right\Vert\\
&\leq
\frac{4p}{\sigma_V^2}\sup_{\vert u\vert \leq n^{1/6}}\left\Vert V_n\left(\theta_u\right) - V\right\Vert\\
&=
o_P\left(1\right).
\end{align*}
Write \({\eta_n} = {\sup_{\vert u\vert \leq n^{1/6}}\Vert\bar{G}_u^\top V_n(\theta_u)^{-1}\bar{G}_u - G^\top V^{-1}G\Vert}\).
With
\begin{align*}
\bar{G}_u^\top V_n\left(\theta_u\right)^{-1}\bar{G}_u - G^\top V^{-1}G
&=
\left(\bar{G}_u - G\right)^\top V_n\left(\theta_u\right)^{-1}\bar{G}_u
+
G^\top\left(V_n\left(\theta_u\right)^{-1} - V^{-1}\right)\bar{G}_u\\
&\quad
+
G^\top V^{-1}\left(\bar{G}_u - G\right)
\end{align*}
and \({\Vert\bar{G}_u\Vert} \leq {\Vert G\Vert + \Vert\bar{G}_u - G\Vert}\),
\begin{align*}
\eta_n
&\leq
\sup_{\vert u\vert \leq n^{1/6}}\left\Vert\bar{G}_u - G\right\Vert\left\Vert V_n\left(\theta_u\right)^{-1}\right\Vert\left\Vert\bar{G}_u\right\Vert
+
\left\Vert G\right\Vert\sup_{\vert u\vert \leq n^{1/6}}\left\Vert V_n\left(\theta_u\right)^{-1} - V^{-1}\right\Vert\left\Vert\bar{G}_u\right\Vert\\
&\quad
+
\left\Vert G\right\Vert\left\Vert V^{-1}\right\Vert\sup_{\vert u\vert \leq n^{1/6}}\left\Vert\bar{G}_u - G\right\Vert\\
&=
o_P\left(1\right).
\end{align*}
Let \({R_n(u)} = {R_n^{(1)}(u) - (1/2)u^\top(\bar{G}_u^\top V_n(\theta_u)^{-1}\bar{G}_u - \Omega^{-1})u}\), so that
\begin{equation*}
\ell\left(\theta_u\right) - \ell\left(\widehat{\theta}\right)
=
-\frac{1}{2}u^\top\Omega^{-1}u + R_n\left(u\right).
\end{equation*}
Then,
\begin{align*}
\sup_{\vert u\vert \leq n^{1/6}}\frac{\left\vert R_n\left(u\right)\right\vert}{1 + \left\vert u\right\vert^2}
&\leq
\sup_{\vert u\vert \leq n^{1/6}}\frac{\left\vert R_n^{(1)}\left(u\right)\right\vert}{1 + \left\vert u\right\vert^2}
+
\frac{1}{2}\sup_{\vert u\vert \leq n^{1/6}}\frac{\left\vert u^\top\left(\bar{G}_u^\top V_n\left(\theta_u\right)^{-1}\bar{G}_u - \Omega^{-1}\right)u\right\vert}{1 + \left\vert u\right\vert^2}\\
&\leq
\varepsilon_n^{\prime\prime}
+
\frac{1}{2}\sup_{\vert u\vert \leq n^{1/6}}\frac{\left\Vert\bar{G}_u^\top V_n\left(\theta_u\right)^{-1}\bar{G}_u - \Omega^{-1}\right\Vert\left\vert u\right\vert^2}{1 + \left\vert u\right\vert^2}\\
&\leq
\varepsilon_n^{\prime\prime} + \frac{1}{2}\eta_n\\
&=
o_P\left(1\right).
\end{align*}

\noindent (iv).
On the event that \({E_n}\) holds, \({\widehat{\theta}} \in {B_{\delta/4}}\), and \({n} \geq {n_*}\), fix \({\theta} \in {B_\delta}\) with \({\vert\theta - \widehat{\theta}\vert} > {n^{-1/3}}\), so that \({\vert h_n(\theta)\vert} \geq {c_G\vert\theta - \widehat{\theta}\vert} > {c_G n^{-1/3}}\) by \eqref{eq:bvm lipschitz}.
If \({\theta} \in {\Theta_{n, 2}}\), then \({\vert\lambda(\theta)\vert} \geq {c_2 n^{-1/3}}\) by \cref{lemma:bvm multiplier}(ii).
With \({c_2 n^{-1/3}} \leq {\bar{\lambda}}\) and \({(\sigma_V/24)c_2^2 n^{-2/3}} \leq {1}\), enlarging \({n_*}\) if necessary, \eqref{eq:bvm psi bound} at \({\lambda(\theta)}\) and \({\log(1 + x)} \geq {x/2}\) for \({x} \in {[0, 1]}\) give
\begin{align*}
A\left(\theta\right) + n\log n
&=
-n\psi_\theta\left(\lambda\left(\theta\right)\right)\\
&\leq
-n\log\left(1 + \frac{\sigma_V}{24}\min\left\{c_2 n^{-1/3}, \bar{\lambda}\right\}^2\right)\\
&=
-n\log\left(1 + \frac{\sigma_V}{24}c_2^2 n^{-2/3}\right)\\
&\leq
-\frac{\sigma_V c_2^2}{48} n^{1/3}.
\end{align*}
If instead \({\theta} \in {\Theta_{n, 1}}\), then \({c_G n^{-1/3}} < {\vert h_n(\theta)\vert} \leq {C_1 n^{-1/3}}\).
With \({\varepsilon_n} \leq {c_G^2/(16C_V(1 + C_1^2))}\), enlarging \({n_*}\) if necessary, \eqref{eq:bvm A fine} at \({\theta}\) gives
\begin{align*}
A\left(\theta\right) + n\log n
&\leq
-\frac{n}{2}h_n\left(\theta\right)^\top V_n\left(\theta\right)^{-1}h_n\left(\theta\right) + \varepsilon_n\left(1 + n\left\vert h_n\left(\theta\right)\right\vert^2\right)\\
&\leq
-\frac{n}{4C_V}\left\vert h_n\left(\theta\right)\right\vert^2 + \varepsilon_n\left(1 + C_1^2\right)n^{1/3}\\
&\leq
-\frac{c_G^2}{4C_V}n^{1/3} + \frac{c_G^2}{16C_V}n^{1/3}\\
&\leq
-\frac{c_G^2}{8C_V} n^{1/3}.
\end{align*}
In both cases, \({A(\theta) + n\log n} \leq {-c_8n^{1/3}}\) with \({c_8} = {\min\{\sigma_V c_2^2/48, c_G^2/(8C_V)\}}\).
With \({A(\widehat{\theta}) + n\log n} \geq {-\tau_n}\) from (i), \({\textnormal{pen}(\theta)} \leq {0}\), and \({\textnormal{pen}(\widehat{\theta})} \geq {-(1 + C_0)\log n - 4\tau_n - C}\) from (ii), \eqref{eq:bvm decomposition} gives
\begin{align*}
\ell\left(\theta\right) - \ell\left(\widehat{\theta}\right)
&=
\left(A\left(\theta\right) + n\log n\right) - \left(A\left(\widehat{\theta}\right) + n\log n\right) + \textnormal{pen}\left(\theta\right) - \textnormal{pen}\left(\widehat{\theta}\right)\\
&\leq
-c_8n^{1/3} + 5\tau_n + \left(1 + C_0\right)\log n + C\\
&\leq
-\frac{c_8}{2}n^{1/3},
\end{align*}
enlarging \({n_*}\) if necessary.
Taking the supremum over \({\theta}\), \eqref{eq:bvm coarse} holds with \({c_7} = {c_8/2}\) on an event of probability tending to \({1}\).
\(\hfill\blacksquare\)

Turning to the theorem, in the notation of \cref{lemma:bvm expansions} set \({U_n} = {\{u \in \mathbb{R}^p : \theta_u \in \Theta\}}\) and
\begin{equation*}
f_n(u)
=
\pi(\theta_u)\exp\left(\ell(\theta_u) - \ell(\widehat{\theta})\right)\mathds{1}\left\{u \in U_n\right\},
\end{equation*}
with \({C_n} = {\int f_n(u) du} = {n^{p/2}\exp(-\ell(\widehat{\theta}))\int_\Theta \pi(\theta)L_{RET}(\theta)d\theta} \in {(0, \infty)}\) almost surely.
By a change of variables, the posterior density of \({n^{1/2}(\theta - \widehat{\theta})}\) is \({f_n(u)/C_n}\).
Similarly, the density of \({N(0, \Omega)}\) is \({\phi(u)/D}\), where
\begin{equation*}
\phi(u)
=
\pi(\theta_0)\exp\left(-\frac{1}{2}u^\top\Omega^{-1}u\right)
\end{equation*}
and \({D} = {\int\phi(u) du} = {\pi(\theta_0)(2\pi)^{p/2}\det(\Omega)^{1/2}} \in {(0, \infty)}\) by \cref{condition:prior}.
The total variation identity gives
\begin{equation*}
\sup_{\mathcal{B}}\left\vert\pi\left(n^{1/2}\left(\theta - \widehat{\theta}\right) \in \mathcal{B} \middle\vert \mathcal{D}_n\right) - N\left(0, \Omega\right)\left(\mathcal{B}\right)\right\vert
=
\frac{1}{2}\int\left\vert\frac{f_n(u)}{C_n} - \frac{\phi(u)}{D}\right\vert du.
\end{equation*}
Adding and subtracting \({\phi(u)/C_n}\), we have
\begin{equation*}
\frac{1}{2}\int\left\vert\frac{f_n(u)}{C_n} - \frac{\phi(u)}{D}\right\vert du
\leq
\frac{J_n}{2C_n} + \frac{\left\vert D - C_n\right\vert}{2C_n}
\leq
\frac{J_n}{C_n},
\end{equation*}
where
\begin{equation*}
J_n = \int\left\vert f_n(u) - \phi(u)\right\vert du
\end{equation*}
and the second inequality uses \({\vert D - C_n\vert} \leq {J_n}\).
If \({J_n} = {o_P(1)}\), then \({C_n} \geq {D - J_n} \geq {D/2}\) with probability approaching \({1}\), so \({J_n/C_n} \leq {(2/D)J_n} = {o_P(1)}\), and the theorem follows.
The remainder of the proof establishes that \({J_n} = {o_P(1)}\).

Fix \({\epsilon, \eta} \in {(0, 1)}\).
Let \({\sigma_\Omega}\) be the smallest eigenvalue of \({\Omega^{-1}}\), and set \({\bar{\pi}} = {\sup_{\theta \in \Theta}\pi(\theta)}\), which is finite by \cref{condition:parameter space,condition:prior}.
Choose \({K} > {0}\) such that
\begin{equation}\label{eq:bvm phi tail}
\int_{\vert u\vert > K}\phi(u) du \leq \frac{\epsilon}{8},
\end{equation}
\begin{equation}\label{eq:bvm envelope tail}
\bar{\pi}
\exp\left(\frac{\sigma_\Omega}{8}\right)
\int_{\vert u\vert > K}
\exp\left(-\frac{3\sigma_\Omega}{8}\left\vert u\right\vert^2\right) du
\leq
\frac{\epsilon}{8},
\end{equation}
and \({N(0, \Omega)(\{u : \vert u\vert \geq K\})} \leq {\eta/32}\); all three requirements hold for all sufficiently large \({K}\).
By the convergence in distribution from \cref{lemma:bvm preliminaries}(ii) and the portmanteau theorem applied to the closed set \({\{u : \vert u\vert \geq K\}}\),
\begin{align*}
\limsup_{n \to \infty}\textnormal{pr}\left(n^{1/2}\left\vert\widehat{\theta} - \theta_0\right\vert > K\right)
&\leq
\limsup_{n \to \infty}\textnormal{pr}\left(n^{1/2}\left\vert\widehat{\theta} - \theta_0\right\vert \geq K\right)\\
&\leq
N\left(0, \Omega\right)\left(\left\{u : \left\vert u\right\vert \geq K\right\}\right)
\leq
\frac{\eta}{32},
\end{align*}
so \({\textnormal{pr}(n^{1/2}\vert\widehat{\theta} - \theta_0\vert > K)} \leq {\eta/16}\) for all sufficiently large \({n}\).
Since \({\widehat{\theta}} \to {\theta_0}\) in probability by \cref{lemma:bvm preliminaries}(ii), we also have \({\textnormal{pr}(\widehat{\theta} \notin B_{\delta/4})} \leq {\eta/16}\) for all sufficiently large \({n}\).
For some \({n_0}\), each of the following events has probability at least \({1 - \eta/8}\) for all \({n} \geq {n_0}\):
\({E^{(1)}}\), on which \({\theta_u} \in {\Theta_{n, 1}}\) for all \({\vert u\vert} \leq {n^{1/6}}\) and \eqref{eq:bvm lan} holds with \({\sup_{\vert u\vert \leq n^{1/6}}\vert R_n(u)\vert/(1 + \vert u\vert^2)} \leq {\sigma_\Omega/8}\), by \cref{lemma:bvm expansions}(iii);
\({E^{(2)}}\), the event in \eqref{eq:bvm coarse}, by \cref{lemma:bvm expansions}(iv);
\({E^{(3)}}\), on which \({n^{1/2}\vert\widehat{\theta} - \theta_0\vert} \leq {K}\) and \({\widehat{\theta}} \in {B_{\delta/4}}\), by the two bounds just established and a union bound;
\({E^{(4)}}\), on which \({\sup_{\vert\theta - \theta_0\vert > \delta/4} n^{-1}\{\ell(\theta) - \ell(\theta_0)\}} \leq {-\epsilon_\delta}\), by \cref{condition:bvm} applied with \({\delta/4}\), which furnishes the fixed constant \({\epsilon_\delta} > {0}\);
and \({E^{(5)}}\), on which \({\int_{\vert u\vert \leq K}\vert f_n(u) - \phi(u)\vert du} \leq {\epsilon/8}\), as verified below.
For \({E^{(5)}}\), first note that
\begin{equation*}
\sup_{\vert u\vert \leq K}\left\vert\theta_u - \theta_0\right\vert
\leq
K n^{-1/2} + \left\vert\widehat{\theta} - \theta_0\right\vert
=
o_P(1),
\end{equation*}
since \({K}\) is fixed and \({\widehat{\theta}} \to {\theta_0}\) in probability by \cref{lemma:bvm preliminaries}(ii).
Since \({\theta_0}\) is interior to \({\Theta}\) by \cref{condition:parameter space}, there is \({r} > {0}\) with \({\{\theta : \vert\theta - \theta_0\vert \leq r\}} \subseteq {\Theta}\).
On the event \({\{\sup_{\vert u\vert \leq K}\vert\theta_u - \theta_0\vert \leq r\}}\), whose probability tends to \({1}\), we have \({\theta_u} \in {\Theta}\) for every \({\vert u\vert} \leq {K}\), that is, \({\{u : \vert u\vert \leq K\}} \subseteq {U_n}\).
Hence, \({\textnormal{pr}(\{u : \vert u\vert \leq K\} \subseteq U_n)} \to {1}\), and on this containment \({f_n(u)} = {\pi(\theta_u)\exp(\ell(\theta_u) - \ell(\widehat{\theta}))}\) for every \({\vert u\vert} \leq {K}\).
Next, \({\pi}\) is uniformly continuous on the compact \({\Theta}\) by \cref{condition:parameter space,condition:prior}.
For every \({\epsilon^\prime} > {0}\) there is \({\delta^\prime} > {0}\) such that \({\vert\pi(\theta) - \pi(\theta^\prime)\vert} \leq {\epsilon^\prime}\) whenever \({\theta, \theta^\prime} \in {\Theta}\) and \({\vert\theta - \theta^\prime\vert} \leq {\delta^\prime}\).
On the event \({\{\sup_{\vert u\vert \leq K}\vert\theta_u - \theta_0\vert \leq \min\{\delta^\prime, r\}\}}\), whose probability tends to \({1}\), every \({\theta_u}\) with \({\vert u\vert} \leq {K}\) lies in \({\Theta}\) and satisfies \({\vert\pi(\theta_u) - \pi(\theta_0)\vert} \leq {\epsilon^\prime}\);
since \({\epsilon^\prime}\) is arbitrary, \({\sup_{\vert u\vert \leq K}\vert\pi(\theta_u) - \pi(\theta_0)\vert} = {o_P(1)}\).
For \({n} \geq {K^6}\) we have \({K} \leq {n^{1/6}}\), so on the event in \cref{lemma:bvm expansions}(iii), whose probability tends to \({1}\), \({\theta_u} \in {\Theta_{n, 1}}\) and \({R_n(u)}\) is defined for every \({\vert u\vert} \leq {K}\), and
\begin{equation*}
\left\vert R_n(u)\right\vert
=
\left(1 + \left\vert u\right\vert^2\right)\frac{\left\vert R_n(u)\right\vert}{1 + \left\vert u\right\vert^2}
\leq
\left(1 + K^2\right)\sup_{\vert v\vert \leq n^{1/6}}\frac{\left\vert R_n(v)\right\vert}{1 + \left\vert v\right\vert^2}.
\end{equation*}
Hence, \({\sup_{\vert u\vert \leq K}\vert R_n(u)\vert} = {o_P(1)}\) by \cref{lemma:bvm expansions}(iii).
On the containment \({\{v : \vert v\vert \leq K\}} \subseteq {U_n}\), for each \({\vert u\vert} \leq {K}\), \eqref{eq:bvm lan} gives
\begin{align*}
\left\vert f_n(u) - \phi(u)\right\vert
&=
\exp\left(-\frac{1}{2}u^\top\Omega^{-1}u\right)\left\vert\pi(\theta_u)\exp\left(R_n(u)\right) - \pi(\theta_0)\right\vert\\
&\leq
\left\vert\pi(\theta_u)\exp\left(R_n(u)\right) - \pi(\theta_0)\right\vert\\
&\leq
\left\vert\pi(\theta_u) - \pi(\theta_0)\right\vert
\exp\left(\left\vert R_n(u)\right\vert\right)
+ \pi(\theta_0)\left\vert\exp\left(R_n(u)\right) - 1\right\vert\\
&\leq
\left\vert\pi(\theta_u) - \pi(\theta_0)\right\vert
\exp\left(\left\vert R_n(u)\right\vert\right)
+
\pi(\theta_0)
\left\vert R_n(u)\right\vert
\exp\left(\left\vert R_n(u)\right\vert\right)\\
&=
\left(
\left\vert
\pi(\theta_u) - \pi(\theta_0)
\right\vert
+ \pi(\theta_0)\left\vert R_n(u)\right\vert
\right)
\exp\left(\left\vert R_n(u)\right\vert\right).
\end{align*}
Taking the supremum over \({\vert u\vert} \leq {K}\) gives
\begin{align*}
&\sup_{\vert u\vert \leq K}\left\vert f_n(u) - \phi(u)\right\vert\\
&\quad\leq
\left(\sup_{\vert u\vert \leq K}\left\vert\pi(\theta_u) - \pi(\theta_0)\right\vert
+
\pi(\theta_0)
\sup_{\vert u\vert \leq K}\left\vert R_n(u)\right\vert\right)
\exp\left(\sup_{\vert u\vert \leq K}\left\vert R_n(u)\right\vert\right),
\end{align*}
and thus \({\sup_{\vert u\vert \leq K}\vert f_n(u) - \phi(u)\vert} = {o_P(1)}\), since the containment holds with probability approaching \({1}\).
Finally, we have
\begin{equation*}
\int_{\vert u\vert \leq K}\left\vert f_n(u) - \phi(u)\right\vert du
\leq
\textnormal{vol}\left(\left\vert u\right\vert \leq K\right)
\sup_{\vert v\vert \leq K}\left\vert f_n(v) - \phi(v)\right\vert,
\end{equation*}
where \({\textnormal{vol}(\cdot)}\) denotes Lebesgue measure.
Hence, \({\int_{\vert u\vert \leq K}\vert f_n(u) - \phi(u)\vert du} = {o_P(1)}\); thus \({\textnormal{pr}(E^{(5)})} \to {1}\).

Now, observe that
\begin{align*}
J_n
&=
\int_{\vert u\vert \leq K}\left\vert f_n(u) - \phi(u)\right\vert du
+
\int_{\vert u\vert > K}\left\vert f_n(u) - \phi(u)\right\vert du\\
&\leq
\int_{\vert u\vert \leq K}\left\vert f_n(u) - \phi(u)\right\vert du
+
\int_{\vert u\vert > K}\phi(u) du
+
\int_{\vert u\vert > K}f_n(u) du,
\end{align*}
where the first term is at most \({\epsilon/8}\) on \({E^{(5)}}\) and the second at most \({\epsilon/8}\) by \eqref{eq:bvm phi tail}.
For the third term, we split the integral over the regions
\begin{align*}
Z_1 &= \left\{u : K < \left\vert u\right\vert \leq n^{1/6}\right\},\\
Z_2 &= \left\{u : n^{1/6} < \left\vert u\right\vert \leq \frac{\delta}{2}n^{1/2}\right\},\\
Z_3 &= \left\{u : \left\vert u\right\vert > \frac{\delta}{2}n^{1/2}\right\}.
\end{align*}
On \({E^{(1)}}\), for \({u} \in {Z_1 \cap U_n}\), \eqref{eq:bvm lan} gives
\begin{align*}
f_n(u)
&=
\pi(\theta_u)\exp\left(-\frac{1}{2}u^\top\Omega^{-1}u + R_n(u)\right)\\
&\leq
\pi(\theta_u)\exp\left(-\frac{\sigma_\Omega}{2}\left\vert u\right\vert^2 + \frac{\sigma_\Omega}{8}\left(1 + \left\vert u\right\vert^2\right)\right)\\
&\leq
\bar{\pi}\exp\left(\frac{\sigma_\Omega}{8}\right)\exp\left(-\frac{3\sigma_\Omega}{8}\left\vert u\right\vert^2\right),
\end{align*}
so \({\int_{Z_1}f_n(u) du} \leq {\epsilon/8}\) by \eqref{eq:bvm envelope tail}.
On \({E^{(2)} \cap E^{(3)}}\), for \({u} \in {Z_2 \cap U_n}\), we have \({\vert\theta_u - \widehat{\theta}\vert} \in {(n^{-1/3}, \delta/2]}\) and
\begin{equation*}
\left\vert\theta_u - \theta_0\right\vert
\leq
\left\vert\theta_u - \widehat{\theta}\right\vert + \left\vert\widehat{\theta} - \theta_0\right\vert
\leq
\frac{\delta}{2} + \frac{\delta}{4}
\leq
\delta,
\end{equation*}
so \({\theta_u}\) lies in the supremum set of \eqref{eq:bvm coarse} and \({f_n(u)} \leq {\bar{\pi}\exp(-c_7 n^{1/3})}\).
Hence,
\begin{align*}
\int_{Z_2}f_n(u) du
&\leq
\bar{\pi}
\exp\left(-c_7 n^{1/3}\right)
\textnormal{vol}\left(\left\vert u\right\vert \leq \frac{\delta}{2}n^{1/2}\right)\\
&=
\bar{\pi}
\textnormal{vol}\left(\left\vert u\right\vert \leq \frac{\delta}{2}\right)
n^{p/2}
\exp\left(-c_7 n^{1/3}\right)\\
&
\to 0,
\end{align*}
so \({\int_{Z_2}f_n(u) du} \leq {\epsilon/8}\) for all sufficiently large \({n}\).
On \({E^{(1)} \cap E^{(3)} \cap E^{(4)}}\), for \({u} \in {Z_3 \cap U_n}\), we have \({\vert\theta_u - \theta_0\vert} \geq {\vert\theta_u - \widehat{\theta}\vert - \vert\widehat{\theta} - \theta_0\vert} > {\delta/4}\), so \({\ell(\theta_u) - \ell(\theta_0)} \leq {-n\epsilon_\delta}\).
Let \({u_0} = {n^{1/2}(\theta_0 - \widehat{\theta})}\), so that \({\vert u_0\vert} \leq {K} \leq {n^{1/6}}\) for \({n} \geq {K^6}\).
Since \({\theta_{u_0}} = {\theta_0}\), \eqref{eq:bvm lan} gives
\begin{align*}
\ell(\theta_0) - \ell(\widehat{\theta})
&=
-\frac{1}{2}u_0^\top\Omega^{-1}u_0 + R_n(u_0)\\
&\leq
R_n(u_0)\\
&\leq
\frac{\sigma_\Omega}{8}\left(1 + K^2\right)\\
&=
C_K.
\end{align*}
Hence, \({\ell(\theta_u) - \ell(\widehat{\theta})} = {\ell(\theta_u) - \ell(\theta_0) + \ell(\theta_0) - \ell(\widehat{\theta})} \leq {C_K - n\epsilon_\delta}\), and
\begin{align*}
\int_{Z_3}f_n(u) du
&=
\int_{Z_3 \cap U_n}f_n(u) du\\
&\leq
\exp\left(C_K - n\epsilon_\delta\right)\int_{Z_3 \cap U_n}\pi(\theta_u) du\\
&\leq
\exp\left(C_K - n\epsilon_\delta\right)\int_{U_n}\pi(\theta_u) du\\
&=
\exp\left(C_K - n\epsilon_\delta\right)n^{p/2}\int_{\Theta}\pi(\theta)d\theta\\
&=
\exp\left(C_K - n\epsilon_\delta\right)n^{p/2}\\
&
\to 0,
\end{align*}
so \({\int_{Z_3}f_n(u) du} \leq {\epsilon/8}\) for all sufficiently large \({n}\).
Collecting the bounds, and enlarging \({n_0}\) if necessary so that the bounds for \({Z_2}\) and \({Z_3}\) also hold for all \({n} \geq {n_0}\), we have \({J_n} \leq {5\epsilon/8} < {\epsilon}\) on \({E^{(1)} \cap \cdots \cap E^{(5)}}\) for all \({n} \geq {n_0}\), while \({\textnormal{pr}(E^{(1)} \cap \cdots \cap E^{(5)})} \geq {1 - 5\eta/8} > {1 - \eta}\).
Hence, \({\textnormal{pr}(J_n > \epsilon)} \leq {5\eta/8} < {\eta}\) for all \({n} \geq {n_0}\).
Since \({\epsilon}\) and \({\eta}\) are arbitrary, \({J_n} = {o_P(1)}\), which completes the proof.
\(\hfill\blacksquare\)

\begin{remark}\label{rmk:bvm centering}
Since the total variation distance is invariant under translation, applying the shift \({n^{1/2}(\widehat{\theta} - \theta_0)}\) conditionally on the data shows that Theorem 3 is equivalent to the statement that
\begin{equation}\label{eq:bvm shifted claim}
\sup_{\mathcal{B}}\left\vert
\pi\left(n^{1/2}\left(\theta - \theta_0\right) \in \mathcal{B} \relmiddle | \mathcal{D}_n\right)
-
N\left(n^{1/2}\left(\widehat{\theta} - \theta_0\right), \Omega\right)\left(\mathcal{B}\right)
\right\vert
\to
0
\end{equation}
in probability.
By the standard expression for the total variation distance between normal distributions with a common positive definite covariance matrix \citep[Theorem~1]{barsov1987estimates},
\begin{align}\label{eq:bvm normal tv}
\begin{split}
\sup_{\mathcal{B}}\left\vert N\left(m_1, \Omega\right)\left(\mathcal{B}\right) - N\left(m_2, \Omega\right)\left(\mathcal{B}\right)\right\vert
&=
2\Phi\left(\frac{\left\vert\Omega^{-1/2}\left(m_1 - m_2\right)\right\vert}{2}\right) - 1\\
&\leq
\left(2\pi\right)^{-1/2}\left\vert\Omega^{-1/2}\left(m_1 - m_2\right)\right\vert,
\end{split}
\end{align}
where \({m_1, m_2} \in {\mathbb{R}^p}\) and \({\Phi}\) denotes the standard normal distribution function.
The inequality holds since \({\Phi}\) has density bounded by \({(2\pi)^{-1/2}}\).
Let
\begin{equation*}
\Delta_n = -\Omega G^\top V^{-1}n^{-1/2}\sum_{i = 1}^n g_i(\theta_0)
=
-G^{-1}n^{1/2}h_n(\theta_0).
\end{equation*}
Combining the inequality in \eqref{eq:bvm normal tv} with \cref{lemma:bvm preliminaries}(ii) yields
\begin{align}\label{eq:bvm center swap}
\begin{split}
\sup_{\mathcal{B}}&\left\vert
N\left(n^{1/2}\left(\widehat{\theta} - \theta_0\right), \Omega\right)\left(\mathcal{B}\right) - N\left(\Delta_n, \Omega\right)\left(\mathcal{B}\right)\right\vert\\
&\leq
\left(2\pi\right)^{-1/2}\left\vert\Omega^{-1/2}\left(n^{1/2}\left(\widehat{\theta} - \theta_0\right) - \Delta_n\right)\right\vert\\
&\leq
\left(2\pi\right)^{-1/2}\left\Vert\Omega^{-1/2}\right\Vert o_P(1)\\
&=
o_P(1).
\end{split}
\end{align}
The triangle inequality, together with \eqref{eq:bvm shifted claim} and \eqref{eq:bvm center swap}, implies that
\begin{equation}\label{eq:bvm delta claim}
\sup_{\mathcal{B}}\left\vert
\pi\left(n^{1/2}\left(\theta - \theta_0\right) \in \mathcal{B} \relmiddle | \mathcal{D}_n\right)
-
N\left(\Delta_n, \Omega\right)\left(\mathcal{B}\right)
\right\vert
\to
0
\end{equation}
in probability.
By the equality in \eqref{eq:bvm normal tv} applied with \({m_1} = {\Delta_n}\) and \({m_2} = {0}\), the total variation distance between \({N(\Delta_n, \Omega)}\) and \({N(0, \Omega)}\) equals \({2\Phi(\vert\Omega^{-1/2}\Delta_n\vert/2) - 1}\).
Since \({\Delta_n}\) converges in distribution to \({N(0, \Omega)}\), this total variation distance converges in distribution to \({2\Phi(\vert Z\vert/2) - 1}\) for a standard normal random vector \({Z}\) in \({\mathbb{R}^p}\).
This limit is nondegenerate and supported on \({(0, 1)}\).
If \eqref{eq:bvm delta claim} also held with \({N(0, \Omega)}\) in place of \({N(\Delta_n, \Omega)}\), the triangle inequality would force the total variation distance between \({N(\Delta_n, \Omega)}\) and \({N(0, \Omega)}\) to converge to zero in probability, contradicting the nondegenerate limit.
Therefore, the random centering \({\Delta_n}\) cannot be replaced by \({0}\) in \eqref{eq:bvm delta claim}; that is, the fixed centering \({\theta_0}\) cannot replace \({\widehat{\theta}}\) in Theorem 3.
The estimator-centered formulation of Theorem 3 follows the convention of \citet[Theorem~2]{yiu2020inference}.
\end{remark}

\begin{remark}\label{rmk:retel reduced}
For \({\textnormal{RETEL}_r}\), write
\begin{equation*}
\widetilde{\ell}(\theta)
=
\log\widetilde{L}_{RET}(\theta)
=
\sum_{i = 1}^n \log p_i(\theta)
=
A(\theta) - n\log(1 + x_\theta),
\end{equation*}
so \eqref{eq:bvm decomposition} holds with the same multiplier \({\lambda(\theta)}\), the same \({A(\theta)}\), and \({\textnormal{pen}(\theta)}\) in \eqref{eq:bvm pen} replaced by \({\textnormal{pen}_r(\theta)} = {-n\log(1 + x_\theta)}\).
Note that \({\textnormal{pen}_r(\theta)} \leq {0}\).
The proof of \cref{lemma:bvm expansions}(ii) gives \({\textnormal{pen}_r(\widehat{\theta})} \geq {-4\tau_n}\) and
\begin{align*}
\sup_{\theta \in \Theta_{n, 1}}\left\vert\textnormal{pen}_r(\theta) - \textnormal{pen}_r(\widehat{\theta})\right\vert
&\leq
\left(8\tau_n t_n + C\tau_n\varrho_n\right) + \left(8\tau_n t_n + C\tau_n\varrho_n\right)\\
&\leq
C\left(\tau_n\varrho_n + \frac{\tau_n^2}{n}\right)\\
&\leq
\varepsilon_n^\prime.
\end{align*}
Every other step of the argument is unchanged.
Consequently, the conclusion of Theorem 3 also holds for the posterior obtained from \({\textnormal{RETEL}_r}\), provided \cref{condition:bvm} holds with \({\widetilde{L}_{RET}}\) in place of \({L_{RET}}\).
For this version, the requirement \({\tau_n^{-1}} = {O(n^{C_0})}\) in \cref{condition:retel asymptotics} is not used, since it enters the argument only through the lower bound for \({\log a_n}\) in the proof of \cref{lemma:bvm expansions}(ii).
\end{remark}

\subsection{Proposition~4}
It follows from the compactness of \({\Theta}\) under \cref{condition:parameter space} and Lemma~1 of \citet{berger2009formal} that \({I(\pi\mid\mathcal{M}_2)} < {\infty}\).
We write
\begin{align*}
I\left(\pi\mid\mathcal{M}_2\right)
&=
\int_{\mathcal{X}}
\int_{\mathcal{X}} D_{KL}
\left(
\pi\left(\cdot \relmiddle | x_1,x_2\right)
\relmiddle \Vert
\pi\left(\cdot\right)
\right)
m\left(x_1,x_2\right)dx_1dx_2 \\
&=
\int_{\mathcal{X}}
\int_{\mathcal{X}}
\int_{\Theta}
\pi
\left(
\theta \relmiddle | x_1,x_2
\right)
\log\left(
\frac{\pi\left(\theta \relmiddle | x_1,x_2\right)}
{\pi\left(\theta\right)}
\right)
d\theta
m\left(x_1,x_2\right)dx_1dx_2 \\
&=
\int_{\mathcal{X}}
\int_{\mathcal{X}}
\int_{\Theta}
\pi\left(\theta\right)
p\left(x_1,x_2 \relmiddle | \theta\right)
\log\left(
\frac{p\left(x_1,x_2 \relmiddle | \theta\right)}
{m\left(x_1,x_2\right)}
\right)
d\theta
dx_1dx_2 \\
&=
\int_{\Theta}
\pi\left(\theta\right)
\int_{\mathcal{X}}
\int_{\mathcal{X}}
p\left(x_1,x_2 \relmiddle | \theta\right)
\log\left(
\frac{p\left(x_1,x_2 \relmiddle | \theta\right)}
{m\left(x_1,x_2\right)}
\right)
dx_1dx_2
d\theta
\end{align*}
and
\begin{equation*}
I\left(\pi\mid\mathcal{M}_1\right)
=
\int_{\Theta}
\pi\left(\theta\right)
\int_{\mathcal{X}}
p\left(x_1 \relmiddle | \theta\right)
\log\left(
\frac{p\left(x_1 \relmiddle | \theta\right)}
{m\left(x_1\right)}
\right)
dx_1
d\theta.
\end{equation*}
Let
\begin{equation*}
A_1
=
\int_{\mathcal{X}}
p\left(x_1 \relmiddle | \theta\right)
\log\left(
\frac{p\left(x_1 \relmiddle | \theta\right)}
{m\left(x_1\right)}
\right)
dx_1
\end{equation*}
and
\begin{equation*}
A_2
=
\int_{\mathcal{X}}
\int_{\mathcal{X}}
p\left(x_1,x_2 \relmiddle | \theta\right)
\log\left(
\frac{p\left(x_1,x_2 \relmiddle | \theta\right)}
{m\left(x_1,x_2\right)}
\right)
dx_1dx_2.
\end{equation*}
It suffices to show that \({A_1} \leq {A_2}\).
To this end,
\begin{align*}
A_2
&=
\int_{\mathcal{X}}
\int_{\mathcal{X}}
p\left(x_2 \relmiddle | x_1, \theta\right)
p\left(x_1 \relmiddle | \theta\right)
\log\left(
\frac{p\left(x_2 \relmiddle | x_1, \theta\right)
p\left(x_1 \relmiddle | \theta\right)}
{m\left(x_2 \relmiddle | x_1\right)
m\left(x_1\right)}
\right)
dx_2dx_1 \\
&=
\int_{\mathcal{X}}
\int_{\mathcal{X}}
p\left(x_2 \relmiddle | x_1, \theta\right)
p\left(x_1 \relmiddle | \theta\right)
\log\left(
\frac{p\left(x_2 \relmiddle | x_1, \theta\right)}
{m\left(x_2 \relmiddle | x_1\right)}
\right)
dx_2dx_1 \\
&\quad +
\int_{\mathcal{X}}
\int_{\mathcal{X}}
p\left(x_2 \relmiddle | x_1, \theta\right)
p\left(x_1 \relmiddle | \theta\right)
\log\left(
\frac{p\left(x_1 \relmiddle | \theta\right)}
{m\left(x_1\right)}
\right)
dx_2dx_1 \\
&=
\int_{\mathcal{X}}
p\left(x_1 \relmiddle | \theta\right)
\left[\int_{\mathcal{X}}
p\left(x_2 \relmiddle | x_1, \theta\right)
\log\left(
\frac{p\left(x_2 \relmiddle | x_1, \theta\right)}
{m\left(x_2 \relmiddle | x_1\right)}
\right)
dx_2\right]
dx_1 + A_1.
\end{align*}
Now, for all \(x_1\) and \(\theta\), the quantity within square brackets, being a Kullback--Leibler divergence, is nonnegative and the result follows.
\(\hfill\blacksquare\)

\section{Quantile-Quantile Plots}
\begin{figure}[!t]
\centering
\includegraphics[width=\linewidth,height=6.4in,keepaspectratio]{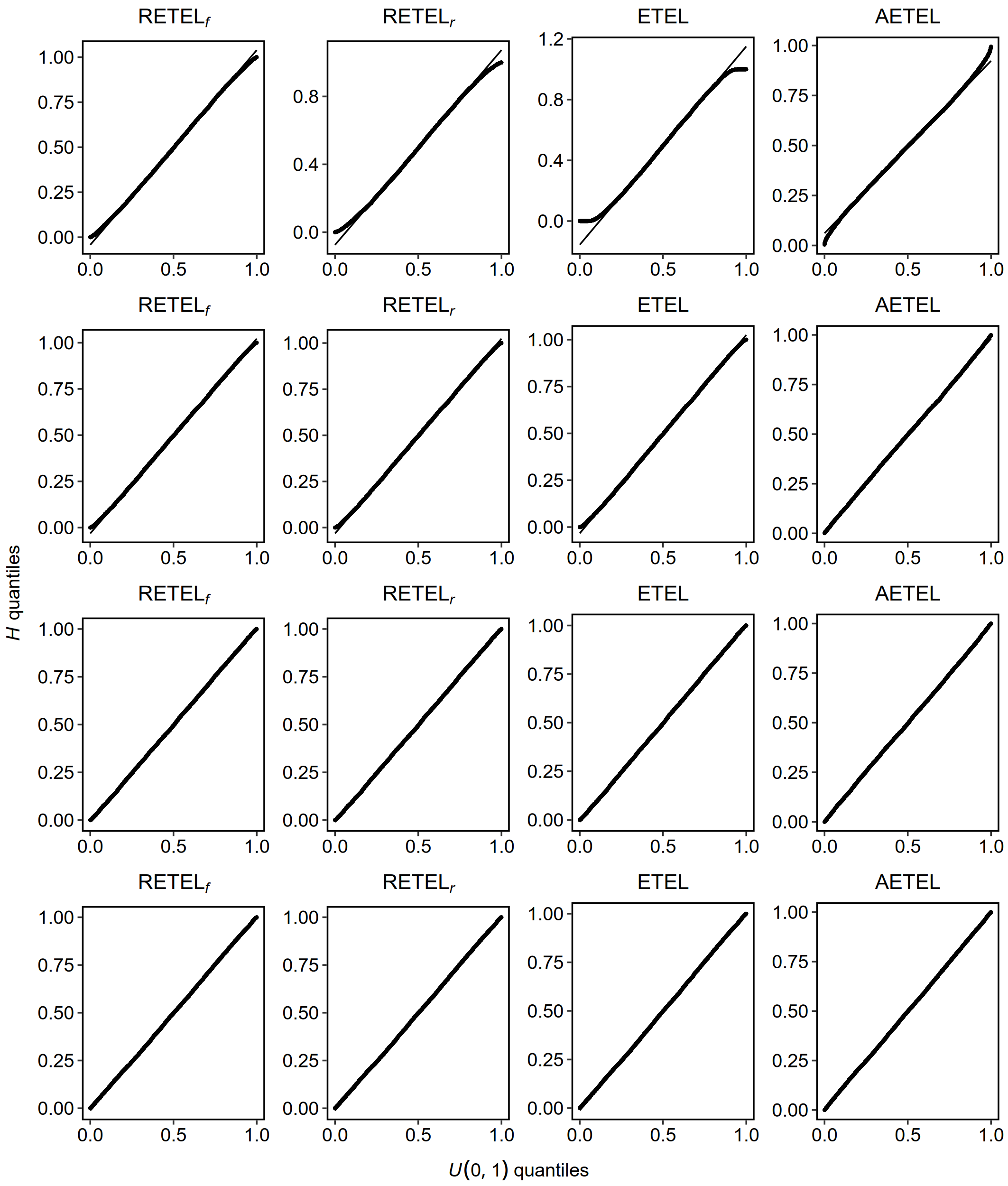}
\caption{
Quantile-quantile plots for the distribution of \({H}\) versus \({U(0, 1)}\) under \({s} = {1}\) and \({\tau_n} = {1}\) for different sample sizes (\({n} = {5}\) in the first row, \({n} = {20}\) in the second row, \({n} = {50}\) in the third row, and \({n} = {100}\) in the fourth row).
}
\label{fig:qq_tau1_s1}
\end{figure}

\begin{figure}[!t]
\centering
\includegraphics[width=\linewidth,height=6.4in,keepaspectratio]{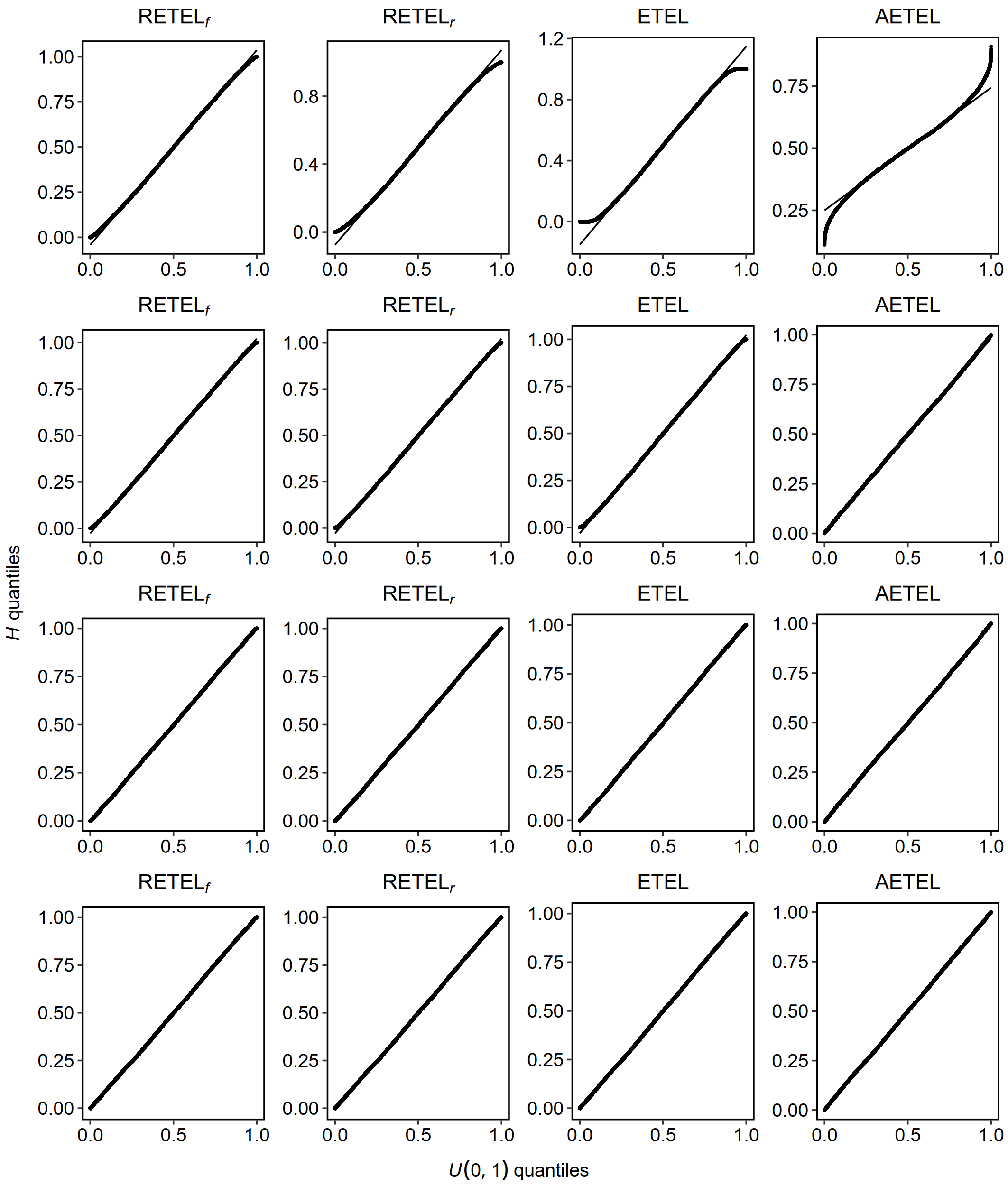}
\caption{
Quantile-quantile plots for the distribution of \({H}\) versus \({U(0, 1)}\) under \({s} = {5}\) and \({\tau_n} = {1}\) for different sample sizes (\({n} = {5}\) in the first row, \({n} = {20}\) in the second row, \({n} = {50}\) in the third row, and \({n} = {100}\) in the fourth row).
}
\label{fig:qq_tau1_s5}
\end{figure}

\begin{figure}[!t]
\centering
\includegraphics[width=\linewidth,height=6.4in,keepaspectratio]{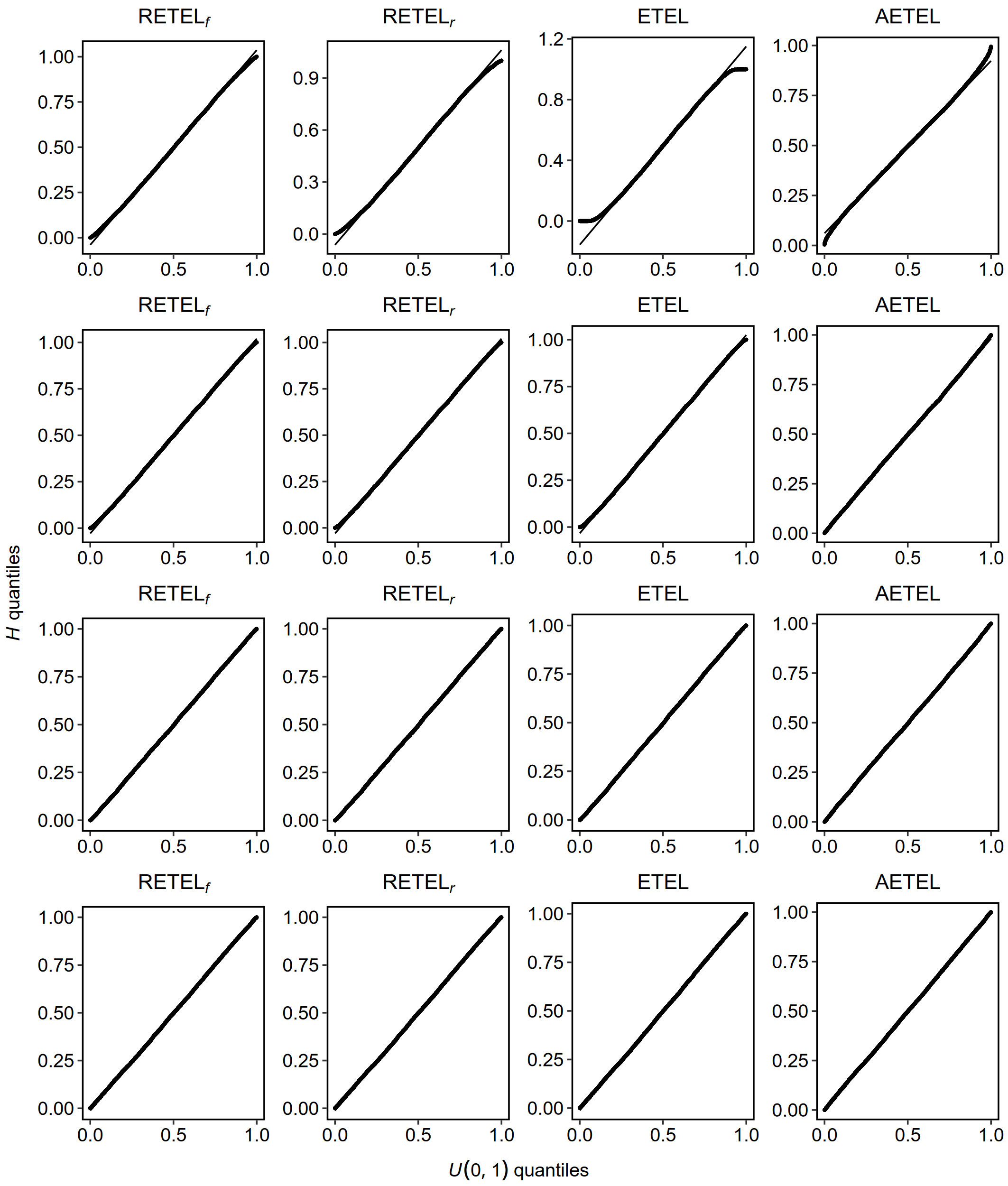}
\caption{
Quantile-quantile plots for the distribution of \({H}\) versus \({U(0, 1)}\) under \({s} = {1}\) and \({\tau_n} = {\log n}\) for different sample sizes (\({n} = {5}\) in the first row, \({n} = {20}\) in the second row, \({n} = {50}\) in the third row, and \({n} = {100}\) in the fourth row).
}
\label{fig:qq_taulogn_s1}
\end{figure}

\begin{figure}[!t]
\centering
\includegraphics[width=\linewidth,height=6.4in,keepaspectratio]{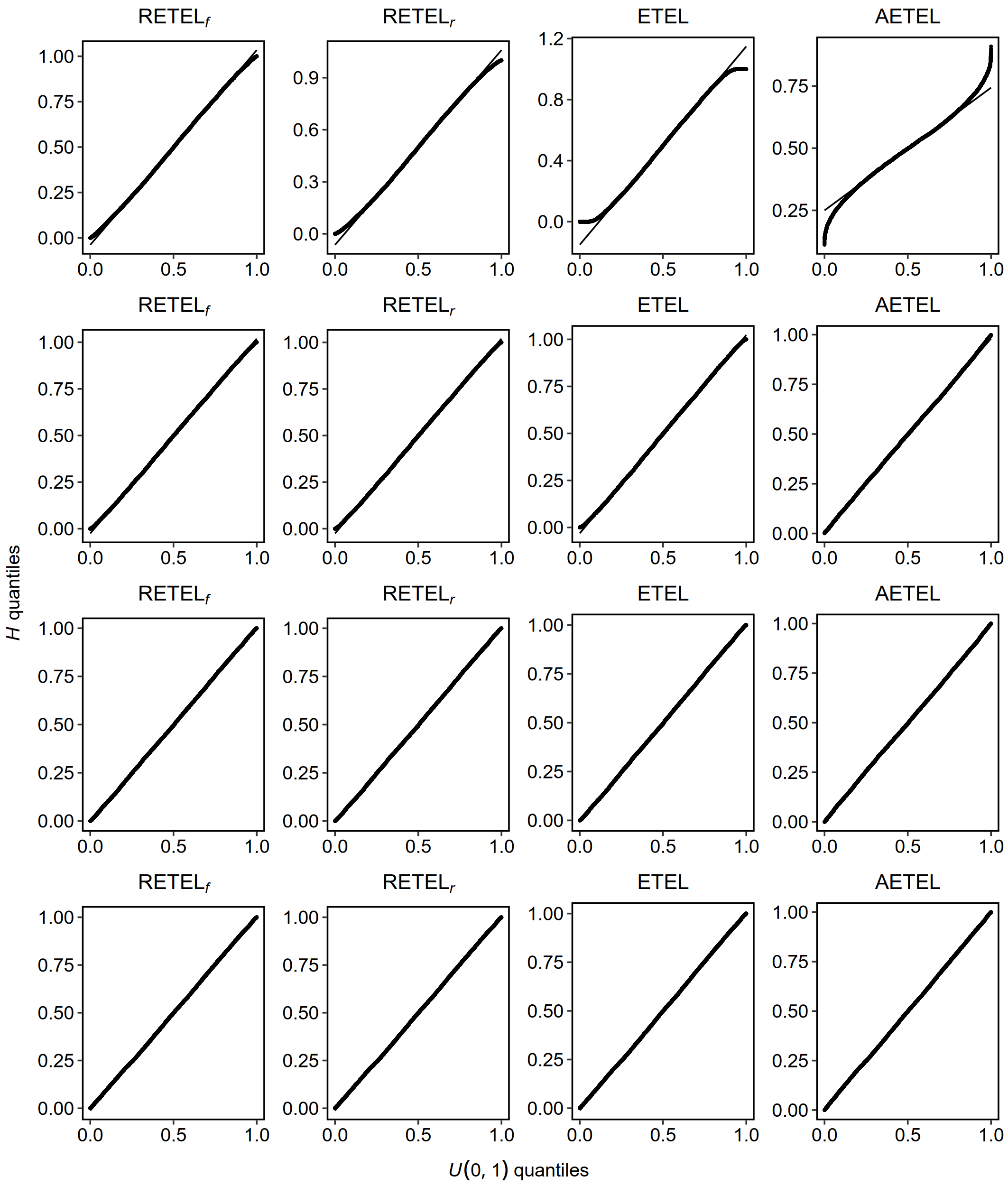}
\caption{
Quantile-quantile plots for the distribution of \({H}\) versus \({U(0, 1)}\) under \({s} = {5}\) and \({\tau_n} = {\log n}\) for different sample sizes (\({n} = {5}\) in the first row, \({n} = {20}\) in the second row, \({n} = {50}\) in the third row, and \({n} = {100}\) in the fourth row).
}
\label{fig:qq_taulogn_s5}
\end{figure}

\begin{figure}[!t]
\centering
\includegraphics[width=\linewidth]{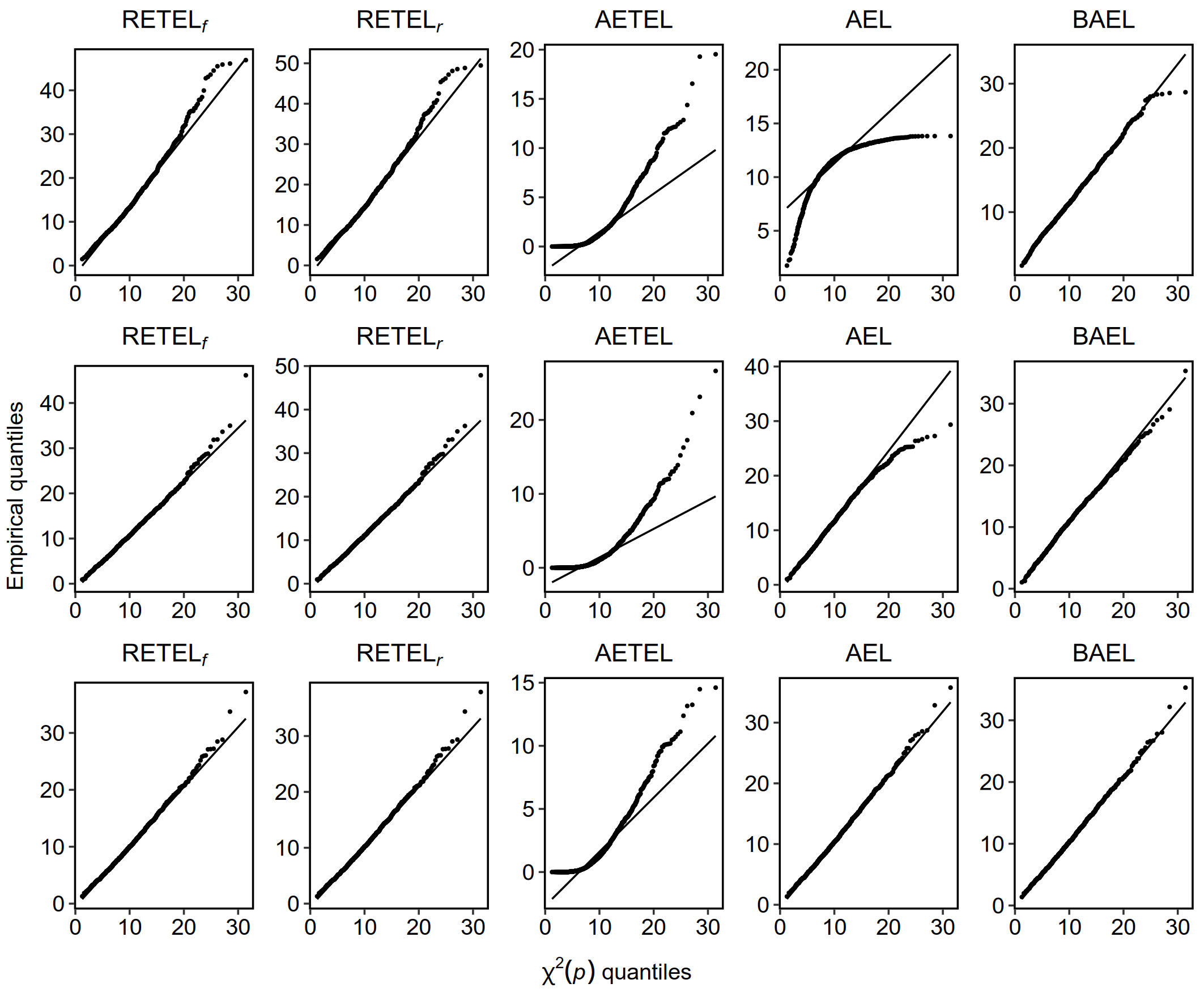}
\caption{
Quantile-quantile plots for the distribution of minus twice the log-likelihood ratio statistics versus \({\chi^2_p}\) under \({p} = {10}\) for different sample sizes (\({n} = {20}\) in the first row, \({n} = {50}\) in the second row, and \({n} = {100}\) in the third row).
}
\label{fig:qq_p10}
\end{figure}

\begin{figure}[!t]
\centering
\includegraphics[width=\linewidth]{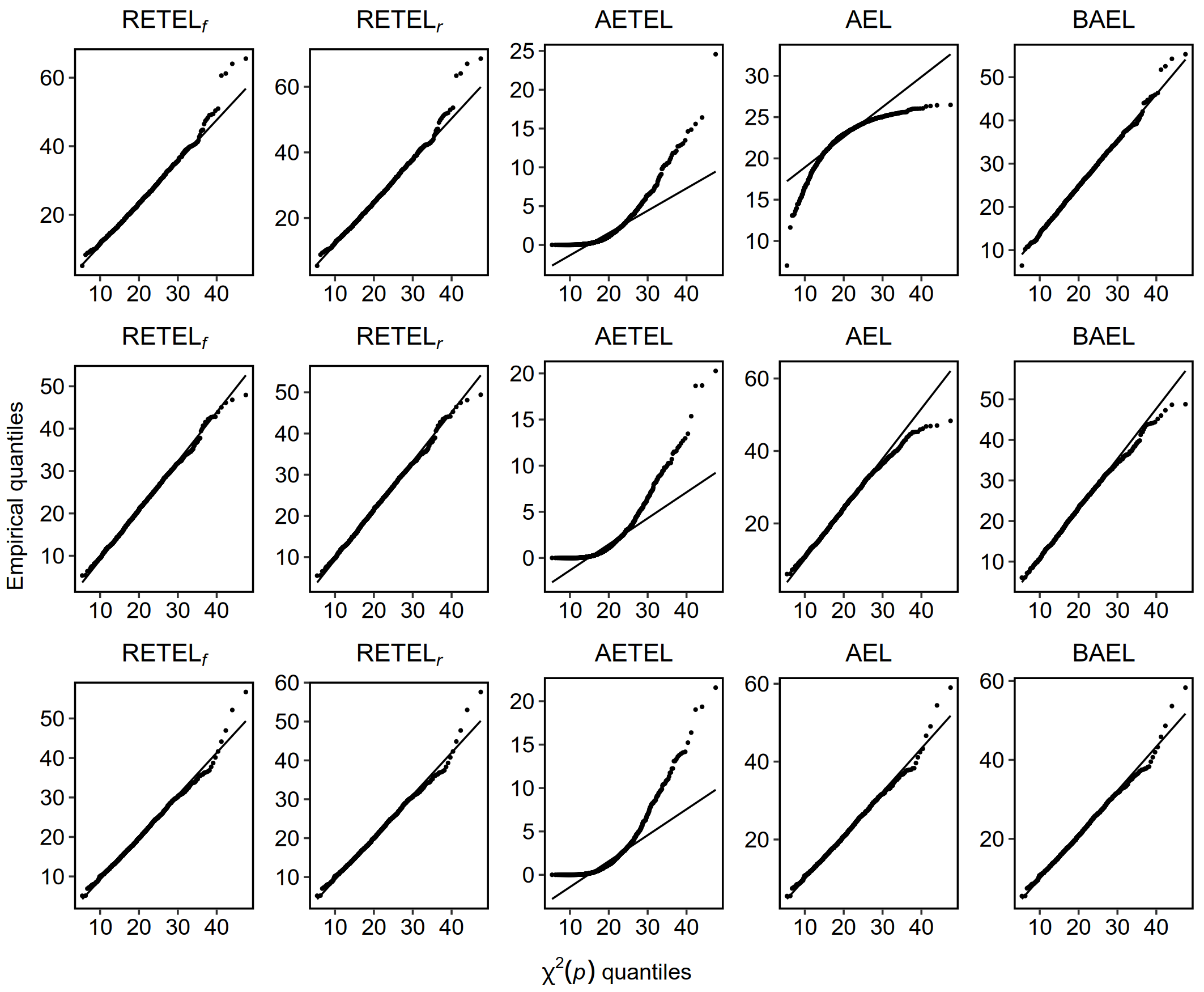}
\caption{
Quantile-quantile plots for the distribution of minus twice the log-likelihood ratio statistics versus \({\chi^2_p}\) under \({p} = {20}\) for different sample sizes (\({n} = {40}\) in the first row, \({n} = {100}\) in the second row, and \({n} = {200}\) in the third row).
}
\label{fig:qq_p20}
\end{figure}

\begin{figure}[!t]
\centering
\includegraphics[width=\linewidth]{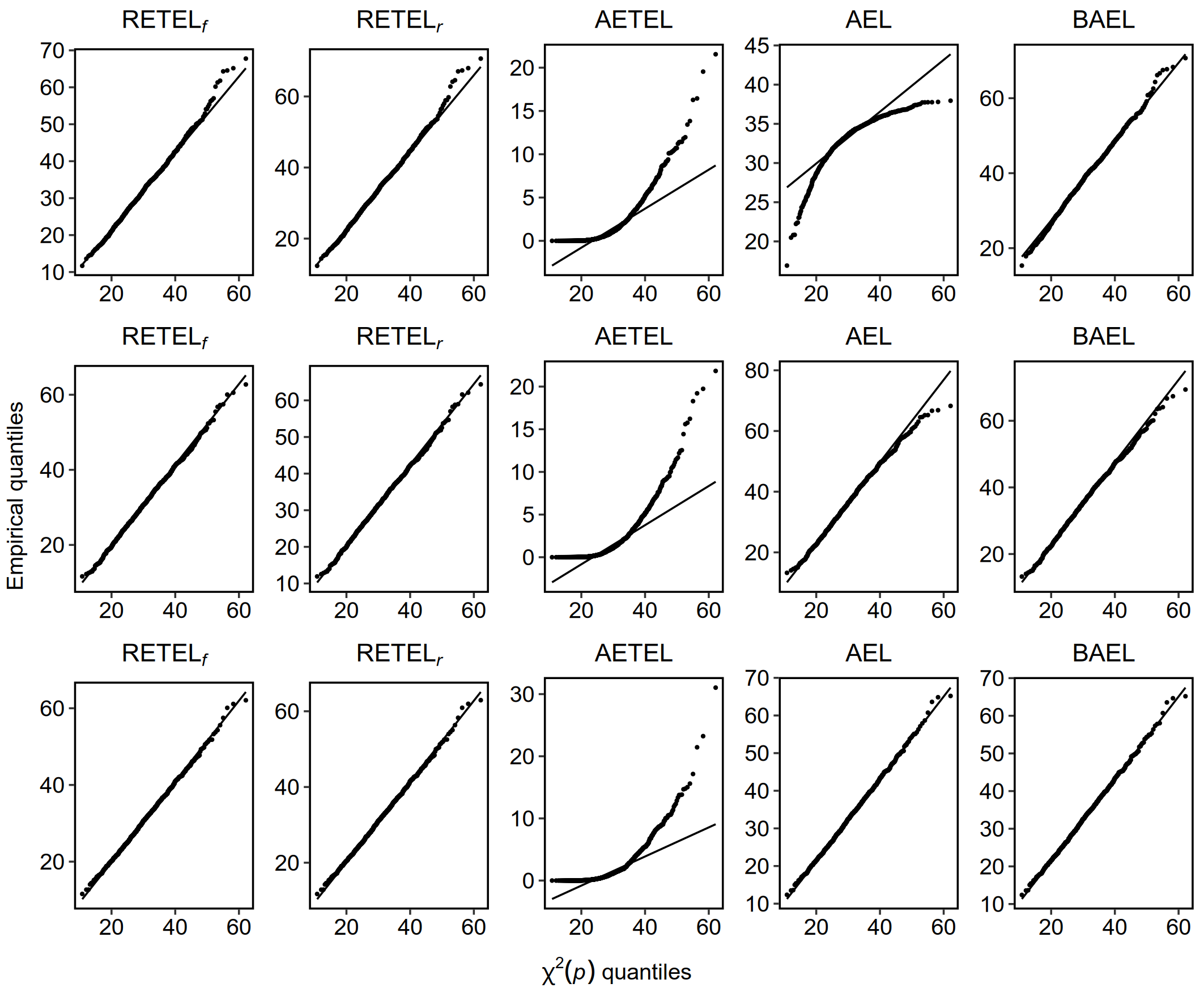}
\caption{
Quantile-quantile plots for the distribution of minus twice the log-likelihood ratio statistics versus \({\chi^2_p}\) under \({p} = {30}\) for different sample sizes (\({n} = {60}\) in the first row, \({n} = {150}\) in the second row, and \({n} = {300}\) in the third row).
}
\label{fig:qq_p30}
\end{figure}

\cref{fig:qq_tau1_s1,fig:qq_tau1_s5,fig:qq_taulogn_s1,fig:qq_taulogn_s5} show quantile-quantile plots comparing the distribution of \({H}\) to \({U(0, 1)}\) from the simulations in Section 5.1 of the main paper.
\cref{fig:qq_p10,fig:qq_p20,fig:qq_p30} show quantile-quantile plots comparing the distribution of minus twice the log-likelihood ratio statistics to \({\chi^2_p}\) from the simulations in Section 5.3 of the main paper.

\clearpage
\section{Computational Environment and Empirical Runtimes}
All numerical illustrations in Sections~5 and~6 of the main paper are conducted on a MacBook Pro equipped with an Apple M2 Pro processor (10-core CPU, up to 3.5 GHz) and 16 GB of unified memory. 
The computational environment uses R version 4.5.0 running on macOS Tahoe 26.2. 
All reported computation times are measured using a single CPU core in a controlled, single-user environment. 
\cref{tab:mb,tab:cr_l0,tab:cr_l2,tab:kl,tab:cr_mvnorm,tab:cr_lm,tab:application} summarize the empirical wall-clock runtimes for the simulation studies (Section~5) and the data application (Section~6).

Throughout, EL and its variants exhibit lower total runtimes than ETEL, including RETEL. 
We emphasize that this disparity primarily reflects differences in software implementation rather than underlying algorithmic efficiency. 
The EL implementation (and its variants) utilizes a compiled Newton--Raphson solver written in \texttt{C++}, 
whereas ETEL (and its variants) relies on \texttt{R}'s L-BFGS optimization routines.
In this relatively low-dimensional setting, the total runtime is dominated by the interfacing overhead of the \texttt{R} interpreter during the iterative optimization process. 
Furthermore, the marginal increase in computation time for RETEL compared to ETEL is attributed to the additional evaluations required for the regularization term within the objective function.
The relatively longer runtimes for EEL in \cref{tab:cr_lm} are due to the composite similarity mapping, which requires repeated EL evaluations along a secant line.

\clearpage
\pagebreak
\vspace*{\fill}
\begin{table}[ht]
\setlength{\tabcolsep}{4.95pt}
\centering
\def~{\hphantom{0}}
\caption{
Runtimes for the simulation study in Section~5.1 (Kolmogorov--Smirnov test).
Entries are reported as the mean (standard deviation) in seconds, based on 10{,}000 replications.
}
{\begin{tabularx}{\textwidth}{cc|YYYYYY}
\toprule
\multicolumn{2}{c}{} &
\multicolumn{2}{c}{\({\tau_n} = {1}\)} &
\multicolumn{2}{c}{\({\tau_n} = {\log n}\)} &
\multicolumn{2}{c}{}\\  
\midrule
\({n}\) & \({s}\) & \({\textnormal{RETEL}_f}\) & \({\textnormal{RETEL}_r}\) & \({\textnormal{RETEL}_f}\) & \({\textnormal{RETEL}_r}\) & ETEL & AETEL\\
\midrule
\multirow{2}{*}{\({5}\)} 
& \({1}\) 
& 1.10 (0.04)
& 1.07 (0.03)
& 1.08 (0.03) 
& 1.11 (0.07)
& 0.88 (0.25) 
& 0.88 (0.03) \\ 
& \({5}\) 
& 1.10 (0.03)
& 1.08 (0.03)
& 1.08 (0.07)
& 1.10 (0.04)
& 0.88 (0.25) 
& 0.88 (0.03) \\[1ex] 
\multirow{2}{*}{\({20}\)} 
& \({1}\) 
& 1.09 (0.03) 
& 1.07 (0.04)
& 1.08 (0.03)  
& 1.09 (0.05)
& 0.92 (0.03) 
& 0.88 (0.04) \\ 
& \({5}\) 
& 1.10 (0.03) 
& 1.08 (0.04) 
& 1.08 (0.04)
& 1.12 (0.08) 
& 0.92 (0.04)
& 0.88 (0.03) \\[1ex]
\multirow{2}{*}{\({50}\)} 
& \({1}\) 
& 1.09 (0.03) 
& 1.09 (0.03)
& 1.08 (0.03) 
& 1.11 (0.07)  
& 0.93 (0.04)
& 0.88 (0.04) \\ 
& \({5}\) 
& 1.09 (0.04)
& 1.09 (0.03)
& 1.07 (0.04)
& 1.10 (0.06) 
& 0.93 (0.05)
& 0.89 (0.06) \\[1ex]
\multirow{2}{*}{\({100}\)} 
& \({1}\) 
& 1.10 (0.04)
& 1.09 (0.04)
& 1.08 (0.03)
& 1.14 (0.08) 
& 0.93 (0.03)
& 0.88 (0.03) \\ 
& \({5}\) 
& 1.07 (0.04)
& 1.10 (0.03) 
& 1.07 (0.03) 
& 1.12 (0.06)  
& 0.93 (0.03)
& 0.88 (0.03) \\
\bottomrule
\end{tabularx}
}
\label{tab:mb}
\end{table}
\vspace*{\fill}

\clearpage
\vspace*{\fill}
\begin{table}[!h]
\def~{\hphantom{0}}
\caption{
Runtimes for the simulation study in Section~5.1 (posterior credible intervals with \({l} = {0}\)).
Entries are reported as the mean (standard deviation) in seconds, based on 10{,}000 replications.
}
{\begin{tabularx}{\textwidth}{cc|YYYY}
\toprule
\({n}\) & \({s}\) & \({\textnormal{RETEL}_f}\) &
\({\textnormal{RETEL}_r}\) &
ETEL &
AETEL\\
\midrule
\multirow{3}{*}{\({5}\)} 
& \({0.5}\) 
& 1.10 (0.04) & 1.10 (0.04) & 0.96 (0.07) & 0.89 (0.05) \\ 
& \({1}\) 
& 1.10 (0.03) & 1.09 (0.04) & 0.95 (0.05) & 0.89 (0.04) \\ 
& \({5}\) 
& 1.11 (0.04) & 1.10 (0.05) & 0.97 (0.08) & 0.89 (0.03) \\[1ex] 
\multirow{3}{*}{\({20}\)} 
& \({0.5}\) 
& 1.10 (0.05) & 1.10 (0.05) & 0.93 (0.05) & 0.88 (0.04) \\ 
& \({1}\) 
& 1.11 (0.04) & 1.09 (0.04) & 0.91 (0.04) & 0.88 (0.04) \\ 
& \({5}\) 
& 1.10 (0.04) & 1.09 (0.05) & 0.92 (0.04) & 0.88 (0.06) \\[1ex] 
\multirow{3}{*}{\({50}\)} 
& \({0.5}\) 
& 1.09 (0.04) & 1.09 (0.04) & 0.93 (0.05) & 0.88 (0.06) \\ 
& \({1}\) 
& 1.09 (0.04) & 1.09 (0.04) & 0.95 (0.07) & 0.89 (0.08) \\ 
& \({5}\) 
& 1.10 (0.04) & 1.09 (0.04) & 0.96 (0.10) & 0.92 (0.08) \\[1ex] 
\multirow{3}{*}{\({100}\)} 
& \({0.5}\) 
& 1.10 (0.03) & 1.10 (0.04) & 0.94 (0.06) & 1.00 (0.12) \\ 
& \({1}\) 
& 1.10 (0.04) & 1.10 (0.05) & 0.96 (0.05) & 1.02 (0.09) \\ 
& \({5}\) 
& 1.09 (0.04) & 1.10 (0.05) & 0.97 (0.07) & 1.05 (0.12)\\
\bottomrule
\end{tabularx}}
\label{tab:cr_l0}
\end{table}

\begin{table}[!ht]
\def~{\hphantom{0}}
\caption{
Runtimes for the simulation study in Section~5.1 (posterior credible intervals with \({l} = {2}\)).
Entries are reported as the mean (standard deviation) in seconds, based on 10{,}000 replications.
}
{\begin{tabularx}{\textwidth}{cc|YYYY}
\toprule
\({n}\) & \({s}\) & \({\textnormal{RETEL}_f}\) &
\({\textnormal{RETEL}_r}\) &
ETEL &
AETEL\\
\midrule
\multirow{3}{*}{\({5}\)} 
& \({0.5}\) 
& 1.14 (0.09) & 1.09 (0.05) & 0.95 (0.05) & 0.99 (0.08) \\ 
& \({1}\) 
& 1.12 (0.07) & 1.09 (0.04) & 0.96 (0.08) & 0.99 (0.10) \\ 
& \({5}\) 
& 1.11 (0.06) & 1.13 (0.07) & 0.94 (0.05) & 0.92 (0.05) \\[1ex] 
\multirow{3}{*}{\({20}\)} 
& \({0.5}\) 
& 1.13 (0.06) & 1.13 (0.07) & 0.97 (0.07) & 0.97 (0.09) \\ 
& \({1}\) 
& 1.21 (0.08) & 1.14 (0.10) & 0.97 (0.07) & 0.98 (0.09) \\ 
& \({5}\) 
& 1.14 (0.07) & 1.12 (0.06) & 0.98 (0.07) & 0.98 (0.09) \\[1ex] 
\multirow{3}{*}{\({50}\)} 
& \({0.5}\) 
& 1.13 (0.06) & 1.09 (0.05) & 0.94 (0.06) & 0.99 (0.08) \\ 
& \({1}\) 
& 1.13 (0.05) & 1.14 (0.07) & 0.99 (0.09) & 0.99 (0.08) \\ 
& \({5}\) 
& 1.11 (0.04) & 1.12 (0.06) & 0.97 (0.07) & 0.97 (0.25) \\[1ex] 
\multirow{3}{*}{\({100}\)} 
& \({0.5}\) 
& 1.13 (0.06) & 1.11 (0.04) & 0.96 (0.11) & 0.92 (0.04) \\ 
& \({1}\) 
& 1.14 (0.06) & 1.10 (0.04) & 0.93 (0.04) & 0.94 (0.05) \\ 
& \({5}\) 
& 1.17 (0.09) & 1.10 (0.03) & 0.93 (0.03) & 0.92 (0.06) \\
\bottomrule
\end{tabularx}}
\label{tab:cr_l2}
\end{table}
\vspace*{\fill}

\clearpage
\vspace*{\fill}
\begin{table}[!ht]
\def~{\hphantom{0}}
\caption{
Runtimes for the simulation study in Section~5.2 (expected Kullback--Leibler divergence).
Entries are reported as the mean (standard deviation) in seconds, based on 1{,}000 replications.
}
\centering
\begin{tabularx}{\textwidth}{c|YYYY}
\toprule
\({n}\) & \({\textnormal{RETEL}_f}\) & \({\textnormal{RETEL}_r}\) & EL & ETEL \\
\midrule
2  & 46.39 (1.02) & 45.71 (1.38) & 7.66 (0.19) & 32.27 (3.95) \\
4  & 45.78 (1.45) & 44.67 (0.44) & 7.40 (0.13) & 35.71 (1.75) \\
6  & 45.76 (1.09) & 44.63 (0.41) & 7.66 (0.34) & 36.67 (1.09) \\
8  & 45.11 (0.86) & 44.59 (0.45) & 7.60 (0.31) & 37.64 (1.11) \\
10 & 44.87 (1.18) & 44.49 (0.52) & 7.42 (0.15) & 39.16 (0.96) \\
\bottomrule
\end{tabularx}
\label{tab:kl}
\end{table}
\vspace*{\fill}

\clearpage
\vspace*{\fill}
\begin{table}[!ht]
\def~{\hphantom{0}}
\caption{
Runtimes for the simulation study in Section~5.3 (confidence regions for the mean).
Entries are reported as the total time in seconds for all 1{,}000 replications.
}
{\begin{tabularx}{\textwidth}{cc|YYYYY}
\toprule
\({p}\) & \({n}\) &
\({\textnormal{RETEL}_f}\) &
\({\textnormal{RETEL}_r}\) &
AETEL &
AEL &
BAEL\\  
\midrule
\multirow{3}{*}{\({10}\)} 
& \({20}\) 
& \({1.50}\) 
& \({1.47}\) 
& \({1.04}\) 
& \({0.43}\) 
& \({0.50}\) \\ 
& \({50}\) 
& \({1.49}\) 
& \({1.51}\) 
& \({1.10}\) 
& \({0.52}\) 
& \({0.55}\) \\ 
& \({100}\) 
& \({1.50}\) 
& \({1.48}\) 
& \({1.20}\) 
& \({0.49}\) 
& \({0.55}\) \\[1ex] 
\multirow{3}{*}{\({20}\)} 
& \({40}\) 
& \({1.80}\) 
& \({1.73}\) 
& \({1.14}\) 
& \({0.51}\) 
& \({0.67}\) \\ 
& \({100}\) 
& \({1.69}\) 
& \({1.69}\) 
& \({1.27}\) 
& \({0.61}\) 
& \({0.77}\) \\ 
& \({200}\) 
& \({1.95}\) 
& \({1.90}\) 
& \({1.54}\) 
& \({0.78}\) 
& \({0.91}\) \\[1ex] 
\multirow{3}{*}{\({30}\)} 
& \({60}\) 
& \({2.03}\) 
& \({2.05}\) 
& \({1.27}\) 
& \({0.66}\) 
& \({0.81}\) \\ 
& \({150}\) 
& \({2.24}\) 
& \({2.26}\) 
& \({1.70}\) 
& \({0.97}\) 
& \({1.23}\) \\ 
& \({300}\) 
& \({2.76}\) 
& \({2.71}\) 
& \({2.23}\) 
& \({1.35}\) 
& \({1.67}\) \\
\bottomrule
\end{tabularx}}
\label{tab:cr_mvnorm}
\end{table}

\begin{table}[!ht]
\def~{\hphantom{0}}
\caption{
Runtimes for the simulation study in Section~5.3 (confidence regions for regression). 
Entries are reported as the total time in seconds for all 1{,}000 replications.
}
{\begin{tabularx}{\textwidth}{cc|YYYYY}
\toprule
\({p}\) & \({n}\) &
\({\textnormal{RETEL}_f}\) &
\({\textnormal{RETEL}_r}\) &
AETEL &
AEL &
EEL\\  
\midrule
\multirow{3}{*}{\({10}\)} 
& \({50}\) 
& \({1.61}\) 
& \({1.77}\) 
& \({1.25}\) 
& \({0.38}\) 
& \({2.43}\) \\ 
& \({100}\) 
& \({1.66}\) 
& \({1.69}\) 
& \({1.40}\) 
& \({0.35}\) 
& \({3.28}\) \\ 
& \({200}\) 
& \({1.83}\) 
& \({1.85}\) 
& \({1.58}\) 
& \({0.42}\) 
& \({3.06}\) \\[1ex] 
\multirow{3}{*}{\({20}\)} 
& \({100}\) 
& \({2.18}\) 
& \({2.18}\) 
& \({1.59}\) 
& \({0.50}\) 
& \({3.51}\) \\ 
& \({200}\) 
& \({2.49}\) 
& \({2.43}\) 
& \({1.86}\) 
& \({0.67}\) 
& \({4.49}\) \\ 
& \({400}\) 
& \({2.97}\) 
& \({3.69}\) 
& \({2.29}\) 
& \({1.11}\) 
& \({6.78}\) \\[1ex] 
\multirow{3}{*}{\({30}\)} 
& \({150}\) 
& \({2.93}\) 
& \({2.84}\) 
& \({2.17}\) 
& \({0.92}\) 
& \({5.51}\) \\ 
& \({300}\) 
& \({3.69}\) 
& \({3.66}\) 
& \({2.96}\) 
& \({1.43}\) 
& \({8.74}\) \\ 
& \({600}\) 
& \({5.92}\) 
& \({5.71}\) 
& \({4.59}\) 
& \({3.12}\) 
& \({14.40~}\) \\
\bottomrule
\end{tabularx}}
\label{tab:cr_lm}
\end{table}
\vspace*{\fill}

\clearpage
\vspace*{\fill}
\begin{table}[!ht]
\centering
\def~{\hphantom{0}}
\caption{Runtimes for the application in Section~6.1 (median income data).}
\begin{tabularx}{\textwidth}{c|YY}
\toprule
Method & Total wall-clock time\({^\dagger}\) & Time per iteration\({^\ddagger}\)\\
\midrule
\({\textnormal{RETEL}_f}\) & 50.7 minutes & 3.04 milliseconds\\
\({\textnormal{RETEL}_r}\) & 50.8 minutes & 3.05 milliseconds\\
EL                         & 11.5 minutes & 0.69 milliseconds\\
ETEL                       & 44.2 minutes & 2.65 milliseconds\\
\bottomrule
\multicolumn{3}{l}{\small \({^\dagger}\) Total time for four chains of 250,000 iterations each, run sequentially on one core.} \\
\multicolumn{3}{l}{\small \({^\ddagger}\) Calculated as the average time per iteration.} 
\end{tabularx}
\label{tab:application}
\end{table}
\vspace*{\fill}

\clearpage
\bibliographystyle{plainnat}
\bibliography{bibliography}